\documentclass[draftclsnofoot,onecolumn,12pt]{IEEEtran}

\usepackage{comment}
\usepackage{color}

\usepackage{textcomp}
\usepackage{tabularx}

\usepackage{graphicx,epsf,psfrag}
\usepackage{amsmath,amssymb}
\usepackage{amsfonts}
\usepackage{mathrsfs}
\usepackage{etoolbox}
\usepackage{cite}

\usepackage{latexsym}

\newcommand{\beq}{\begin{equation}}
\newcommand{\eeq}{\end{equation}}
\newcommand{\be}{\begin{equation}}
\newcommand{\ee}{\end{equation}}
\newcommand{\eps}{\epsilon}

\newcommand{\bi}{\begin{itemize}}
\newcommand{\ei}{\end{itemize}}

\newcommand{\calA}{\mathcal{A}}
\newcommand{\calB}{\mathcal{B}}
\newcommand{\calC}{\mathcal{C}}
\newcommand{\calD}{\mathcal{D}}
\newcommand{\calE}{\mathcal{E}}

\newcommand{\calG}{\mathcal{G}}

\newcommand{\calN}{\mathcal{N}}

\newcommand{\calP}{\mathcal{P}}
\newcommand{\calQ}{\mathcal{Q}}
\newcommand{\calR}{\mathcal{R}}
\newcommand{\calS}{\mathcal{S}}
\newcommand{\calT}{\mathcal{T}}

\newcommand{\calV}{\mathcal{V}}
\newcommand{\calW}{\mathcal{W}}
\newcommand{\calX}{\mathcal{X}}
\newcommand{\calY}{\mathcal{Y}}

\newcommand{\bR}{\mathbf{R}}

\newcommand{\bw}{\mathbf{w}}
\newcommand{\bW}{\mathbf{W}}
\newcommand{\bx}{\mathbf{x}}
\newcommand{\bX}{\mathbf{X}}
\newcommand{\by}{\mathbf{y}}
\newcommand{\bY}{\mathbf{Y}}

\newcommand{\bbE}{\mathbb{E}}

\newcommand{\bbN}{\mathbb{N}}

\newcommand{\bbP}{\mathbb{P}}

\newcommand{\bbR}{\mathbb{R}}

\DeclareMathAlphabet{\mathbsf}{OT1}{cmss}{bx}{n}
\DeclareMathAlphabet{\mathssf}{OT1}{cmss}{m}{sl}

\DeclareSymbolFont{bsfletters}{OT1}{cmss}{bx}{n}  
\DeclareSymbolFont{ssfletters}{OT1}{cmss}{m}{n}
\DeclareMathSymbol{\bsfGamma}{0}{bsfletters}{'000}
\DeclareMathSymbol{\ssfGamma}{0}{ssfletters}{'000}
\DeclareMathSymbol{\bsfDelta}{0}{bsfletters}{'001}
\DeclareMathSymbol{\ssfDelta}{0}{ssfletters}{'001}
\DeclareMathSymbol{\bsfTheta}{0}{bsfletters}{'002}
\DeclareMathSymbol{\ssfTheta}{0}{ssfletters}{'002}
\DeclareMathSymbol{\bsfLambda}{0}{bsfletters}{'003}
\DeclareMathSymbol{\ssfLambda}{0}{ssfletters}{'003}
\DeclareMathSymbol{\bsfXi}{0}{bsfletters}{'004}
\DeclareMathSymbol{\ssfXi}{0}{ssfletters}{'004}
\DeclareMathSymbol{\bsfPi}{0}{bsfletters}{'005}
\DeclareMathSymbol{\ssfPi}{0}{ssfletters}{'005}
\DeclareMathSymbol{\bsfSigma}{0}{bsfletters}{'006}
\DeclareMathSymbol{\ssfSigma}{0}{ssfletters}{'006}
\DeclareMathSymbol{\bsfUpsilon}{0}{bsfletters}{'007}
\DeclareMathSymbol{\ssfUpsilon}{0}{ssfletters}{'007}
\DeclareMathSymbol{\bsfPhi}{0}{bsfletters}{'010}
\DeclareMathSymbol{\ssfPhi}{0}{ssfletters}{'010}
\DeclareMathSymbol{\bsfPsi}{0}{bsfletters}{'011}
\DeclareMathSymbol{\ssfPsi}{0}{ssfletters}{'011}
\DeclareMathSymbol{\bsfOmega}{0}{bsfletters}{'012}
\DeclareMathSymbol{\ssfOmega}{0}{ssfletters}{'012}

\newcommand{\tilR}{\tilde{R}}

\newcommand{\hatw}{\hat{w}}
\newcommand{\hatW}{\hat{W}}
\newcommand{\tilw}{\tilde{w}}
\newcommand{\tilW}{\tilde{W}}

\newcommand{\tilX}{\tilde{X}}

\newcommand{\tily}{\tilde{y}}
\newcommand{\tilY}{\tilde{Y}}

\ifcsmacro{theorem}{}{
\newtheorem{theorem}{Theorem}
\newtheorem{lemma}[theorem]{Lemma}
\newtheorem{proposition}[theorem]{Proposition}

\newtheorem{definition}{Definition} 
 
\newtheorem{remark}{Remark}

}

\newcommand{\qednew}{\nobreak \ifvmode \relax \else
      \ifdim\lastskip<1.5em \hskip-\lastskip
      \hskip1.5em plus0em minus0.5em \fi \nobreak
      \vrule height0.75em width0.5em depth0.25em\fi}

\newcommand{\uy}{\underline{y}}

\newcommand{\uw}{\underline{w}}
\newcommand{\uX}{\underline{X}}
\newcommand{\uY}{\underline{Y}}
\newcommand{\uZ}{\underline{Z}}
\newcommand{\uW}{\underline{W}}

\newcommand{\edit}[1]{#1}
\newcommand{\oin}{\edit{=}}
\newcommand{\nedit}[1]{#1}
\renewcommand{\calC}{\nedit{\mathcal{C}}}

\allowdisplaybreaks

\title{Strong Converses Are Just Edge Removal Properties}

\author{Oliver Kosut, \IEEEmembership{Member, IEEE} and J{\"o}rg Kliewer, \IEEEmembership{Senior Member, IEEE}
\thanks{O.~Kosut is with the School of Electrical, Computer and Energy Engineering, Arizona State University, Tempe, AZ 85287 USA (email: \hbox{okosut@asu.edu}).}
\thanks{J. Kliewer is with the Department of Electrical and Computer Engineering, New Jersey Institute of Technology, Newark, NJ 07102 USA (email: \hbox{jkliewer@njit.edu}).}
\thanks{This work was presented in part at the 2016 IEEE International Symposium on Information Theory.}
\thanks{This material is based upon work supported by the National Science
  Foundation under Grant No. CCF-1439465, CCF-1440014, CNS-1526547, CCF-1453718.}
}

\begin{document}

\maketitle


\begin{abstract}
This paper explores the relationship between two ideas in network information theory: edge removal and strong converses. Edge removal properties state that if an edge of small capacity is removed from a network, the capacity region does not change too much. Strong converses state that, for rates outside the capacity region, the probability of error converges to 1 \edit{as the blocklength goes to infinity}. Various notions of edge removal and strong converse are defined, depending on how edge capacity and error probability scale with blocklength, and relations between them are proved. Each class of strong converse implies a specific class of edge removal. The opposite directions are proved for deterministic networks. Furthermore, a technique based on a novel, causal version of the blowing-up lemma is used to prove that for discrete memoryless networks, the weak edge removal property---that the capacity region changes continuously as the capacity of an edge vanishes---is equivalent to the exponentially strong converse---that outside the capacity region, the probability of error goes to 1 exponentially fast. This result is used to prove exponentially strong converses for several examples, including the discrete 2-user interference channel with strong interference, with only a small variation from traditional weak converse proofs.

\end{abstract}

\section{Introduction}

Consider a general network communication scenario given an arbitrary
collection of sources and sinks connected via an arbitrary network channel. The
sources are independent and each source is demanded by a subset of sinks,
where this subset can be different for each sink. A general interest in
network information theory is to determine the capacity of such networks,
defined as  the set of achievable rates for each source. As this problem is
known to be challenging, we consider the simpler problem of how the capacity of these
networks change if only a single edge is removed from the network. This problem has
first been studied by \cite{ho10,jalali11}. The authors
have shown that for acyclic noiseless networks and a variety of demand types
for which the cut-set bound is tight,  removing an edge of capacity $\delta$
reduces the capacity of each min-cut by at most $\delta$ in each
dimension. Further, in \cite{lee13:_outer} it has been shown for a
noiseless multiple multicast demand that this edge removal property also holds for
generalized network sharing  outer bound \nedit{\cite{kamath11:_gener}}; \edit{for the linear \nedit{programming}  outer bound \nedit{\cite{yeung97}}, \cite{lee13:_outer} shows that  removing an edge of capacity $\delta$ reduces the capacity by at most $K\delta$, where $K$ depends only on the network}. In addition, the existence of the edge removal property has
for example been tied to the problem whether a network coding instance
allows a reconstruction with $\epsilon$ or zero error
\cite{langberg11:_networ_codin,Chan2014}, respectively. Another example is the
connection of edge removal to the equivalency between a network coding
instance and a corresponding index coding problem
\cite{Wong2013}. Recently, it has been shown that for a multiple-access channel with a so
called ``cooperation facilitator'' \cite{Noorzad2014,Noorzad2015,Noorzad2016,Noorzad2016a,Noorzad2018} the edge
removal property does not hold. In particular, for this setting the authors
show the surprising result that adding a small capacity edge can lead to a
significant increase in network capacity. These results have also been extended
to networks with state \cite{Noorzad2017a} and to edges which can carry only
a single bit over all times under the maximal error criterion \cite{Langberg2016}.
However, despite the significant progress that has been made to
understand scenarios in which the edge removal property holds, the solution to the
general problem is open.

In this work, we address the connection of edge removal to the existence of strong converses for networks subject to an average probability of error constraint. As far as we know, this connection has been explored in the literature only briefly in \cite[Chap.~3, \edit{p.~48}]{Gu2009}. The strong
converse theorem states that the error probability converges to 1 for large
blocklengths $n$ if the rate exceeds the capacity. This is in contrast
to a weak converse which only indicates that the error probability is
bounded away from zero if we operate at a rate beyond capacity. The benefit
of a strong converse is that it strengthens the interpretation of capacity
as a sharp phase transition in \edit{achievable} probability of error. \edit{It also allows for} the following interesting
interpretation: if a strong converse exists for a given network instance,
$\epsilon$ reliable codes (i.e., codes which allow reconstruction with
$\epsilon$ error) must have rate tuples within the capacity region for
$\epsilon\in[0,1)$ and large $n$. \edit{Thus, a strong converse refines a capacity (or first-order) result, which provides only the limiting behavior as the probability of error vanishes and the blocklength goes to infinity. However, a strong converse does not provide as much refinement as a second-order (or dispersion) result \cite{Polyanskiy2010}, which clarifies the (usually $O(1/\sqrt{n})$) backoff from capacity for small blocklengths and fixed probability of error.  Therefore, strong converses constitute ``one-and-a-half-th order'' results.}
Strong converses have been 
established for numerous problems, including point-to-point settings, e.g., for discrete memoryless
channels \edit{\cite{Wolfowitz1957}} and quantum channels \edit{\cite{Winter1999,Ogawa1999}}. Recently it has been shown that a strong
converse holds for a discrete memoryless networks with tight
cut-set bounds \cite{Fong2017}.
\edit{There has also been work establishing \emph{exponentially strong converses}, which state that for any rate vector outside the asymptotically-zero error capacity region, the error probability approaches 1 exponentially fast. Exponentially strong converses have been considered for point-to-point channels in \cite{Arimoto1973,Dueck1979}, and for several network problems in \cite{Oohama2015a,Oohama2015b,Oohama2016,Oohama2016a}.
}

In the following, we categorize the notions of edge removal and strong
converses into different classes depending on how edge capacity and
error probability, resp., scale with blocklength, and demonstrate
relations between these instances. See Fig.~\ref{fig:properties} for a
summary of our results. In particular, our contributions are as follows:
\begin{enumerate}
\item We show that each specific class of
strong converse always implies a specific class of edge removal. This
implication holds in great generality: whether the network channel model is
deterministic or probabilistic, discrete or continuous, or even whether it
has memory. 
\item We show that implications in the opposite direction \edit{(edge removal implies strong converse)} hold
in some cases. In particular, we show that each opposite direction holds for
deterministic networks. \edit{However, these opposite directions do not always hold; for example, for a simple discrete memoryless point-to-point channel, each edge removal property holds, but the strongest form of the strong converse---the \emph{extremely strong converse}---does not hold.}
\item We further show that for \emph{all} discrete
memoryless stationary networks, the \emph{exponentially strong converse} is equivalent to the \emph{weak edge removal} property. The weak edge removal property states that \edit{if} a small edge
with rate growing sublinear in the blocklength \edit{is removed}, the \edit{asymptotically-}zero error capacity
region does not change. The proof is based on a novel, \emph{causal} version of the
blowing-up lemma \cite{Marton1986}.
\item We demonstrate that for networks composed of independent point-to-point links with acyclic topology, a similar equivalence
holds for weaker conditions---between the ordinary strong converse and what
we call the \emph{very weak edge removal} property, wherein the edge carries an unbounded number of bits that grows very slowly with blocklength.
\item These results, particularly the equivalence between weak edge removal and the exponentially strong converse, enable us to, without much effort, strengthen many existing computable outer bounds or weak converses to prove that they hold in an exponentially strong sense. We demonstrate this for the cut-set bound, reproducing the result of \cite{Fong2017} to show that for rates outside the region defined by cut-set bound, the probability of error converges to 1 exponentially fast. We also prove exponentially strong converses for discrete broadcast channels, and for the discrete 2-user interference channel with strong interference.
\end{enumerate}

All the above
  mentioned reduction results between edge removal and strong converses
  reveal the surprising fact that for many cases, satisfying edge removal---a condition related only to first-order capacity---implies a seemingly \emph{stronger} ``one-and-a-half-th order'' property, namely the existence of a specific version of a strong converse indicated by the leftward arrows in
  Fig.~\ref{fig:properties}. This highlights again the power of the edge
  removal property.

This paper is organized as follows. We first introduce the model and
definitions of various strong converse and edge removal properties in
Sec.~\ref{sec:model}. After that, in Sec.~\ref{sec:ER_from_SC} we prove that
strong converses imply edge removal properties. The opposite directions for
deterministic networks is then proven in Sec.~\ref{sec:deterministic}. Then,
in Sec.~\ref{sec:DMN} we prove one of the main results in this paper, namely
equivalence between weak edge removal and the exponentially strong converse
for discrete stationary memoryless. We then show equivalence between very
weak edge removal and the ordinary strong converse for networks of
independent point-to-point links in Sec.~\ref{sec:p2p}. After that, in
Sec.~\ref{sec:applications} we derive several applications of our results,
including the cut-set bound, broadcast channels, and interference
channel. Finally, Sec.~\ref{sec:conclusion} offers the conclusions.

\section{Model and Definitions}\label{sec:model}

\edit{We begin by introducing notation to be used throughout the paper. Subsequently we introduce our network model, and formally define the notions of strong converse and edge removal that will be the main focus, while proving some simple properties of these definitions.} \nedit{There are number of subtly different definitions of rate regions: we summarize them in Table~\ref{table:defs} for convenience.}

\emph{Notation:} For an integer $k$ we define $[1:k]=\{1,\ldots,k\}$. All logarithms and exponentials have base $2$. \edit{The notation $(a_n)_n$ represents an infinite sequence of values $a_n$ for each positive integer $n$.} For sequences $\edit{(a_n)_n,(b_n)_n}$, we write $a_n\doteq b_n$ if $\log(a_n)/n$ and $\log(b_n)/n$ have the same limit as $n\to\infty$. Given two probability \edit{distributions} $P$ and $Q$ on the same alphabet $\calX$, the relative entropy \edit{(for discrete distributions)} is given by 
\be
D(P\|Q) = \edit{\sum_{x\in\calX} P(x)\log \frac{P(x)}{Q(x)}}.
\ee
\edit{Given conditional distributions $P_{Y|X}$ and $Q_{Y|X}$, and marginal distribution $R_{X}$, the conditional relative entropy is given by
\be
D(P_{Y|X}\|Q_{Y|X}|R_X)=\sum_{x,y} R_X(x) P_{Y|X}(y|x)\log \frac{P_{Y|X}(y|x)}{Q_{Y|X}(y|x)}.
\ee
}The total variational distance \edit{(for discrete distributions)} is given by
\be
d_{\text{TV}}(P,Q)= \edit{\frac{1}{2} \sum_{x\in\calX} |P(x)-Q(x)|}.
\ee
The Hamming distance between two sequences $x^n,y^n\in\calX^n$ is denoted
\be
d_{\text{H}}(x^n,y^n)=|\{t\in[1:n]: x_t\ne y_t\}|.
\ee
For a set \edit{$\calA\subseteq\bbR^n$}, \edit{$\overline{\calA}$} indicates the closure of \edit{$\calA$} \edit{with respect to the Euclidean distance}. \edit{We denote the set of nonnegative real numbers by $\bbR_+$.} \edit{Given a vector $\bx=(x_1,\ldots,x_n)\in\bbR^n$ and a scalar $\gamma\in\bbR$, we denote the vector-scalar sum as
\be\label{eq:vector_scalar_sum}
\bx+\gamma=(x_1+\gamma,\ldots,x_n+\gamma).
\ee
Given a sets $\calA,\calB\subseteq\bbR^n$ we denote the set sum as
\be
\calA+\calB=\{\bx+\by:\bx\in\calA,\ \by\in\calB\}.
\ee}

\subsection{Network Model}

\begin{table}
\nedit{
\caption{Summary of capacity region definitions}
\label{table:defs}
\renewcommand{\arraystretch}{1.2}
\begin{tabularx}{\linewidth}{r | X}
$\calR_{\calV}(\calN,n,\eps,k)$ & Finite blocklength rate region for network $\calN$\\
\hline
$n$ & Blocklength\\
$\eps$ & Average probability of error\\
$k$ & Number of bits carried by edge $(a,b)$ in the modified network as shown in Fig.~\ref{fig:edge_removal}. If omitted then the network is unmodified (i.e., $k=0$)\\
$\calV$ & Set of nodes in $\calN$ connected to extra nodes $a$ and $b$. If omitted then $\calV=[1:d]$; i.e., $a$ and $b$ connect to all nodes\\
\hline \hline
$\calC_{\calV}(\calN,(\eps_n)_n,(k_n)_n)$ & Asymptotic capacity region for network $\calN$\\
\hline
$(\eps_n)_n$ & Probability of error sequence as a function of blocklength $n$. If replaced by $0^+$ then asymptotically vanishing error probability \\
$(k_n)_n$ & Bit-capacity sequence of edge $(a,b)$ as a function of blocklength $n$. If omitted then the network is unmodified (i.e., $k_n=0$ for all $n$)\\
$\calV$ & See above
\end{tabularx}}
\end{table}

We begin with \edit{a network model for an arbitrary causal network channel}. Many of our results apply only for discrete memoryless networks or deterministic networks, but some basic results apply in much more generality.

Consider a network consisting of $d$ nodes, where node $i\in[1:d]$ wishes to convey a message $W_i$ at rate $R_i$ to a set of destination nodes $\calD_i\subseteq[1:d]$.\footnote{We assume for simplicity that at most one message originates at each node; all results can be easily generalized to the scenario in which multiple messages originate at each node.} The channel model consists of:
\begin{itemize}
\item An input alphabet $\calX_i$ for each $i\in[1:d]$,
\item An output alphabet $\calY_i$ for each $i\in[1:d]$,
\item For each time step $t$, a conditional probability measure
\beq\label{eq:channel_distribution}
\edit{P_{Y_{1t},\ldots,Y_{dt}|Y_{1}^{t-1},\ldots,Y_d^{t-1},X_1^t,\ldots,X_d^t}}.
\eeq
\end{itemize}
\edit{Note that the channel outputs at time $t$ depend on all previous inputs up to time $t$, and all previous outputs up to time $t-1$.}

\begin{definition}A network is \emph{memoryless and stationary} if the probability measure in \eqref{eq:channel_distribution} can be written \nedit{as}
\be
\edit{P_{Y_{1t},\ldots,Y_{dt}|X_{1t},\ldots,X_{dt}}}
\ee
and these distributions are the same for all $t$.
\end{definition}
\begin{definition}
A network is \emph{deterministic} if \edit{the channel outputs at time $t$ are fixed given the channel inputs up to time $t$; i.e., the conditional probability distribution in \eqref{eq:channel_distribution} takes values only in $\{0,1\}$.}
\end{definition}
\begin{definition}
A network is \emph{discrete} if all input and output alphabets are finite sets.\footnote{\edit{While this is technically an incorrect use of ``discrete'', we use it to mean ``finite alphabet'' as this is the usual convention in the literature; see for example \cite[p.~39]{ElGamalKim:11Book}.}}
\end{definition}

For any $\bR=(R_1,\ldots,R_d)\in\bbR_+^d$, an $(\bR,n)$ code consists of\edit{:}
\begin{itemize}
\item \edit{For} each node $i\in[1:d]$ and time $t\in[1:n]$, an encoding function
\be\label{eq:encoding_function}
\phi_{it}:[1:2^{nR_i}]\times\calY_i^{t-1}\to \calX_i,
\ee
\item \edit{For} each $i,j\in[1:d]$ where $j\in\calD_i$, a decoding function
\be\label{eq:decoding_function}
\psi_{ij}:[1:2^{nR_j}]\times \calY_j^n \to  [1:2^{nR_i}].
\ee
\end{itemize}
Assume messages $W_i$ for $i=1,\ldots,d$ are independent and each uniformly distributed over $[1:2^{nR_i}]$. The channel input from node $i$ at time $t$ is given by $X_{it}=\phi_{it}(W_i,Y_i^{t-1})$. For $j\in\calD_i$, the estimate of $W_i$ at node $j$ is given by $\hat{W}_{ij}=\psi_{ij}(W_j,Y_j^n)$. We write $\bW$ for the complete vector of messages, and $\hat{\bW}$ for the complete vector of message estimates. Given \edit{an} $(\bR,n)$ code, the average probability of error is
\be
\nedit{\mathrm{P}_{\mathrm{e}}^{(n)}}=\bbP(\hat{\bW}\ne\bW)
\ee
where $\hat{\bW}\ne\bW$ denotes the event that \edit{there exists a node $i$ and a message index $j$ such that node $i$ decodes message $j$ incorrectly; that is,} $\hatW_{ij}\ne W_j$ for \edit{any} $i\in[1:d]$, $j\in\calD_i$. For blocklength $n$ and $\eps\in[0,1]$, let \edit{$\calR(\calN,n,\eps)\subseteq\bbR_+^d$} be the set of rates $\bR$ for which there exists an $(\bR,n)$ code with \edit{average} probability of error at most $\eps$.\footnote{\edit{We allow for any $\eps\in[0,1]$ in our definitions for maximum generality, even though $\eps=1$ is a trivial case in which the rate region is unbounded.}} \edit{Given a sequence $(\eps_n)_n$ where $\eps_n\in[0,1]$ for all $n\in\bbN$,} we say a rate vector $\bR$ is \emph{achievable with respect to }\edit{$(\eps_n)_n$} if there exists an integer $n_0$ such that for all $n\ge n_0$, \edit{$\bR\in\calR(\calN,n,\eps_{n})$}. The capacity region \edit{$\calC(\calN,(\eps_n)_n)$} is given by the closure of the set of all achievable rate vectors with respect to $(\eps_n)_n$. \edit{Alternatively, we may define
\be\label{eq:capacity_region_def}
\calC(\calN,(\eps_n)_n)=\overline{\bigcup_{n_0\in\bbN}\ \bigcap_{n\ge n_0} \calR(\calN,n,\eps_{n})}.
\ee}%
\nedit{Throughout the paper, we use $\calR$ to denote a finite blocklength region, and $\calC$ to denote an asymptotic region. (Table~\ref{table:defs} summarizes this notation.)}
Note that $\calR(\calN,n,\eps)$ is defined as a function of the single value $\eps$, whereas \edit{$\calC(\calN,(\eps_n)_n)$} is a function of the infinite sequence \edit{$(\eps_n)_n$}.

In principle \edit{$\calC(\calN,(\eps_n)_n)$} is defined for any sequence \edit{$(\eps_n)_n$}. However, it will be useful to restrict ourselves to sequences for which $-\frac{1}{n}\log(1-\eps_n)$ has a limit; the following proposition, proved in Appendix~\ref{appendix:error_sequences}, shows that we may do this without loss of generality for memoryless stationary networks.

\begin{proposition}\label{prop:error_sequences}
Let $\calN$ be any memoryless stationary network. For any $\alpha>0$, let \edit{$(\eps_n)_n$ and $(\tilde\eps_n)_n$} be two sequences where
\be\label{eq:liminf_equality}
\alpha=\liminf_{n\to\infty} -\frac{1}{n}\log(1-\eps_n)=\liminf_{n\to\infty} -\frac{1}{n}\log(1-\tilde\eps_n).
\ee
Then \edit{$\calC(\calN,(\eps_n)_n)=\calC(\calN,(\tilde\eps_n)_n)$}.
\end{proposition}

As consequence of Proposition~\ref{prop:error_sequences}, for any sequence \edit{$(\eps_n)_n$} where $\alpha=\liminf_{n\to\infty} -\frac{1}{n}\log (1-\eps_n)>0$, \edit{$\calC(\calN,(\eps_n)_n)=\calC(\calN,(1-\exp\{-n\alpha\})_n)$}. Thus, it is enough to focus on sequences \edit{$(\eps_n)_n$} where either $\eps_n=1-\exp\{-n\alpha\}$ for some $\alpha>0$, or $-\log(1-\eps_n)\oin o(n)$. Note that the latter includes any sequence converging to a constant in $[0,1)$.

For fixed $\eps$, \edit{$\calC(\calN,(\eps)_n)$} denotes the capacity region with asymptotic error probability $\eps$. With some abuse of notation, define the usual asymptotically-zero-error capacity region as
\be\label{eq:zero_error_def1}
\calC(\calN,0^+)=\bigcap_{\eps>0} \calC(\calN,\edit{(\eps)_n}).
\ee
Equivalently we may write
\be\label{eq:zero_error_def2}
\calC(\calN,0^+)=\bigcup_{\eps_n\oin o(1)} \calC(\calN,\edit{(\eps_n)_n}).
\ee

\begin{remark}
Using average probability of error rather than maximal probability of error in our definition of capacity region is not merely convenient; it is critical to many of our results. Indeed, it is illustrated in \cite{Langberg2016,Noorzad2018} that edge removal characteristics are very different with maximal probability of error rather than average, and thus the relationship between edge removal and strong converses in the maximal probability of error context is likely to be different.
\end{remark}

\begin{figure}
\begin{center}
\includegraphics[scale=1.25]{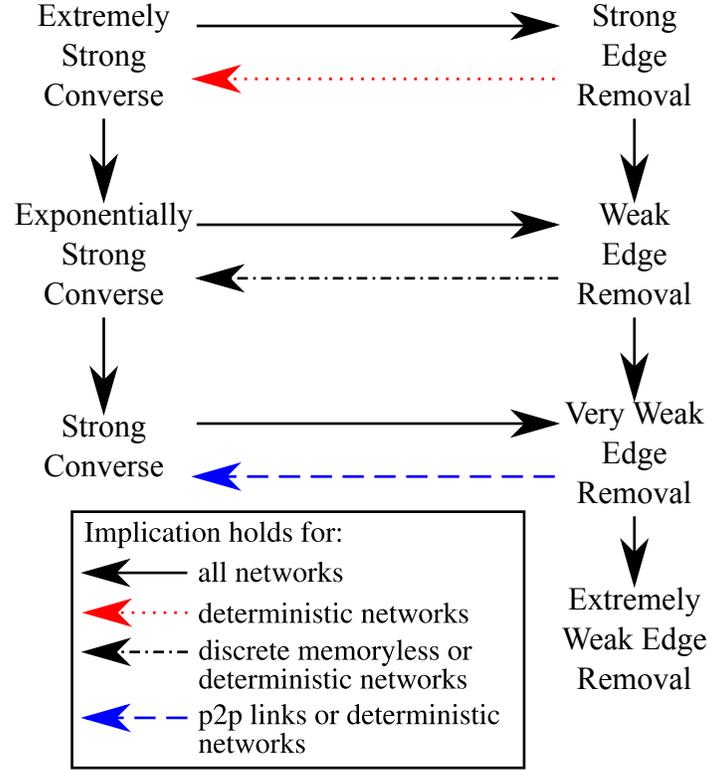}
\end{center}
\caption{Diagram showing the relationships between various strong converses and edge removal properties. Solid black lines represent implications that always hold (Remarks~\ref{remark:strong_converse_ordering} and~\ref{remark:edge_removal_ordering}, and Theorem~\ref{thm:1direction}). All the dashed or dotted lines hold for deterministic networks (Theorem~\ref{thm:deterministic}) but do not hold in general. The red dotted line does not hold even for noisy memoryless stationary networks (Remark~\ref{remark:extremely}). The black dash-dotted line holds for discrete memoryless stationary networks (Theorem~\ref{thm:DMN}). The blue dashed line holds for discrete memoryless stationary networks made up of independent point-to-point links (Theorem~\ref{thm:p2p}), and we conjecture that it holds for all discrete memoryless stationary networks.}
\label{fig:properties}
\end{figure}

We proceed to define 7 different properties: 3 notions of a strong converse and 4 notions of the edge removal property. The relationships that we will prove among these properties are shown in Fig.~\ref{fig:properties}.

\subsection{Strong Converses}

\begin{definition}
Strong converses are defined in terms of whether, for a given constant $\gamma>0$ and a sequence \edit{$(\eps_n)_n$},
\be\label{eq:strong_converse_condition}
\calC(\calN,\edit{(\eps_n)_n})\subseteq\calC(\calN,0^+)+\edit{[0,\gamma]^d}.
\ee
We say network $\calN$ satisfies:
\begin{itemize}
\item the \emph{extremely strong converse} if for all $\gamma>0$, \eqref{eq:strong_converse_condition} holds if $-\log(1-\eps_n)=\frac{\gamma n}{K}$, where $K$ is a positive constant depending only on the network.
\item the \emph{exponentially strong converse} if for all $\gamma>0$, \eqref{eq:strong_converse_condition} holds for some \edit{$(\eps_n)_n$} where $-\log(1-\eps_n)\oin \Theta(n)$.
\item the \emph{strong converse} if for all $\gamma>0$, \eqref{eq:strong_converse_condition} holds for some \edit{$(\eps_n)_n$} where $-\log(1-\eps_n)\to\infty$.
\end{itemize}
\end{definition}

\edit{
\begin{remark}
Statements similar to \eqref{eq:strong_converse_condition} will occur throughout this paper; this condition may be alternatively written as follows: for any $\bR\in \calC(\calN,(\eps_n)_n)$, there exists $\bR'\in\calC(\calN,0^+)$ such that $R_i\le R'_i+\gamma$ for all $i\in[1:d]$.
\end{remark}
}

\begin{remark}\label{remark:strong_converse_ordering}
One can see immediately that the strong converses are ordered by strength; i.e., the extremely strong converse implies the exponentially strong converse, which in turn implies the ordinary strong converse.
\end{remark}

The following proposition gives some equivalent definitions for each of these strong converse properties. It is proved in Appendix~\ref{appendix:strong_converse_defs}.
\begin{proposition}\label{prop:strong_converse_defs}
\begin{enumerate}
\item Network $\calN$ satisfies the extremely strong converse if and only if there exists a constant $K$ depending only on $\calN$ such that either of the following hold:
\begin{enumerate}
\item For any $\bR\notin\calC(\calN,0^+)$, any sequence of $(\bR,n)$ codes has probability of error \edit{$(\eps_n)_n$} satisfying
\be\label{eq:extremely_strong_converse_alternative}
\liminf_{n\to\infty} -\frac{1}{n}\log(1-\eps_n)\ge \frac{\beta}{K}
\ee
where $\beta$ is the smallest number such that $\bR\in\calC(\calN,0^+)+\beta$.
\item For any \edit{sequence $(\eps_n)_n$} where $1-\eps_n\doteq 2^{-n\alpha}$, $\calC(\calN,\edit{(\eps_n)_n})\subseteq\calC(\calN,0^+)+\edit{[0,K\alpha]^d}$.
\end{enumerate}
\item Network $\calN$ satisfies the exponentially strong converse if and only if either of the following hold:
\begin{enumerate}
\item For all $\bR\notin\calC(\calN,0^+)$, any sequence of $(\bR,n)$ codes has probability of error approaching 1 exponentially fast.
\item For any sequence \edit{$(\eps_n)_n$} for which $-\log (1-\eps_n)\oin o(n)$, $\calC(\calN,\edit{(\eps_n)_n})\edit{\subseteq}\calC(\calN,0^+)$.
\end{enumerate}
\item Network $\calN$ satisfies the strong converse if and only if any of the following hold:
\begin{enumerate}
\item For all $\bR\notin\calC(\calN,0^+)$, any sequence of  $(\bR,n)$ codes has probability of error approaching 1 as $n\to\infty$. 
\item For all $\eps\in(0,1)$, $\calC(\calN,\edit{(\eps)_n})=\calC(\calN,0^+)$.
\item There exists a sequence \edit{$\edit(\eps_n)_n$ where $\eps_n\to 1$} and $\calC(\calN,\edit{(\eps_n)_n})=\calC(\calN,0^+)$.
\end{enumerate}
\end{enumerate}
\end{proposition}

\begin{remark}\label{remark:extremely}
Exponential bounds on the probability of success for rates above capacity for point-to-point channels were first considered in \cite{Arimoto1973}. Later, \cite{Dueck1979} exactly characterized the optimal exponent of the success probability for rates above capacity. \edit{Similar results have been found for network problems in \cite{Oohama2015a,Oohama2015b,Oohama2016,Oohama2016a}. For point-to-point channels,} \cite{Dueck1979} showed that for a discrete-memoryless point-to-point channel $P_{Y|X}$ with capacity $C$, for all $R>C$ the optimal probability of error $\eps_n$ satisfies $1-\eps_n\doteq 2^{-\alpha(R) n}$ where
\be\label{eq:above_capacity_exponent}
\alpha(R)=\edit{\min_{Q_{X,Y}} \Big[ D\big(Q_{Y|X}\|P_{Y|X}|Q_X\big)+|R-I_{Q_{X,Y}}(X;Y)|^+\Big]}
\ee
where \edit{$Q_X$ and $Q_{Y|X}$ are the marginal and conditional distributions derived from $Q_{X,Y}$ respectively,} \edit{$I_{Q_{X,Y}}(X;Y)$} is the mutual information between $X$ and $Y$ where $(X,Y)\sim \edit{Q_{X,Y}}$, and $|\cdot|^+$ represents the positive part. Intuitively, \edit{$Q_{Y|X}$} represents an empirical conditional distribution; correct decoding is possible if the channel behaves like one with capacity greater than $R$ (i.e. when the second term in \eqref{eq:above_capacity_exponent} is zero), and the first term in \eqref{eq:above_capacity_exponent} is  the exponential rate of the probability that channel \edit{$P_{Y|X}$} behaves like \edit{$Q_{Y|X}$} with input distribution \edit{$Q_X$}.

This result constitutes an exponentially strong converse in our terminology, since $\alpha(R)>0$ for all $R>C$, but interestingly it is \emph{not} an extremely strong converse for \edit{many} noisy channels. Note that an extremely strong converse is equivalent to $\edit{\frac{d\alpha(R)}{dR}\big|_{R=C}}>0$. However, as we show in the following proposition (proved in Appendix~\ref{appendix:above_capacity}) this holds \edit{only for very specialized channels}.

\begin{proposition}\label{prop:above_capacity}
\edit{Consider a discrete-memoryless point-to-point channel $P_{Y|X}$ with capacity $C$. Let $P_Y$ be the (unique) capacity-achieving output distribution. If
\be\label{eq:above_capacity_condition}
\log \frac{P_{Y|X}(y|x)}{P_Y(y)}\le C\text{ for all }x,y
\ee
then $\alpha(R)=R-C$. Otherwise, $\frac{d\alpha(R)}{dR}\big|_{R=C}=0$.}
\end{proposition}
\edit{Examples of point-to-point channels that satisfy \eqref{eq:above_capacity_condition} include:
\begin{itemize}
\item essentially noiseless channels, i.e., where $C=\log \min\{|\calX|,|\calY|\}$,
\item completely noisy channels, i.e., where $Y$ is independent of $X$,
\item noisy typewriter channels, i.e., where $Y=X+Z$ with summation over some group $\calG$, where $Z$ is uniform on a subset of $\calG$ and independent of $X$.
\end{itemize}
\nedit{Note also that \eqref{eq:above_capacity_condition} implies that the channel dispersion is 0 (cf.~\cite[Thm.~49]{Polyanskiy2010}), but the converse is not true. In particular, the channel dispersion is 0 if and only if there exists a capacity-achieving input distribution $P_X$ such that $\log \frac{P_{Y|X}(y|x)}{P_Y(y)}\le C$ for all $y$ and all $x$ with $P_X(x)>0$. However, \eqref{eq:above_capacity_condition} can fail to hold if  $\log \frac{P_{Y|X}(y|x)}{P_Y(y)}> C$ for some pair $x,y$ even if $P_X(x)=0$ for all capacity-achieving input distributions $P_X$. (For example, this is the case for channels termed \emph{exotic} in \cite{Polyanskiy2010}.)}

However, most channels of interest do not satisfy \eqref{eq:above_capacity_condition}, including binary symmetric channels and binary erasure channels.} Thus, while we are able to show equivalence between the extremely strong converse and the strong edge removal property for deterministic networks (see Fig.~\ref{fig:properties}), this equivalence cannot hold for \edit{many} noisy networks, as the extremely strong converse simply does not hold.
\end{remark}

\subsection{Edge Removal Properties}

\begin{figure}
\begin{center}
\includegraphics{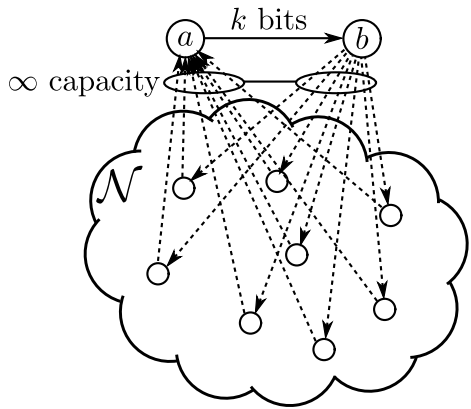}
\end{center}
\caption{The modified network for edge removal properties. Nodes $a$ and $b$ are connected to \nedit{nodes in $\calV$ (usually $\calV$ is the set of all nodes)} by infinite capacity links, while the link between them is limited to only \edit{$k$} bits. Edge removal properties hold when the capacity region of this network is unchanged when the link between $a$ and $b$ is removed.}
\label{fig:edge_removal}
\end{figure}

For a subset of nodes $\calV\subseteq[1:d]$ and an integer  \edit{$k$}, we define a modified network $\calN(\calV,\edit{k})$, illustrated in Fig.~\ref{fig:edge_removal}, as follows: Start with $\calN$, and add two nodes denoted $a$ and $b$.\footnote{These are special nodes in that messages do not originate at them. Thus the capacity region of $\calN(\calV,\edit{k})$ has the same dimension as that of $\calN$.} For each node $i\in\calV$, add an infinite capacity link from $i$ to $a$, and an infinite capacity link from $b$ to $i$. Finally, add a bit-pipe from $a$ to $b$ that can noiselessly transmit \edit{$k$} bits total \edit{across the $n$-length coding block.} In the case that \edit{$k$} is not an integer multiple of $n$, this bit-pipe cannot be modeled as a stationary memoryless channel. Instead, we assume that the \edit{$k$} bits are scheduled such that after $t$ timesteps, $\lfloor \frac{\edit{k}}{n}\,t\rfloor$ have been transmitted; that is, at time $t$, the link is allowed to transmit exactly
\be\label{eq:bit_pipe_schedule}
\left\lfloor \frac{\edit{k}}{n}\,t\right\rfloor-
\left\lfloor \frac{\edit{k}}{n}\,(t-1)\right\rfloor
\ee
bits.\footnote{One could imagine other models, such as where the bit transmission schedule is flexible but chosen in advance by the code, or where the schedule can be chosen at run-time. These model variations are unlikely to impact results, but here we adopt the more restrictive model.} \edit{Let $\calR_{\calV}(\calN,n,\eps,k)$ be the set of rate vectors $\bR$ such that there exists an $(\bR,n)$ code on $\calN(\calV,k)$ with average probability at most $\eps$. That is, $\calR_{\calV}(\calN,n,\eps,k)=\calR(\calN(\calV,k),n,\eps)$. Given sequences $(\eps_n)_n$ and $(k_n)_n$ where $\eps_n\in[0,1]$ and $k_n\in\bbN$, we define $\calC_{\calV}(\calN,(\eps_n)_n,(k_n)_n)$ to be the capacity region of the sequence of networks $(\calN(\calV,k_n))_n$ where $(k_n)_n$ determines the dependence between the capacity of the edge $(a,b)$ and the blocklength. Formally, we define
\be\label{eq:edge_capacity_region_def}
\calC_{\calV}(\calN,(\eps_n)_n,(k_n)_n)=\overline{\bigcup_{n_0\in\bbN}\ \bigcap_{n\ge n_0} \calR_{\calV}(\calN,n,\eps_{n},k_n)}.
\ee}%
%
%
For the most part we are interested in the case that $\calV=[1:d]$, so we define for convenience \edit{$\calR(\calN,n,\eps,k)=\calR_{[1:d]}(\calN,n,\eps,k)$ and} $\calC(\calN,\edit{(\eps_n)_n,(k_n)_n})=\calC_{[1:d]}(\calN,\edit{(\eps_n)_n,(k_n)_n})$. \edit{We further define $\calC_{\calV}(\calN,0^+,(k_n)_n)$ and $\calC(\calN,0^+,(k_n)_n)$ analogously to \eqref{eq:zero_error_def1}--\eqref{eq:zero_error_def2}.} For any \edit{$(k_n)_n$}, it is certainly true that $\calC(\calN,\edit{(\eps_n)_n})\subseteq\calC(\calN,\edit{(\eps_n)_n,(k_n)_n})$. Note also that 
$
\calC(\calN,\edit{(\eps_n)_n,(0)_n})=\calC(\calN,\edit{(\eps_n)_n}).
$

Roughly, edge removal properties state that for small \edit{$k$}, the capacity of network $\calN(\calV,\edit{k})$ is not too different from that of $\calN$. To be precise, we define four different versions of this property as follows.

\begin{definition}\label{def:edge_removal}
Edge removal properties are defined in terms of whether, for a given constant $\gamma>0$ and a sequence $\edit{(k_n)_n}$,
\be\label{eq:edge_removal_condition}
\calC(\calN,0^+,\edit{(k_n)_n})\subseteq \calC(\calN,0^+)+\edit{[0,\gamma]^d}.
\ee
We say network $\calN$ satisfies:
\begin{itemize}
\item the \emph{strong edge removal property} if for all $\gamma>0$, \eqref{eq:edge_removal_condition} holds for $k_n=\frac{\gamma n}{K}$, where $K$ is a positive constant depending only on the network.
\item the \emph{weak edge removal property} if for all $\gamma>0$, \eqref{eq:edge_removal_condition} holds for some $k_n\oin \Theta(n)$.
\item the \emph{very weak edge removal property} if for all $\gamma>0$, \eqref{eq:edge_removal_condition} holds for some $k_n\to\infty$.
\item the \emph{extremely weak edge removal property} if for all $\gamma>0$, \eqref{eq:edge_removal_condition} holds for all bounded $k_n$.
\end{itemize}
\end{definition}

\begin{remark}\label{remark:edge_removal_ordering}
One can again see immediately that the edge removal properties are ordered by strength; i.e., the strong property implies the weak property, which implies the very weak property, which implies the extremely weak property.
\end{remark}

The following proposition gives several alternative definitions of each of the edge removal properties. It is proved in Appendix~\ref{appendix:edge_removal_defs}.

\begin{proposition}\label{prop:edge_removal_defs}
\begin{enumerate}
\item The strong edge removal property holds if and only if there exists a finite positive constant $K$ depending only on the network $\calN$ such that for all $\delta>0$, 
\be\label{eq:strong_alternative}
\calC(\calN,0^+,\edit{(\delta n)_n})\subseteq \calC(\calN,0^+)+\edit{[0,K\delta]^d}.
\ee
\item The weak edge removal property holds if and only if, 
\be\label{eq:weak_alternative1}
\bigcap_{\delta >0} \calC(\calN,0^+,\edit{(\delta n)_n})=\calC(\calN,0^+)
\ee
and also if and only if
\be\label{eq:weak_alternative2}
\bigcup_{k_n\oin o(n)} \calC(\calN,0^+,\edit{(k_n)_n})=\calC(\calN,0^+).
\ee
\item The very weak edge removal property holds if and only if 
\be\label{eq:very_weak_alternative1}
\bigcap_{k_n:k_n\to \infty} \calC(\calN,0^+,\edit{(k_n)_n})=\calC(\calN,0^+)
\ee
and also if and only if 
\be\label{eq:very_weak_alternative2}
\bigcap_{\eps>0}\ \overline{\bigcup_{k\in\bbN} \calC(\calN,\edit{(\eps)_n},\edit{(k)_n})}=\calC(\calN,0^+).
\ee
\item The extremely weak edge removal property holds if and only if
\be\label{eq:extremely_weak_alternative}
\bigcup_{k\in\bbN} \calC(\calN,0^+,\edit{(k)_n})=\calC(\calN,0^+).
\ee
\end{enumerate}
\end{proposition}

\begin{remark}
\edit{Most works on the edge removal problem (e.g., \cite{ho10,jalali11}) consider removing an arbitrary edge from the network, rather than the specific topology shown in Fig.~\ref{fig:edge_removal}. Most similar to this topology is the notion of a \emph{super-source network} in \cite{Gu2008}, which was defined for source coding problems as a network containing a node that can view all sources, and has links to each other node. Another similar notion from the literature is that of the \emph{cooperation facilitator} \cite{Noorzad2014,Noorzad2015,Noorzad2016,Noorzad2016a,Noorzad2018,Noorzad2017a}, which connects to the transmitting nodes (but not the receiving node) in a multiple-access network. We choose the topology in Fig.~\ref{fig:edge_removal} because it ensures that the link that is added/removed is at least as useful as any other link. That is,} when $\calV=[1:d]$, then node $a$ has complete knowledge of every signal sent in the network, so the link $(a,b)$ can be used to simulate any other small-capacity link. In particular, for any network $\calN'$ consisting of $\calN$ supplemented by a link (or multiple links) with total capacity at most $k_n$ bits, then $\calC(\calN',\edit{(\eps_n)_n})\subseteq \calC(\calN,\edit{(\eps_n)_n},\edit{(k_n)_n})$. One example of such a network $\calN'$ is one that allows for rate-limited feedback. For this reason, one consequence of edge removal results are outer bounds on networks with rate-limited feedback. 
\end{remark}

\begin{remark}
The extremely weak edge removal property, wherein the extra edge carries a bounded number of bits as the blocklength grows, appears in none of our results proving relationships to strong converses. Nevertheless, we have chosen to include  this definition because it is a natural one, and indeed the property seems tantalizingly likely to be true for all realistic systems. However, it was shown in \cite{Langberg2016} that for maximal error probability, there exists a network where the extremely weak property does \emph{not} hold. This again points to the contrast between average and maximal error probability. In light of our other results, the extremely weak property also presents an interesting question: namely, is it  equivalent to \emph{some} version of a strong converse? Based on our results that for some networks, the very weak edge removal property is equivalent to the ordinary strong converse, if there is an equivalent converse to the extremely weak property, it appears that it would need to be \emph{weaker} than the ordinary strong converse, but perhaps stronger than the ordinary weak converse. No such property has occurred to us.
\end{remark}

\section{Deriving Edge Removal Properties from Strong Converses}\label{sec:ER_from_SC}

The following theorem states that each of the three strong converse properties implies one of the edge removal properties. This result holds for \edit{any causal network channel given by \eqref{eq:channel_distribution}}.

\begin{theorem}\label{thm:1direction}
For any network $\calN$, the following hold:
\begin{enumerate}
\item The strong converse implies very weak edge removal.
\item The exponentially strong converse implies weak edge removal.
\item The extremely strong converse implies strong edge removal.
\end{enumerate}
\end{theorem}

Statement (2) of this theorem was proved for noiseless networks in \cite[Sec.~3.3]{Gu2009}. Our proof \edit{uses essentially the same principle as theirs, namely converting a code on a network with an extra edge to a code on a network without one by fixing a value sent along this edge, and assuming at all other nodes that this value was sent. The following lemma provides a refined version of this argument, relating the achievable rate regions for the network with and without the extra edge at finite blocklengths.
}


\begin{lemma}\label{lemma1}
For any integers $n$ and $k$ and any $\eps\in[0,1]$,
\beq\label{eq:prop1}
\edit{\calR(\calN,n,\eps,k)\subseteq \calR(\calN,n,1-(1-\eps)2^{-k}).}
\eeq
\end{lemma}
\begin{IEEEproof}
Let $\bR\in\edit{\calR(\calN,n,\eps,k)}$, so there is an $n$-length code with rate vector $\bR$ and probability of error at most $\eps$ on network $\calN([1:d],k)$. We convert this code to one on network $\calN$ as follows. Under the code on $\calN([1:d],k)$, let $X_{ab}$ be the message sent on the link from node $a$ to node $b$. Recall that $X_{ab}\in\{0,1\}^{k}$. Let $\calE$ be the overall error event for network $\calN([1:d],k)$. We have
\be
1-\eps\le \bbP(\calE^c)=\sum_{x_{ab}\in\{0,1\}^{k}} \bbP(X_{ab}=x_{ab}) \bbP(\calE^c|X_{ab}=x_{ab}).
\ee
There must be some $x^*_{ab}\in\{0,1\}^{k}$ for which
\be
\bbP(X_{ab}=x^*_{ab}) \bbP(\calE^c|X_{ab}=x^*_{ab})\ge  (1-\eps) 2^{-k}.
\ee
Construct a code for network $\calN$ that behaves exactly like the original code on network $\calN([1:d],k)$, except that all nodes assume that node $b$ received the signal $x^*_{ab}$. Let $\nedit{\mathrm{P}_\mathrm{e}}$ be the probability of error for this code. Note that with probability $\bbP(X_{ab}=x^*_{ab})$, the code's behavior will be just as if the code on $\calN([1:d],k)$ were in effect. Thus
\be
1-\nedit{\mathrm{P}_\mathrm{e}}\ge \bbP(X_{ab}=x^*_{ab})\bbP(\calE^c|X_{ab}=x^*_{ab}) \ge (1-\eps) 2^{-k}.
\ee
Therefore $\bR\in \edit{\calR(\calN,n,1-(1-\eps)2^{-k})}$.
\end{IEEEproof}

\begin{IEEEproof}[Proof of Theorem~\ref{thm:1direction}]
We first show statement (1). Assume the strong converse holds. Thus
\begin{align}
\bigcap_{\eps>0}\ \overline{\bigcup_{k\in\bbN}\calC(\calN,\edit{(\eps)_n,(k)_n})}
&\subseteq \bigcap_{\eps\in(0,1)}\ \overline{\bigcup_{k\in\bbN} \calC(\calN,\edit{(1-(1-\eps)2^{-k})_n})}\label{eq:t1}
\\&= \bigcap_{\eps>0}\ \overline{\bigcup_{k\in\bbN}\calC(\calN,0^+)}\label{eq:t2}
\\&=\calC(\calN,0^+)\label{eq:t3}
\end{align}
where \eqref{eq:t1} follows from Lemma~\ref{lemma1}; \eqref{eq:t2} follows from the strong converse, because $1-(1-\eps)2^{-k}\in(0,1)$ for any $\eps\in(0,1)$ and $k\in\bbN$; and \eqref{eq:t3} follows because $\calC(\calN,0^+)$ is closed. Therefore, very weak edge removal holds by the equivalent definition in \eqref{eq:very_weak_alternative2} of Proposition~\ref{prop:edge_removal_defs}.

We now prove statement (2). Assume the exponentially strong converse holds. For any $k_n\oin o(n)$, we have
\begin{align}
\calC(\calN,0^+,\edit{(k_n)_n})
&=\bigcap_{{\eps}>0} \calC(\calN,\edit{(\eps)_n,(k_n)_n})\nonumber
\\&\subseteq \bigcap_{\eps>0} \calC(\calN,\edit{(1-(1-\eps)2^{-k_n})_n})\label{eq:t3a}
\\&\subseteq \bigcup_{\eps_n:-\log(1-\eps_n)\oin o(n)} \calC(\calN,\edit{(\eps_n)_n})\label{eq:t4}
\\&\subseteq \calC(\calN,0^+)\label{eq:t5}
\end{align}
where \eqref{eq:t3a} follows from Lemma~\ref{lemma1}, \eqref{eq:t4} from the fact that $k_n\oin o(n)$, and \eqref{eq:t5} from the exponentially strong converse. Therefore  weak edge removal holds.

We now prove statement (3). Assume the extremely strong converse holds. For any $\delta>0$ we have
\begin{align}
\calC(\calN,0^+,\edit{(\delta n)_n})
&=\bigcap_{{\eps}>0} \calC(\calN,\edit{(\eps)_n},\edit{(\delta n)_n})\nonumber
\\&\subseteq \bigcap_{{\eps}>0} \calC(\calN,\edit{(1-(1-\eps)2^{-\delta n})_n})\label{eq:t6}
\end{align}
where \eqref{eq:t6} follows from Lemma~\ref{lemma1}. Note that $(1-\eps)2^{-\delta n}\doteq 2^{-\delta n}$. Thus if $\bR\in\calC(\calN,0^+,\delta n)$, then, by the extremely strong converse, $\bR-K\delta\in \calC(\calN,0^+)$ for some constant $K$. Therefore strong edge removal holds.
\end{IEEEproof}

\section{Deterministic Networks}\label{sec:deterministic}

The following theorem states that for deterministic networks, each implication of Theorem~\ref{thm:1direction} is also an equivalence.

\begin{theorem}\label{thm:deterministic}
For any deterministic network $\calN$, the following hold:
\begin{enumerate}
\item The very weak edge removal property holds if and only if the strong converse holds.
\item The weak edge removal property holds if and only if the exponentially strong converse holds.
\item The strong edge removal property holds if and only if the extremely strong converse holds.
\end{enumerate}
\end{theorem}

To prove Theorem~\ref{thm:deterministic}, we begin with several lemmas. The first is the well-known reverse Markov inequality, which will be instrumental in proving that edge removal properties imply strong converses.

\begin{lemma}\label{lemma:reverse_markov}
Let $X$ be a real-valued random variable where $X\le x_{\max}$ a.s. For any $\tau\le \bbE X$,
\be
\bbP(X>\tau)\ge \frac{\bbE X - \tau}{x_{\max}-\tau}.
\ee
\end{lemma}

The following lemma provides the core result that is needed to prove Theorem~\ref{thm:deterministic}. The proof is adapted from that of \cite[Lemma 2]{Langberg2012}.

\begin{lemma}\label{lemma_det}
Let $\calN$ be a deterministic network. \edit{For any $\eps\in[0,1)$, any $n\in\bbN$, and any $\tilde\eps\in(0,1)$, }
\be\label{eq:lemma_det}
\edit{\calR(\calN,n,\eps)\subseteq\calR(\calN,n,\tilde{\eps},\eta(\tilde{\eps},d)-3d\log(1-\eps))}
\ee
where
\be\label{eq:eta_def}
\edit{\eta(\tilde\eps,d)=3d(d+1)+3d\log\ln\frac{4d}{\tilde\eps}.}
\ee
\end{lemma}
\begin{IEEEproof}
Let $\bR\in \edit{\calR(\calN,n,\eps)}$. That is, there exists a code with rate vector $\bR$ and blocklength $n$ achieving probability of error $\eps$. The key to the proof is to show that if the rates are reduced slightly from those in $\bR$, then an extra edge allows achieving arbitrarily small probability of error. In particular, given a target probability of error $\tilde\eps$, define a rate vector $\tilde{\bR}=(\tilR_1,\ldots,\tilR_d)$ given by
\be
\tilR_i=\begin{cases} R_i-\frac{k}{n}, & R_i\ge\frac{2k}{n}\\
0, & R_i< \frac{2k}{n}\end{cases}
\ee
where we choose with hindsight (recall $d$ is the number of messages in the network)
\be\label{eq:k_def}
k=\left\lceil d+\log \ln \frac{4d}{\tilde\eps}-\log(1-\eps)\right\rceil.
\ee
We will proceed prove that
\be\label{eq:lemma_det_dk}
\tilde{\bR}\in \edit{\calR(\calN,n,\tilde\eps,dk)}
\ee
by constructing a code of rate $\tilde{\bR}$ on network $\calN([1:d],dk)$. However, to prove the lemma we need to show that $\bR$, rather than $\tilde{\bR}$, is contained in the \edit{right-hand side (RHS)} of \eqref{eq:lemma_det}. Given \eqref{eq:lemma_det_dk} and that $nR_i-n\tilR_i\le 2k$, we may simply expand the edge from node $a$ to $b$ to carry $2dk$ additional bits, adding $2k$ bits for each message, which implies
\be\label{eq:lemma_det_to_prove}
\bR\in\edit{\calR(\calN,n,\tilde\eps,3dk)}.
\ee
This is now enough to prove the lemma, since $3dk\le \edit{\eta(\tilde\eps,d)-3d\log(1-\eps)}$ \edit{where $\eta(\tilde\eps,d)$ is defined in \eqref{eq:eta_def}.}

We now prove \eqref{eq:lemma_det_dk}. For $i=1,\ldots,d$, let $\calW_i=[2^{nR_i}]$ be the message set for the $i$th message of the original code of rate $\bR$ and probability of error $\eps$, and let
\be
\calW=\prod_{i=1}^d \calW_i
\ee
be the set of complete message vectors $\bw=(w_1,\ldots,w_d)$. Let $R=\sum_i R_i$, so $|\calW|=2^{nR}$. Since the network is deterministic \edit{and the code is fixed}, whether or not an error occurs depends entirely on the message vector $\bw\in\calW$ that is chosen. Let $\Gamma$ be the subset of $\calW$ of message vectors that do not lead to errors. Thus the probability of error is precisely $1-2^{-nR} |\Gamma|$. By the assumption that the probability of error is at most $\eps$, we have that
\be
|\Gamma|\ge |\calW| (1-\eps) = 2^{nR}(1-\eps).
\ee

Recall that $\tilR_i=0$ if $nR_i<2k$, so this message is not significant. For ease of notation, we assume for now that $nR_i\ge 2k$ for all messages $i$, so that $\tilR_i=R_i-\frac{k}{n}$. We employ a \edit{version of a} random binning argument. For each $i$, randomly choose \edit{the sets
\be
\edit{\calP_i(1),\ldots,\calP_i(2^{n\tilR_i})}
\ee
to be a partition of $\calW_i$ where $|\calP_i(\tilw_i)|=2^k$ for all $\tilw_i\in[1:2^{n\tilR_i}]$, such that all such partitions are equally likely.} \edit{Furthermore, let $\calP(\tilde\bw)$ for $\tilde\bw=(\tilw_1,\ldots,\tilw_d)$ be the set of message vectors $\bw\in\calW$ such that $w_i\in\calP_i(\tilw_i)$ for all $i\in[1:d]$.} Given these partitions, the code proceeds as follows. Messages $\tilW_1,\ldots,\tilW_d$ are all transmitted to node $a$. Node $a$ then \edit{chooses a message vector $\bW=(W_1,\ldots,W_d)$ from the set $\Gamma\cap\calP(\tilde\bW)$ in an arbitrary manner. If this set is empty, then we declare an error.}
For each $i$, let $I_i\in\{1,\ldots,2^k\}$ be the index of $W_i$ in the set $\edit{\calP}_i(\tilW_i)$. Node $a$ determines $I_i$ for each $i$ and transmits $(I_1,\ldots,I_d)$ to node $b$. Note that the number of bits required is $dk$.

At the originating source node for message $i$, $W_i$ can be determined from $\tilW_i$ and $I_i$. Subsequently, the code proceeds as if $\bW$ were the true message vector. When a destination node $j$ produces a message estimate $\hat{W}_{ij}$, it constructs the final message estimate as the $\widehat{\tilW}_{ij}\in[1:2^{n\tilR_i}]$ such that $\hat{W}_{ij}\in \edit{\calP}_i\left(\widehat{\tilW}_{ij}\right)$. Since by assumption $\bW\in\Gamma$, there is no error as long as \edit{$\Gamma\cap \calP(\tilde\bW)$} is not empty.

For $\tilde{\bw}=(\tilw_1,\ldots,\tilw_d)$ let
\be\label{eq:q_tildebw_def}
q(\tilde{\bw})\triangleq\bbP\left(\Gamma\cap \edit{\calP(\tilde\bw)}=\emptyset\right)
\ee
where the probability is with respect to the random choice of partitions $\edit{\calP}_i$. We proceed to show that $q(\tilde{\bw})\le\tilde\eps$ for all $\tilde\bw$. \edit{Thus, the probability of error averaged over both the message vector $\bW$ and the random choice of partitions is at most $\tilde\eps$.} This proves that there exists at least one \edit{deterministic} code with \edit{average} probability of error $\tilde{\eps}$.

For each $i\in[1:\edit{d-1}]$, define for all $w_1,\ldots,w_{i-1}$, the set
\be
\edit{\calA}_i(w_1,\ldots,w_{i-1})=\Big\{w_i: |\{(w_{i+1},\ldots,w_d):(w_1,\ldots,w_d)\in\Gamma\}|\ge (1-\eps) 2^{n(R_{i+1}+\cdots+R_d)-i}\Big\}.
\ee
\edit{Moreover, define
\be\label{eq:Ad_def}
\calA_d(w_1,\ldots,w_{d-1})=\{w_d: (w_1,\ldots,w_d)\in\Gamma\}.
\ee
We claim that for all $i\in[1:d]$, if $w_1,\ldots,w_{i-1}$ is such that $w_{i-1}\in\calA_{i-1}(w_1,\ldots,w_{i-2})$, then
\be\label{eq:A_bound}
|\edit{\calA}_i(w_1,\ldots,w_{i-1})|\ge (1-\eps) 2^{nR_i-i}.
\ee
To prove this for $i\in[1:d-1]$, assume}
$w_{i-1}\in \edit{\calA}_{i-1}(w_1,\ldots,w_{i-2})$. Define the random variable
\be
X(w_1,\ldots,w_{i-1})=|\{(w_{i+1},\ldots,w_d): (w_1,\ldots,w_{i-1},W_i,w_{i+1},\ldots,w_d)\in\Gamma\}|.
\ee
where as usual $W_i$ is uniformly distributed on $[1:2^{nR_i}]$. Note that
\begin{align}
\bbE X(w_1,\ldots,w_{i-1}) &= 2^{-nR_i} \sum_{w_i} |\{(w_{i+1},\ldots,w_d): (w_1,\ldots,w_d)\in\Gamma\}|
\\&= 2^{-nR_i} |\{(w_{i},\ldots,w_d):(w_1,\ldots,w_d)\in\Gamma\}|
\\&\ge (1-\eps) 2^{n(R_{i+1}+\cdots+R_d)-(i-1)}\label{eq:exp_X_bound}
\end{align}
where the inequality follows from the assumption that $w_{i-1}\in \edit{\calA}_{i-1}(w_1,\ldots,w_{i-2})$. Hence
\begin{align}
|\edit{\calA}_i(w_1,\ldots,w_{i-1})|
&=2^{nR_i} \bbP\Big(X(w_1,\ldots,w_{i-1})\ge (1-\eps) 2^{n(R_{i+1}+\cdots+R_d)-i}\Big)
\\& \ge 2^{nR_i}\frac{\bbE X(w_1,\ldots,w_{i-1})-(1-\eps) 2^{n(R_{i+1}+\cdots+R_d)-i}}{2^{n(R_{i+1}+\cdots+R_d)}-(1-\eps) 2^{n(R_{i+1}+\cdots+R_d)-i}}\label{eq:A_bound1}
\\& \ge 2^{nR_i} \frac{(1-\eps) 2^{n(R_{i+1}+\cdots+R_d)-i}}{2^{n(R_{i+1}+\cdots+R_d)}}\label{eq:A_bound2}
\\ & = (1-\eps) 2^{nR_i-i}\label{eq:A_bound3}
\end{align}
where \eqref{eq:A_bound1} follows from Lemma~\ref{lemma:reverse_markov} and the fact that $X\nedit{(\cdot)}\le 2^{n(R_{i+1}+\cdots+R_d)}$, and \eqref{eq:A_bound2} follows from \eqref{eq:exp_X_bound}. \edit{This proves \eqref{eq:A_bound} for $i\in[1:d-1]$. For $i=d$, note that if $w_{d-1}\in \calA_{d-1}(w_1,\ldots,w_{d-2})$, then by the definitions of $\calA_{d-1}$ and $\calA_d$,
\be\label{eq:Ad_bound}
|\calA_d(w_1,\ldots,w_{d-1})|=|\{w_d:(w_1,\ldots,w_d)\in\Gamma\}|\ge (1-\eps) 2^{nR_d-(d-1)}>(1-\eps) 2^{nR_d-d}.
\ee
This proves \eqref{eq:A_bound} for $i=d$.}

Fix $\tilde{\bw}=(\tilw_1,\ldots,\tilw_d)$. For each $i=1,\ldots,d$, define
\be
\edit{\calQ}_i=\{(w_1,\ldots,w_i): w_j\in \edit{\calP}_j(\tilw_j)\cap \edit{\calA}_j(w_1,\ldots,w_{j-1})\text{ for all } j\le i\}.
\ee
Note that \edit{for $\bw\in\calQ_d$, certainly $w_i\in\calP_i(\tilw_i)$ for all $i\in[1:d]$, so $\bw\in\calP(\tilde\bw)$. Moreover, since $w_d\in\calA_{d}(w_1,\ldots,w_{d-1})$, by definition $\bw\in\Gamma$. Thus $\calQ_d\subseteq\Gamma\cap \calP(\nedit{\tilde{\bw}})$}, so
\be
q(\tilde{\bw})\le \bbP(\edit{\calQ}_d=\emptyset)
\le \sum_{i=1}^d \bbP(\edit{\calQ}_i=\emptyset|\edit{\calQ}_{i-1}\ne\emptyset).
\ee
\edit{To upper bound $\bbP(\calQ_i=\emptyset|\calQ_{i-1}\ne\emptyset)$, suppose $\calQ_{i-1}\ne\emptyset$, so there exists some $(w_1,\ldots,w_{i-1})\in \edit{\calQ}_{i-1}$. If $\calQ_i$ is empty, then $\calP_i(\tilw_i)\cap\calA_i(w_1,\ldots,w_{i-1})=\emptyset$. Recall that $\calP_i(\tilw_i)$ is one set of a random partition of $\calW_i$, which is chosen independently of $w_1,\ldots,w_{i-1}$. In particular, $\calP_i(\tilw_i)$ is chosen uniformly among all subsets of $\calW_i=[1:2^{nR_i}]$ of size $2^k$, so
\be\label{eq:Qi_bd1}
\bbP(\calP_i(\tilw_i)\cap\calA_i(w_1,\ldots,w_{i-1})=\emptyset)=\frac{\binom{2^{nR_i}-|\edit{\calA}_i(w_1,\ldots,w_{i-1})|}{2^k}}{\binom{2^{nR_i}}{2^k}}.
\ee
Since by assumption $(w_1,\ldots,w_{i-1})\in\calQ_{i-1}$, we have $w_{i-1}\in\calA_{i-1}(w_1,\ldots,w_{i-2})$, so we may apply \eqref{eq:A_bound} to bound
\be\label{eq:Qi_bd2}
\bbP(\edit{\calQ}_i=\emptyset|\edit{\calQ}_{i-1}\ne\emptyset)\le \frac{\binom{2^{nR_i}-(1-\eps)2^{nR_i-i}}{2^k}}{\binom{2^{nR_i}}{2^k}}.
\ee
} Thus
\begin{align}
q(\tilde{\bw})
&\le \sum_{i=1}^d \frac{\binom{2^{nR_i}-(1-\eps)2^{nR_i-i}}{2^k}}{\binom{2^{nR_i}}{2^k}}
\\&=\sum_{i=1}^d \frac{(2^{nR_i}-(1-\eps)2^{nR_i-i})!}{(2^{nR_i}-\nedit{(1-\eps)2^{nR_i-i}}-2^k)!}\cdot\frac{(2^{nR_i}-2^k)!}{(2^{nR_i})!}
\\&\le \sum_{i=1}^d \frac{(2^{nR_i}-(1-\eps)2^{nR_i-i})^{2^k}}{(2^{nR_i}-2^k)^{2^k}}\label{eq:q_bd0}
\\&= \sum_{i=1}^d \edit{\frac{(1-(1-\eps)2^{-i})^{2^k}}{(1-2^{k-nR_i})^{2^k}}}
\\&\le \sum_{i=1}^d \edit{\frac{e^{-(1-\eps)2^{k-d}}}{(1-2^{k-nR_i})^{2^k}}}
\label{eq:q_bd1}
\\&\le \sum_{i=1}^d \frac{\tilde\eps}{4d} \edit{(1-2^{k-nR_i})^{-2^k}}
\label{eq:q_bd2}
\\&\le \sum_{i=1}^d \frac{\tilde\eps}{4d} \edit{(1-2^{-k})^{-2^k}}
\label{eq:q_bd3}
\\&\le \tilde\eps\label{eq:q_bd4}
\end{align}
where \eqref{eq:q_bd0} follows since $a!/b!\le a^{a-b}$ for integers $a,b$, \eqref{eq:q_bd1} follows since \edit{$(1+k)\le e^x$}, \eqref{eq:q_bd2} follows from the choice of $k$ in \eqref{eq:k_def}, \eqref{eq:q_bd3} follows by the assumption that $R_i\ge \frac{2k}{n}$ for all $i$, and \eqref{eq:q_bd4} follows since \edit{$(1-2^{-k})^{-2^k}\le 4$ for any $k\ge 1$. This last fact can be seen by noting that $f(x)=-x\ln(1-x^{-1})$ is decreasing in $x$, which holds because its derivative is given by
\be
f'(x)=-\ln(1-x^{-1})-\frac{1}{x-1}=\ln\left(1+\frac{1}{x-1}\right)-\frac{1}{x-1}\le 0.
\ee
}
\end{IEEEproof}

\begin{IEEEproof}[Proof of Theorem~\ref{thm:deterministic}]
Theorem~\ref{thm:1direction} proves that each strong converse property implies the corresponding edge removal property, so we only need to prove the opposite directions.

Suppose the very weak edge removal property holds. For any constant $\eps$, applying Lemma~\ref{lemma_det} gives
\begin{align}
\calC(\calN,\edit{(\eps)_n})
&\subseteq \bigcap_{\tilde{\eps}>0} \calC(\calN,\edit{(\tilde{\eps})_n,(\eta(\tilde{\eps},d)-3d\log(1-\eps))_n})
\\&\subseteq \bigcap_{\tilde{\eps}>0}\ \overline{\bigcup_{k\in\bbN} \calC(\calN,\edit{(\tilde{\eps})_n,(k)_n})}.
\\&= \calC(\calN,0^+)
\end{align}
where the last equality holds by very weak edge removal. Therefore the strong converse holds.

Now suppose the weak edge removal property holds. For any sequence \edit{$(\eps_n)_n$} where $-\log(1-\eps_n)\oin o(n)$, applying Lemma~\ref{lemma_det} gives
\begin{align}
\calC(\calN,\edit{(\eps_n)_n})
&\subseteq\bigcap_{\tilde{\eps}>0} \calC(\calN,\edit{(\tilde{\eps})_n,(\eta(\tilde{\eps},d)-3d\log(1-\eps_n))_n})
\\&\subseteq\bigcap_{\tilde{\eps}>0} \calC(\calN,\edit{(\tilde{\eps})_n,(\sqrt{n}-3d\log(1-\eps_n))_n})\label{eq:weak_ER_det1}
\\&= \calC(\calN,0^+,\edit{(\sqrt{n}-3d\log(1-\eps_n))_n})
\\&=\calC(\calN,0^+)\label{eq:weak_ER_det3}
\end{align}
where \eqref{eq:weak_ER_det1} follows since \edit{for any $\tilde\eps$ and $d$, $\eta(\tilde\eps,d)\le\sqrt{n}$} for sufficiently large $n$; and \eqref{eq:weak_ER_det3} follows from  weak edge removal, since $\sqrt{n}-\edit{3d}\log(1-\eps_n)\oin o(n)$. Therefore the exponentially strong converse holds. 

Finally, suppose the strong edge removal property holds. For any $\alpha>0$, let $\eps_n$ where $1-\eps_n\doteq 2^{-n\alpha}$. Applying Lemma~\ref{lemma_det} gives
\begin{align}
\calC(\calN,\edit{(\eps_n)_n})
&= \calC(\calN,\edit{(1-2^{-n\alpha})_n})\label{eq:det_strong1}
\\&\subseteq \bigcap_{\tilde{\eps}>0}\calC(\calN,\edit{(\tilde{\eps})_n,(\eta(\tilde{\eps},d)+3d\alpha n)_n})\label{eq:det_strong2}
\\&\subseteq \bigcap_{\tilde{\eps}>0}\calC(\calN,\edit{(\tilde{\eps})_n,((3d+1)\alpha n)_n})\label{eq:det_strong3}
\\&= \calC(\calN,0^+,\edit{((3d+1)\alpha n)_n})\label{eq:det_strong4}
\\&\edit{\subseteq \calC(\calN,0^+)+[0,K(3d+1)\alpha]^d}\label{eq:det_strong5}
\end{align}
where \eqref{eq:det_strong1} follows from Prop.~\ref{prop:error_sequences}, \eqref{eq:det_strong2} follows from Lemma~\ref{lemma_det},  \eqref{eq:det_strong3} follows because \edit{$\eta(\tilde\eps,d)\le \alpha n$ for sufficiently large $n$, \eqref{eq:det_strong4} follows by the definition of $\calC(\calN,0^+,(k_n)_n)$, and \eqref{eq:det_strong5} follows by the equivalent form of the strong edge removal property in \eqref{eq:strong_alternative}, where $K$ is a finite positive constant depending only on the network. Therefore, this network satisfies equivalent form of the extremely strong converse in Prop.~\ref{prop:strong_converse_defs} part (1b).}
\end{IEEEproof}

\section{Discrete Stationary Memoryless Networks}\label{sec:DMN}

The following is our main theorem for discrete stationary memoryless networks, connecting the exponentially strong converse to the weak edge removal property. In addition, we show that both these properties are equivalent to an even weaker form of the weak edge removal property---namely, where \emph{the nodes $a$ and $b$ connect only to transmitting nodes}; i.e. those nodes $i$ where $\calX_i\ne\emptyset$. (Recall the definition $\calC_\calV(\calN,\edit{(\eps_n)_n,(k_n)_n})$ being the capacity region of the network with nodes $a$ and $b$ connected only to nodes in $\calV$.)  This is a generalization of the ``cooperation facilitator'' model from \cite{Noorzad2014,Noorzad2015,Noorzad2016,Noorzad2016a,Noorzad2018,Noorzad2017a}, which connected only to the transmitters in a multiple-access channel, but not the receiver.  The intuition behind connecting only to transmitting nodes is that the extra edge is useful \nedit{\emph{when encoding but not decoding}. The reason is that when decoding, a node attempts to reconstruct a message, which is available exactly at the message's source node. Thus, any small amount of information sent from the omniscient node $a$ could equally well be sent from the source node. However, when encoding, the ``ideal'' transmission may be a function of multiple messages, which are simultaneously available only at the ominscient node $a$. Therefore, even a small capacity link from $a$ to $b$ could in principle provide significant rate gain by connecting to an encoding node. However, if a node does not transmit, it only decodes and never encodes, so the connection from nodes $a$ and $b$ is not helpful.
}



\begin{theorem}\label{thm:DMN}
For any discrete stationary memoryless network $\calN$, the following three statements are equivalent:
\begin{enumerate}
\item The exponentially strong converse holds.
\item The weak edge removal property holds.
\item \edit{For all $\gamma>0$,
\be\label{eq:V_weak_ER}
\calC_{\calV}(\calN,0^+,(k_n)_n)\subseteq \calC(\calN,0^+)+[0,\gamma]^d
\ee
for some sequence $k_n=\Theta(n)$,} where $\calV$ is the set of nodes $i$ such that $\calX_i\ne\emptyset$.
\end{enumerate}
\end{theorem}

Observe that statement 1 of the theorem implies statement 2 by Theorem~\ref{thm:1direction}. \edit{Note that statement 3 is identical to the definition of the weak edge removal, except that the left-hand side (LHS) of \eqref{eq:V_weak_ER} is $\calC_{\calV}(\calN,0^+,(k_n)_n)$ instead of $\calC(\calN,0^+,(k_n)_n)$ as in \eqref{eq:edge_removal_condition}; i.e., in the modified network, nodes $a$ and $b$ connect only to the set $\calV$ of transmitting nodes rather than all nodes. Since for any $\calV\subseteq[1:d]$, $\calC_{\calV}(\calN,0^+,(k_n)_n)\subseteq\calC(\calN,0^+,(k_n)_n)$, statement 2 of the theorem implies statement 3.}
Hence it remains only to show that statement 3 implies statement 1. The main tool in doing so will be a modified version of the blowing-up lemma. The blowing-up lemma, originally proved in \cite{Ahlswede1976} (see also \cite{Marton1986,Raginsky2013}), has been used in the proof of numerous strong converse results. In some sense our result is a generalization of this technique. The traditional blowing-up lemma is stated as follows.

\begin{lemma}\label{lemma:blowing_up}
Let $X^n\in\calX^n$ be a sequence of independent random variables. Fix $\edit{\calA}\subseteq \calX^n$ where $P_{X^n}(\edit{\calA})=\exp\{-n\gamma_n\}$ for a sequence $\gamma_n\to 0$. For any $\ell$, define the \emph{blown-up} version of $\edit{\calA}$ as
\be
\edit{\calA}_\ell = \left\{x^n: d_\text{H}(x^n,y^n)\le \ell\text{ for some }y^n\in \edit{\calA}\right\}
\ee
where $d_\text{H}$ is the Hamming distance. There \edit{exists} a sequence $\delta_n\to 0$ where
\be
P_{X^n}(\edit{\calA}_{n\delta_n})\to 1.
\ee
\end{lemma}

The following is a \emph{causal} version of the blowing-up lemma. It is stronger than the usual blowing-up lemma, but it follows from a \edit{slight} modification of Marton's proof of the blowing-up lemma in \cite{Marton1986}. \edit{One may view this lemma as a causal version of a transportation-cost inequality \cite{Raginsky2013}.}

\begin{lemma}\label{lemma:causal_blowing_up}
Let $X^n\in\calX^n$ be a random sequence, not necessarily independent. Fix $\edit{\calA}\subseteq\calX^n$. There exists a sequence of conditional distributions $P_{Z_t|Y_t,Z^{t-1}}$ for $t=1,\ldots,n$ such that, if we let $Y^n\in\calX^n,Z^n\in\calX^n$ have joint distribution
\begin{equation}\label{eq:YZ_distribution}
P_{Y^n,Z^n}(y^n,z^n) = \prod_{t=1}^n P_{X_t|X^{t-1}}(y_t|z^{t-1}) P_{Z_t|Y_t,Z^{t-1}}(z_t|y_t,z^{t-1})
\end{equation}
then $Z^n\in \edit{\calA}$ almost surely, and
\begin{equation}\label{eq:distance_bound}
\bbE d_\text{H}(Y^n,Z^n)\le \sqrt{\frac{n}{2\log e}\log \frac{1}{P_{X^n}(\edit{\calA})}}.
\end{equation}
\end{lemma}
\begin{IEEEproof}
Let $\tilX^n$ be a random sequence with distribution that of $X^n$ conditioned on the set $\edit{\calA}$. That is,
\be\label{eq:tilX_def}
P_{\tilX^n}(x^n)=\begin{cases} \frac{P_{X^n}(x^n)}{P_{X^n}(\edit{\calA})} & x^n\in \edit{\calA}\\ 0 & x^n\notin \edit{\calA}.\end{cases}
\ee
For any $t\in[1:n]$ and $z^{t-1}\in\calX^{t-1}$, by \cite[Theorem 1]{Strassen1965} there exists a pair of random variables $X_t(z^{t-1}),\tilX_t(z^{t-1})$ with joint distribution $P_{X_t(z^{t-1}),\tilX_t(z^{t-1})}$ such that the marginal distributions satisfy
\begin{align}
P_{X_t(z^{t-1})}&=P_{X_t|X^{t-1}=z^{t-1}},\label{eq:Xt_dist}\\
P_{\tilX_t(z^{t-1})} &= P_{\tilX_t|\tilX^{t-1}=z^{t-1}}\label{eq:Xt_tilde_dist}
\end{align}
and their joint distribution satisfies
\beq\label{eq:TV_equality}
\bbP(X_{\edit{t}}(z^{t-1})\ne \tilX_t(z^{t-1})) = d_{\text{TV}}\big(P_{X_t|X^{t-1}=z^{t-1}},\,P_{\tilX_t|\tilX^{t-1}=z^{t-1}}\big).
\eeq
We now define
\be\label{eq:Zt_dist}
P_{Z_t|Y_t,Z^{t-1}}(z_t|y_t,z^{t-1}) = P_{\tilX_t(z^{t-1})|X_t(z^{t-1})} (z_t|y_t).
\ee
Let $Y^n,Z^n$ have distribution given by \eqref{eq:YZ_distribution}\edit{, where $P_{Z_t|Y_t,Z^{t-1}}$ is defined in \eqref{eq:Zt_dist}}. Note that
\edit{
\begin{align}
P_{Y_t,Z_t|Z^{t-1}}(y_t,z_t|z^{t-1})
&=P_{X_t|X^{t-1}}(y_t|z^{t-1}) P_{Z_t|Y_t,Z^{t-1}}(z_t|y_t,z^{t-1})\label{eq:dist_match1}
\\&=P_{X_t(z^{t-1})}(y_t) P_{\tilX_t(z^{t-1})|X_t(z^{t-1})}(z_t|y_t)\label{eq:dist_match2}
\\&=P_{X_t(z^{t-1}),\tilX_t(z^{t-1})}(y_t,z_t)\label{eq:dist_match3}
\end{align}
where \eqref{eq:dist_match1} follows from \eqref{eq:YZ_distribution}, \eqref{eq:dist_match2} follows from \eqref{eq:Xt_dist} and \eqref{eq:Zt_dist}, and \eqref{eq:dist_match3} follows from simple rules about joint distributions. Thus
\begin{align}
P_{Z_t|Z^{t-1}}(z_t|z^{t-1}) &= \sum_{y_t} P_{Y_t,Z_t|Z^{t-1}}(y_t,z_t|z^{t-1})
\\&= \sum_{y_t} P_{X_t(z^{t-1}),\tilX_t(z^{t-1})}(y_t,z_t)\label{eq:dist_match4}
\\&= P_{\tilX_t(z^{t-1})}(z_t)\label{eq:dist_match5}
\\&= P_{\tilX_t|\tilX^{t-1}}(z_t|z^{t-1})\label{eq:dist_match6}
\end{align}
where \eqref{eq:dist_match4} holds by \eqref{eq:dist_match3},  \eqref{eq:dist_match5} holds simply because the summation in \eqref{eq:dist_match4} represents the marginal distribution of $\tilX_t(z^{t-1})$, and \eqref{eq:dist_match6} holds by \eqref{eq:Xt_tilde_dist}.} Thus $Z^n$ and $\tilX^n$ have the same distribution. In particular, \edit{since by construction $\tilX^n\in\calA$ almost surely, also} $Z^n\in \edit{\calA}$ almost surely. We now have
\begin{align}
\bbE d_\text{H}(Y^n,Z^n) &= \sum_{t=1}^n \bbP(Y_t\ne Z_t)\label{eq:BU00}
\\&= \sum_{t=1}^n \sum_{z^{t-1}} P_{Z^{t-1}}(z^{t-1}) \sum_{y_t\ne z_t} \edit{P_{Y_t,Z_t|Z^{t-1}}(y_t,z_t|z^{t-1})}
\\&= \sum_{t=1}^n \sum_{z^{t-1}} P_{Z^{t-1}}(z^{t-1}) \sum_{y_t\ne z_t} P_{X_t(z^{t-1}),\tilX_t(z^{t-1})}(y_t,z_t)\label{eq:BU0}
\\&= \sum_{t=1}^n \sum_{z^{t-1}} P_{Z^{t-1}}(z^{t-1}) \bbP(X_t(z^{t-1})\ne \tilX_t(z^{t-1}))
\\&= \sum_{t=1}^n \sum_{z^{t-1}} P_{Z^{t-1}}(z^{t-1})\, d_{\text{TV}}\big(P_{X_t|X^{t-1}=z^{t-1}},\,P_{\tilX_t|\tilX^{t-1}=z^{t-1}}\big)\label{eq:BU1}
\\&\le \sum_{t=1}^n \sum_{z^{t-1}} P_{Z^{t-1}}(z^{t-1})\sqrt{\frac{1}{2\edit{\log e}} D(P_{\tilX_t|\tilX^{t-1}=z^{t-1}}\|P_{X_t|X^{t-1}=z^{t-1}})}\label{eq:BU2}
\\&\le n \sqrt{\frac{1}{\edit{(2\log e)}n} \sum_{t=1}^n \sum_{z^{t-1}} P_{Z^{t-1}}(z^{t-1})D(P_{\tilX_t|\tilX^{t-1}=z^{t-1}}\|P_{X_t|X^{t-1}=z^{t-1}})}\label{eq:BU3}
\\&= \sqrt{\frac{n}{2\edit{\log e}} \sum_{t=1}^n \sum_{z^{t-1}} P_{\tilX^{t-1}}(z^{t-1}) D(P_{\tilX_t|\tilX^{t-1}=z^{t-1}}\|P_{X_t|X^{t-1}=z^{t-1}})}\label{eq:BU4}
\\&= \sqrt{\frac{n}{2\log e} D(P_{\tilX^n}\|P_{X^n})}\label{eq:BU4a}
\\&= \sqrt{\frac{n}{2\log e} \log \frac{1}{P_{X^n}(\edit{\calA})}}\label{eq:BU5}
\end{align}
where \edit{\eqref{eq:BU0} holds by \eqref{eq:dist_match3}}, \eqref{eq:BU1} holds by \eqref{eq:TV_equality}, \eqref{eq:BU2} holds by Pinsker's inequality, \eqref{eq:BU3} holds by concavity of the square root,  \eqref{eq:BU4} holds because $Z^n$ and $\tilX^n$ have the same distribution, \eqref{eq:BU4a} holds by the chain rule for relative entropy, and \eqref{eq:BU5} holds because, by \eqref{eq:tilX_def},
\be
\frac{P_{\tilX^n}(\tilX^n)}{P_{X^n}(\tilX^n)}=\frac{1}{P_{X^n}(\edit{\calA})}\quad \text{a.s.}
\ee
\end{IEEEproof}

\begin{remark}
Lemma~\ref{lemma:blowing_up} can be derived from Lemma~\ref{lemma:causal_blowing_up} as follows. If in Lemma~\ref{lemma:causal_blowing_up}, $X^n$ is a sequence of independent random variables, then by \eqref{eq:YZ_distribution}, $Y^n$ has the same distribution as $X^n$. Thus
\begin{align}
P_{X^n}(\edit{\calA}_\ell)& = P_{Y^n}(\edit{\calA}_\ell)
\\&\ge \bbP( d_\text{H}(Y^n,Z^n)\le \ell)\label{eq:blow1}
\\&\ge 1-\frac{1}{\ell} \bbE d_\text{H}(Y^n,Z^n)\label{eq:blow2}
\\&\ge 1-\frac{1}{\ell} \sqrt{\frac{n}{2\log e} \log\frac{1}{P_{X^n}(\edit{\calA})}}\label{eq:blow3}
\end{align}
where \eqref{eq:blow1} holds because $Z^n\in \edit{\calA}$ almost surely, \eqref{eq:blow2} holds by Markov's inequality, and in \eqref{eq:blow3} we have applied \eqref{eq:distance_bound}. Assuming $P_{X^n}(\edit{\calA})=\exp\{-n\gamma_n\}$ where $\gamma_n\to 0$, if we choose, for example, $\delta_n=\gamma_n^{1/4}$, we have $\delta_n\to 0$ and
\be
P_{X^n}(\edit{\calA}_{n\delta_n})\ge 1-\frac{\gamma_n^{1/4}}{\sqrt{2\log e}}\to 1.
\ee
This proves Lemma~\ref{lemma:blowing_up}.
\end{remark}

With Lemma~\ref{lemma:causal_blowing_up} in hand, we complete the proof of Theorem~\ref{thm:DMN} with the following lemma.

\begin{lemma}\label{lemma:DMN}
For any discrete stationary memoryless network $\calN$, \edit{statement 3 of Theorem~\ref{thm:DMN} implies statement 1.}
\end{lemma}

\begin{IEEEproof}
\edit{By the same argument as in the proof of Proposition~\ref{prop:edge_removal_defs}, statement 3 of Theorem~\ref{thm:DMN} is equivalent to
\be
\bigcap_{\delta>0} \calC_{\calV}(\calN,0^+,(\delta n)_n)=\calC(\calN,0^+).
\ee
where again $\calV$ is the set of transmitting nodes. By Proposition~\ref{prop:strong_converse_defs}, the exponentially strong converse holds if and only if, for any sequence $(\eps_n)_n$ where $-\log(1-\eps_n)=o(n)$, $\calC(\calN,(\eps_n)_n)\subseteq\calC(\calN,0^+)$. Thus, to prove the lemma it is enough to show that for any $(\eps_n)_n$ where $-\log(1-\eps_n)=o(n)$, and any $\delta>0$,} $\calC(\calN,\edit{(\eps_n)_n})\subseteq \calC_\calV(\calN,0^+,\edit{(\delta n)_n})$. Let $\bR$ be achievable with respect to $\eps_n$. Thus for sufficiently large $n$ there exists an $n$-length code with average probability of error at most $\eps_n$. Let $(\phi_{it},\psi_{ij})$ be the encoding/decoding functions for this code (see \eqref{eq:encoding_function}--\eqref{eq:decoding_function}).  We describe a new code, illustrated in Fig.~\ref{fig:code_summary},  achieving the same rate vector with vanishing probability of error on the network $\calN(\calV,\delta n)$. Note that for any $i\in\calV^c$, we have $\calX_i=\emptyset$, so if $R_i>0$ the probability of success would be exponentially small; thus we must have $R_i=0$.

\begin{figure}
\begin{center}
\includegraphics[scale=.95]{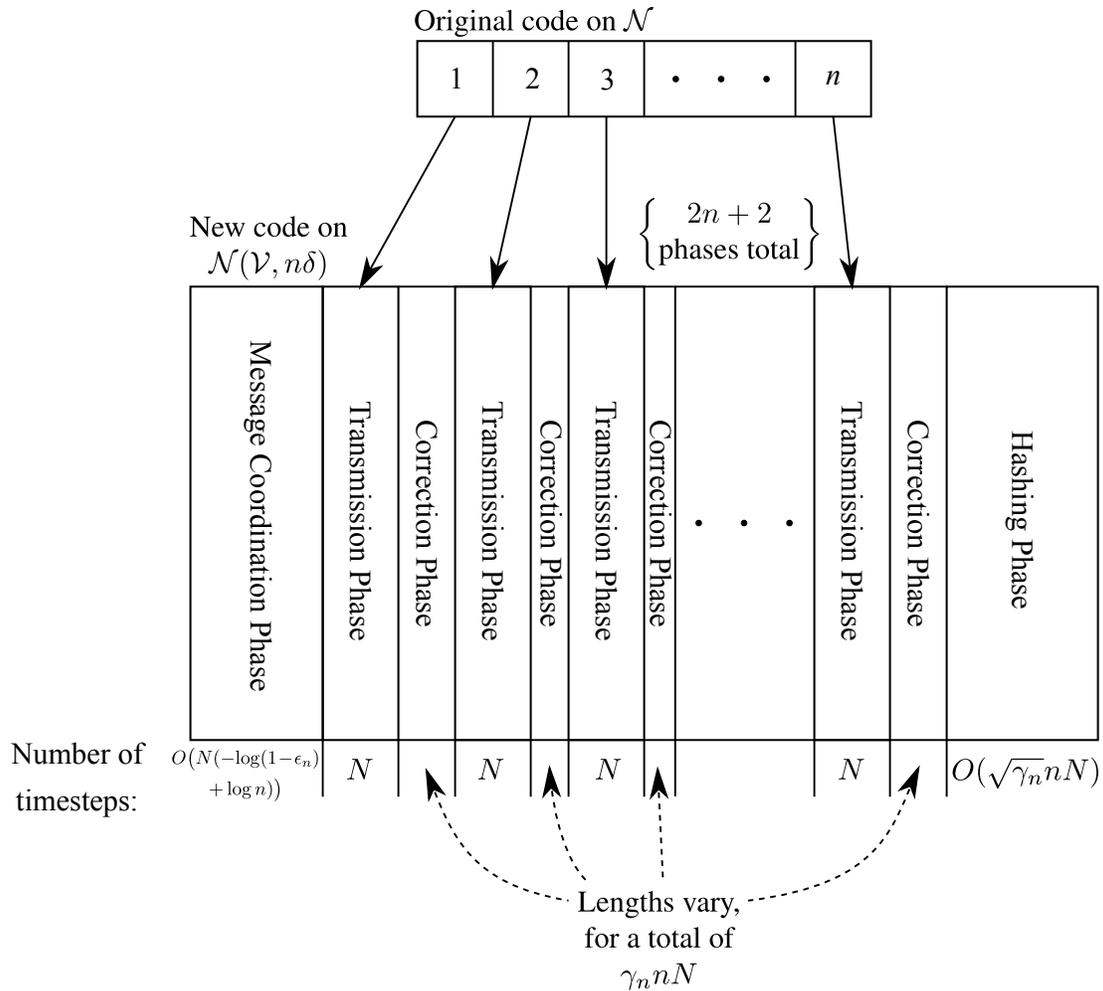}
\end{center}
\caption{Summary of the procedure to convert a code with probability of error $\eps_n$ to one with vanishing probability of error on the network with an extra edge. Each timestep of the original code is copied $N$ times into a transmission phase, followed by a subsequent correction phase that replaces some of the received signals. Prior to the $n$ transmission and correction phases, a message coordination phase ensures that only ``good'' message vectors are used; subsequently a hashing phase is used to ensure all nodes can decode.}
\label{fig:code_summary}
\end{figure}

\emph{Network stacking:} We adopt the notion of \emph{network stacking} from \cite{Koetter2011}. The motivation for our use of network stacking is that it allows us to convert an arbitrary coding operation at a single time instance into a coding operation across a long block, thereby taking advantage of the law of large numbers. In particular, we construct $N$ \edit{independent} copies of the original $n$-length code, \edit{each with its own messages,} using a total of $nN$ channel uses. Each copy is referred to as a ``layer'', indexed by an integer $\ell\in[1:N]$. Unlike a block Markov approach \edit{\cite{Cover1979}}, in which one would transmit an $n$-length block corresponding to the original code in sequence, in the network stacking approach we transmit $N$ copies of a single time instance $t\in[1:n]$ of the original code before moving on to the next one. Thus coding can be done ``across the layers'', using the fact that the $N$ copies of any symbol are i.i.d., while maintaining the causal structure of the original code.

We use underlines to indicate symbols on the stacked network. In particular, $\underline{X}_{it}(\ell)$ is the transmitted symbol from node $i$ at time $t$ in layer $\ell$; $\underline{X}_i^n(\ell)$ refers to the $n$-length sequence of symbols in layer $\ell$; $\underline{X}_{it}$ refers to the $N$-length sequence of symbols at time $t$ in all layers; $\underline{X}_i^n$ refers to the full $nN$-length sequence of all layers and time instances. We define $\underline{Y}_{it}(\ell)$, etc. similarly. Moreover, $\underline{W}_i(\ell)$ is the message originating at node $i$ in layer $\ell$, and $\underline{W}_i$ is the complete vector of messages originating at node $i$ across all $N$ layers.

\emph{Code phases:} Given the original $n$-length code, we construct an $N$-fold stacked code as follows, where the precise dependence between $n$ and $N$ is to be determined. The code consists of $2n+2$ phases, each consisting of a number of timesteps. These phases are visualized in Fig.~\ref{fig:code_summary}. First we have a \emph{message coordination} phase, followed by $n$ \emph{transmission} phases alternating with $n$ \emph{correction} phases, and concluded with a \emph{hashing} phase. In the message coordination phase, nodes coordinate to choose a message vector in each layer with a relatively large probability of success; this is done in exactly the same manner as for deterministic networks in Lemma~\ref{lemma_det}. Each transmission phase corresponds to one timestep $t\in[1:n]$ in the original code:  the layers act independently, each performing the coding functions from the original code at time $t$. In the following correction phase, node $a$ transmits data to node $b$, describing replacements for certain received data in sub-network $\calV$.  Node $b$ then disperses this data to the nodes in $\calV$; in subsequent transmission phases, nodes in $\calV$ use this replaced data in their coding operations. In the final hashing phase, hashes of all messages are dispersed to all nodes, which allows nodes in $\calV^c$ to decode. This last phase is necessary because nodes $a$ and $b$ do not connect directly to nodes in $\calV^c$; thus the correction approach applied to the rest of the network does not work here, since node $a$ does not know what signals were received in $\calV^c$. Instead, hashes are used to correct any remaining errors in messages decoded in $\calV^c$.

The message coordination phase consists of $O(N(-\log(1-\eps_n)+\log n))$ timesteps. Each transmission phase consists of exactly $N$ timesteps, since each layer transmits exactly once. Correction phases have variable lengths, depending on how much correction data is required, but a total of $Nn\gamma_n$ timesteps are allocated for all correction phases, where 
\beq\label{eq:gamma_n_def}
\gamma_n = \left(\frac{-\log\frac{1-\eps_n}{4}}{n}\right)^{1/4}.
\eeq
The hashing phase consists of $O(\sqrt{\gamma_n}nN)$ timesteps. Note that in total, the transmission phases consist of $nN$ timesteps. Recalling that $-\log(1-\eps_n)\oin o(n)$, $\gamma_n\to 0$ as $n\to\infty$, so all other phases consist of a negligible number of timesteps.

\emph{Message coordination phase:} For each message vector $\bw$ of the original code, let $P_c(\bw)$ be the probability of correctly decoding $\bw$. Let
\be
\Gamma=\left\{\bw:P_c(\bw)\ge \frac{1-\eps_n}{2}\right\}.
\ee 
Defining $R=\sum_{i=1}^d R_i$, we may lower bound the cardinality of $\Gamma$ by
\begin{align}
|\Gamma|
&=2^{nR}\, \bbP\left(P_c(\bW)\ge \frac{1-\eps_n}{2}\right)
\\&\ge 2^{nR}\, \frac{\bbE P_c(\bW)-\frac{1-\eps_n}{2}}{1-\frac{1-\eps_n}{2}}\label{eq:gamma_bd1}
\\&\ge 2^{nR}\, \left[(1-\eps_n)-\frac{1-\eps_n}{2}\right]\label{eq:gamma_bd2}
\\&=2^{nR}\,\frac{1-\eps_n}{2}
\end{align}
where \eqref{eq:gamma_bd1} holds by Lemma~\ref{lemma:reverse_markov} and the fact that $P_c(\bW)\le 1$, and \eqref{eq:gamma_bd2} holds since the average probability of error is at most $\eps_n$.

In the message coordination phase, we use an identical outer code as in Lemma~\ref{lemma_det} to ensure that\edit{, with high probability,} only message vectors in $\Gamma$ are ever used. By the same binning argument as in the proof of Lemma~\ref{lemma_det}, this requires only $O(-\log(1-\eps_n)+\log n)$ bits on the link $(a,b)$ for each layer. Note that  nodes $a$ and $b$ are only required to contact the nodes in $\calV$, since nodes in $\calV^c$ have no message originating at them. We may therefore assume throughout the rest of this argument that $\underline{\bW}(\ell)\in\Gamma$ for each $\ell\in[1:N]$.

\emph{Correction codebook:} \edit{Let $P_c(\bw,y_{\calV}^n)$ be the probability of correct decoding given message vector $\bw$, and channel outputs $y_{\calV}^n$ at nodes $\calV$. That is,
\be\label{eq:Pc_W_Y}
P_c(\bw,y_{\calV}^n)=\bbP(\hat{\bW}=\bw|\bW=\bw,Y_{\calV}^n=y_{\calV}^n)
\ee
where again $\hat{\bW}$ is the complete vector of message estimates.} Since \edit{encoding and decoding functions are assumed to be deterministic (cf. \eqref{eq:encoding_function}--\eqref{eq:decoding_function}),} channel inputs $X_{\calV}^n$ are deterministic functions of $Y_{\calV}^n$ \edit{and $\bW$}. \edit{Thus,} the only randomness in the probability in \eqref{eq:Pc_W_Y} are the channel outputs $Y_{\calV^c}^n$ given the inputs $X_{\calV}^n$. Recalling that $\calX_i=\emptyset$ for $i\in\calV^c$, $Y_{\calV^c}^n$ is an independent sequence given $X_{\calV}^n$. For each message vector $\bw$ of the original $n$-length code, let 
\edit{\be\label{eq:calQ_def}
\calQ(\bw)=\left\{y_{\calV}^n: P_c(\bw,y_{\calV}^n)\ge \frac{1-\eps}{4}\right\}.
\ee}%
 Note that for any $\bw\in\Gamma$,
\begin{align}
\edit{\bbE(P_c(\bw,Y_{\calV}^n)|\bW=\bw)}&=\bbP(\hat{\bW}=\bw|\bW=\bw)
\\&= P_c(\bw)
\\&\ge \frac{1-\eps_n}{2}.
\end{align}
Thus, applying Lemma~\ref{lemma:reverse_markov} to the random variable \edit{$P_c(\bw,Y_{\calV}^n)$} gives
\be\label{eq:calQ_bd}
P_{Y_\calV^n|\bW=\bw}(\calQ(\bw))
\ge \frac{1-\eps_n}{4}.
\ee
We now apply Lemma~\ref{lemma:causal_blowing_up} to the distribution  $P_{Y_\calV^n|\bW=\bw}$ and the set $\calQ(\bw)$ to find conditional distributions $P_{Z_{\calV,t}|Y_{\calV,t},Z_{\calV,t}}$ for all $t=[1:n]$. Note that these distributions depend on the message vector $\bw$. For each $y_{\calV,t}\in\calY_\calV$ and $z^{t-1}\in\calY_\calV^{t-1}$, independently draw
\be
f_t(\bw,y_{\calV,t},z_{\calV}^{t-1})\sim P_{Z_{\calV,t}|Y_{\calV,t},Z_{\calV}^{t-1}}.
\ee
These functions constitute a codebook known to all nodes.

\emph{Hashing codebook:} For each $i\in\calV$ and each $\underline{w}_i\in[1:2^{nR_i}]^N$, independently and uniformly draw $g_i(\underline{w}_i)$ from $[1:2^{nN\sqrt{\gamma_n}}]$. These hashing functions also constitute a codebook known to all nodes.

\emph{Transmission phases:} Before the \edit{transmission phase at time $t$}, each node $i\in\calV$ has determined $\uZ_{i}^{t-1}\in\calY_i^{t-1}$, which represent the corrected versions of its received signals (see description below of the correction phases). For each $\ell\in[1:N]$, node $i$ determines and transmits
\be
\uX_{i,t}(\ell)=\phi_{it}(\uW_i(\ell),\uZ_i^{t-1})
\ee
For each $i\in[1:d]$, let $\uY_{i,t}(\ell)$ be the corresponding received signals.

\emph{Correction phases:} In the correction phase after the \edit{transmission phase at time $t$}, node $a$ learns $\uY_{i,t}$ from each $i\in\calV$, and determines, for each $\ell\in[1:N]$,
\beq\label{eq:correction_codebook}
\underline{Z}_{\calV,t}(\ell)=f_t(\underline{\bW}(\ell),\uY_{\calV,t}(\ell),\uZ_{\calV}^{t-1}(\ell)).
\eeq
For each $\ell$ for which $\uZ_{\calV,t}(\ell)\ne \uY_{\calV,t}(\ell)$, node $a$ transmits to node $b$ a bit string with $0$ followed by $\lceil \log N|\calY|\rceil$ bits identifying the layer $\ell\in[1:N]$ as well as the value of $\uZ_{\calV,t}(\ell)\in\calY_\calV$. After doing this for each layer where $\uZ_{\calV,t}(\ell)\ne \uY_{\calV,t}(\ell)$, node $a$ transmits the stop bit $1$, signaling that all nodes should proceed to the next transmission phase. Node $b$ then forwards this data to each node $i\in\calV$. For all layers $\ell$ for which no correcting signal was sent, each node $i\in\calV$ simply sets $\uZ_{it}(\ell)=\uY_{it}(\ell)$.

\emph{Hashing phase:} Node $a$ computes $g_i=g_i(\underline{w}_i)$ for all $i\in\calV$, and transmits these values to node $b$, which subsequently disperses them to nodes in $\calV$.\footnote{One could also compute the hash for message $i$ directly at node $i$, and distribute the hash to all decoder nodes from there. We choose to compute the hash at node $a$ makes merely to make distribution of the hashes simpler to describe.} Note that these hashes consist of a total of $d\sqrt{\gamma_n} nN$ bits, which is sub-linear in $nN$. Thus they can be transmitted over the link $(a,b)$ as long as $\delta>0$. For each node $i\in\calV^c$, if there exists a node $j\in\calV$ where the point-to-point channel from $X_j$ to $Y_i$ has positive capacity, then we use a point-to-point channel code to transmit the hashes from node $j$ to node $i$. If there is no such node $j\in\calV$, \edit{then all received signals at node $i$ are independent of the rest of the network, so node $i$ cannot decode any messages; in particular,} if $i\in\calD_k$ for any $k\in[1:d]$, it must be that $R_k=0$. Since the hashes occupy a sub-linear number of bits, transmitting these hashes to each node in $\calV^c$ takes a sub-linear number of timesteps, and can be done with arbitrarily small probability of error.

\emph{Decoding:} For each $i,j\in\calV$ where $j\in\calD_i$ and each $\ell\in[1:N]$, node $j$ determines
\be
\hat{\uW}_{ij}(\ell) = \psi_{ij}(\uW_j(\ell),\uZ_j^n(\ell)).
\ee
Now consider $i\in[1:d]$ and  $j\in\calV^c\cap\calD_i$ and each $i\in[1:d]$ where $j\in\calD_i$. Given $\underline{Y}_j^n$ and $g_i$, find the unique $\underline{\hatw}_i$ where $g_i=g_i(\underline{\hatw}_i)$ and there exists $\underline{\tily}_i^n$ where $\psi_{ij}(\uW_j(\ell),\underline{\tily}_j^n(\ell))=\hat{\uw}_i(\ell)$ for each $\ell\in[1:N]$ and 
\be\label{eq:Ac_hamming}
d_{\text{H}}(\underline{Y}_j^n,\underline{\tily}_j^n)\le Nn\gamma_n.
\ee
If there is no such $\underline{\hatw}_i$ or more than one, declare an error.

\emph{Probability of error analysis:} Consider the following error events
\be
\calE_1=\{\text{number of timesteps used in correction phases exceeds }Nn\gamma_n\}
\ee
and, for $i\in[1:d]$ and $j\in\calV^c\cap\calD_i$,
\begin{align}
\calE_{2ij}&=\big\{\psi_{ij}(\uW_j(\ell),\tilde{\uy}_j^n(\ell))\edit{\ne}\uW_i(\ell)\text{ for \edit{some} }\ell\in[1:N],\text{ for \edit{all} }\tilde{\uy}_j^n\nonumber\\
&\qquad\text{ where }d_{\text{H}}(\uY_j^n,\tilde{\uy}_j^n)\le Nn\gamma_n\big\},\label{eq:event2}\\
\calE_{3ij}&=\big\{\psi_{ij}(\uW_j(\ell),\tilde{\uy}_j^n(\ell))=\uw'_i(\ell)\text{ for all }\ell\in[1:N],\text{ for some }\uw'_i\ne \uW_i\nonumber\\
&\text{ where } g_i(\uw'_i)=g_i(\uW_i)
\text{ and } \tilde{\uy}_j^n
\text{ where }d_{\text{H}}(\uY_j^n,\tilde{\uy}_j^n)\le Nn\gamma_n\big\}.
\end{align}
Note that as long as $\calE_1$ does not occur, then by Lemma~\ref{lemma:causal_blowing_up}, $\uZ_\calV^n(\ell)\in\calQ(\underline{\bW}(\ell))$ for all $\ell$. By the definition of $\calQ(\bw)$, this ensures that $W_{ji}=w_i$ for all $j\in[1:d]$ and $i\in\calV$. Events $\calE_{2ij},\calE_{3ij}$ cover all errors that can occur at nodes in $\calV^c$. Hence the probability of error of the overall code, averaged over random coding choices, is
\begin{align}
\nedit{\mathrm{P}_\mathrm{e}}&\le \bbP\left(\calE_1\cup\bigcup_{i\in[1:d],j\in\calV^c\cap\calD_i} (\calE_{2ij}\cup \calE_{3ij})\right)
\\&\le \bbP(\calE_1)+ \sum_{i\in[1:d],j\in\calV^c\cap\calD_i} \big[\bbP(\calE_{2ij}|\calE_1^c)+ \bbP(\calE_{3ij}|\calE_1^c)\big].
\end{align}

We first consider $\calE_1$. The number of bits transmitted across link $(a,b)$ during the \edit{correction phase at time $t$} is
\be\label{eq:bits_tth_correction}
d_{\text{H}}(\uY_{\calV,t}, \uZ_{\calV,t}) (\lceil \log N|\calY_\calV|\rceil+1)+1
\ee
where the final $+1$ accounts for the stop bit. Thus the number of bits transmitted during all $n$ correction phases is
\be
d_\text{H}(\uY_{\calV}^n,\uZ_{\calV}^n)(\lceil \log N|\calY_\calV|\rceil+1)+n.
\ee
Recall link $(a,b)$ has capacity $\delta>0$, meaning it can transmit a bit roughly every $1/\delta$ timesteps (cf. \eqref{eq:bit_pipe_schedule}). Thus we can bound $\calE_1$ by
\begin{align}
\bbP(\calE_1)
&=\bbP\left(\frac{1}{\delta}\nedit{\Big[}d_\text{H}(\uY_{\calV}^n,\uZ_{\calV}^n)(\lceil \log N|\calY_\calV|\rceil+1)+n\nedit{\Big]}
>N n\gamma_n\right)\label{eq:BUerr1}
\\&\le \frac{\sum_{\ell=1}^N \bbE d_\text{H}(\uY_{\calV}^n(\ell),\uZ_{\calV}^n(\ell)) (\lceil \log N|\calY_\calV|\rceil+1)+n}{\delta N n\gamma_n}\label{eq:BUerr2}
\\&\le \frac{\sum_{\ell=1}^N \bbE \sqrt{-n \log P_c(\bW(\ell))} (\lceil \log N|\calY_\calV|\rceil+1)+n}{\delta N n\gamma_n}\label{eq:BUerr3}
\\&\le \frac{N \sqrt{-n\log \frac{1-\eps_n}{2}} (\lceil \log N|\calY_\calV|\rceil+1)+n}{\delta N n\gamma_n}\label{eq:BUerr4}
\\&\le \frac{1}{\delta} \gamma_n (\lceil \log N|\calY_\calV|\rceil+1) + \frac{1}{\delta N \gamma_n}\label{eq:BUerr5}
\end{align}
where \eqref{eq:BUerr2} follows from Markov's inequality, \eqref{eq:BUerr3} follows from Lemma~\ref{lemma:causal_blowing_up}, where we have dropped the constant $\frac{1}{2\log e}$ since it is less than $1$, \eqref{eq:BUerr4} from the assumption that $\bW(\ell)\in\Gamma$ for all $\ell$, and  \eqref{eq:BUerr5} from the definition of $\gamma_n$ in \eqref{eq:gamma_n_def}. If we choose $N=\gamma_n^{-2}$, then 
\begin{align}
\bbP(\calE_1)&\le \frac{1}{\delta} \gamma_n\left(\left\lceil \log \frac{1}{\gamma_n^2} |\calY|\right\rceil+1\right)+\frac{\gamma_n}{\delta}
\\&\le\frac{\gamma_n}{\delta} (-2\log \gamma_n +\log|\calY|+3)
\end{align}
which vanishes since $-\gamma_n\log \gamma_n\to 0$ as $\gamma_n\to 0$. 

Now we consider events $\calE_{2ij},\calE_{3ij}$. Recall that if $\calE_1$ does not occur, then $\uZ_\calV^n(\ell)\in\calQ(\underline{\bW}(\ell))$ for all $\ell$. By the definition of $\calQ(\bw)$ in \eqref{eq:calQ_def}, we have, for any $y_{\calV}^n\in\calQ(\bw)$ 
\begin{align}
\frac{1-\eps_n}{4}
&\le \edit{P_c(\bw,y_{\calV}^n)}
\\&= \sum_{y_{\calV^c}^n} P_{Y_{\calV^c}^n|Y_\calV^n=y_\calV^n,\bW=\bw}(y_{\calV^c})\,
 1(\psi_{ij}(y_{j}^n)=w_i\text{ for all }i\in\calV,j\in\calV^c\cap\calD_i).
\end{align}
Note that given $Y_\calV^n=y_\calV^n$ and $\bW=\bw$, $X_\calV^n$ is determined \edit{since coding functions are deterministic}. Since $\calX_i=\emptyset$ for all $i\in\calV^c$, this conditioning also determines $X_{1:d}^n$. Thus, the distribution $P_{Y_{\calV^c}^n|Y_\calV^n=y_\calV^n,\bW=\bw}$ is independent. Applying the blowing up lemma to this distribution and the set of  $y_{\calV^c}$ that cause all messages to be decoded correctly in $\calV^c$, there exists a random sequence $Z_{\calV^c}^n\in\calY_{\calV^c}^n$ that causes all messages to be decoded correctly, and
\be
\bbE d_\text{H}(Y_{\calV^c}^n,Z_{\calV^c}^n)\le \sqrt{-n \log \frac{1-\eps_n}{4}}=n\gamma_n^2.
\ee
In particular, if we produce $N$ copies of this $Z_{\calV^c}^n$ sequence for each layer, then  Markov's inequality gives
\be
\bbP\left( d_\text{H}(\underline{Y}_{\calV^c}^n,\underline{Z}_{\calV^c}^n)>Nn\gamma_n\right)
\le \frac{Nn\gamma_n^2}{Nn\gamma_n}
=\gamma_n.
\ee
In particular, for each $i\in[1:d]$ and $j\in\calV^c\cap\calD_i$, with probability at least $1-\gamma_n$, there exists $\underline{\tily}_{j}^n$ that satisfies the Hamming distance condition \eqref{eq:Ac_hamming}, and is decoded correctly to $w_i$. Thus $\bbP(\calE_{2ij}|\calE_1^c)$ vanishes. We now consider $\calE_{3ij}$. The number of messages $\uw_j'$ that are considered is upper bounded by the number of sequences $\underline{\tily}^n$ satisfying \eqref{eq:Ac_hamming}, which is given by
\be
\sum_{k=0}^{\lfloor Nn\gamma_n\rfloor} \binom{Nn}{k} |\calY_i|^{k}
\le \exp\{nN (H(\gamma_n)+\gamma_n \log |\calY_i|)\}
\ee
where $H(\cdot)$ is the binary entropy function. The probability that any given $\uw_j'\ne \uW_j$ agrees with the hash value $g_j$ is $2^{-nN\sqrt{\gamma_n}}$, so 
\begin{align}
\bbP(\calE_{3ij}|\calE_1^c)&\le\exp\{nN (H(\gamma_n)+\gamma_n \log |\calY_i|)-nN\sqrt{\gamma_n}\}
\\&\le \exp\{-nN\sqrt{\gamma_n}/2\}\label{eq:hash1}
\\&=\exp\{-n \gamma^{-3/2}/2\}\label{eq:hash2}
\end{align}
where \eqref{eq:hash1} holds for sufficiently large $n$, since $\gamma_n\to 0$ and $\lim_{p\to 0} H(p)/\sqrt{p}=0$, and \eqref{eq:hash2} holds again by the choice $N=\gamma_n^{-2}$. Since $n\gamma^{-3/2}\to\infty$ as $n\to\infty$, $\bbP(\calE_{3ij}|\calE_1^c)$ vanishes.
\end{IEEEproof}

\begin{remark}
The blowing-up lemma does not appear to be strong enough to prove that the very weak edge removal property implies the ordinary strong converse. Were we to apply the same argument above to the case $\eps_n=\eps\in(0,1)$, in the key application of the blowing-up lemma in \eqref{eq:BUerr3}, we would have
\be
\bbE d_\text{H}(Y_{\calV}^n,Z_\calV^n) \le \sqrt{-\frac{n}{2}\log\frac{1-\eps}{2}}.
\ee
This suggests that at least $O(\sqrt{n})$ bits per layer would be required on the extra link. However, very weak edge removal requires that we achieve the same capacity region using \emph{any} $k_n$ sequence of bits converging to infinity, which includes sequences growing smaller than $\sqrt{n}$.
\end{remark}

\section{Networks of Independent Point-to-Point Links}\label{sec:p2p}

We now consider the setting of \emph{network equivalence} \cite{Koetter2011}, in
which $\calN$ consists of a stationary memoryless network made up of
independent point-to-point (noisy) links. Let $\bar\calN$ be the same
network in which each noisy point-to-point link is replaced by a noiseless
bit-pipe of the same capacity. The basic result of network equivalence
states that $\calC(\calN,0^+)=\calC(\bar\calN,0^+)$.  Theorem~\ref{thm:DMN}
already asserts that for such networks, the weak edge removal property holds
if and only if the exponentially strong converse holds. The following
theorem proves that, for such networks with acyclic topology, the same holds
for the ``lower level'' in Fig.~\ref{fig:properties}; i.e., the very weak edge removal property and the ordinary strong converse. The proof, given in Appendix~\ref{appendix:p2p}, makes use of the network equivalence principle to connect codes on $\calN$ to codes on $\bar\calN$, and then applies Theorem~\ref{thm:deterministic} on $\bar\calN$.

\begin{theorem}\label{thm:p2p}
For a discrete stationary memoryless network $\calN$ consisting of independent point-to-point links with acyclic topology, the very weak edge removal property holds if and only if the strong converse holds.
\end{theorem}

\section{Applications}\label{sec:applications}

\subsection{Outer Bounds}

Consider any outer bound $\calR_{\text{out}}(\calN)$ for the memoryless stationary network $\calN$; i.e. where $\calC(\calN,0^+)\subseteq \calR_{\text{out}}(\calN)$. Suppose we could show
\be\label{eq:outer_bound_continuity}
\bigcup_{k_n\oin o(n)} \calC_{\calV}(\calN,0^+,\edit{(k_n)_n})\subseteq\calR_{\text{out}}(\calN)
\ee
where as usual $\calV$ is the set of nodes $i$ where $\calX_i\ne\emptyset$. In other words, the outer bound is continuous with respect to the capacity of the extra edge; that is, the outer bound satisfies a weak edge removal property. Then, applying Lemma~\ref{lemma:DMN}, we immediately find
\be
\bigcup_{\eps_n:-\log(1-\eps_n)\oin o(n)} \calC(\calN,\edit{(\eps_n)_n})
\subseteq\calR_{\text{out}}(\calN).
\ee
This suggests that the outer bound holds in an \emph{exponentially strong} sense; that is, for any rate vector outside $\calR_{\text{out}}(\calN)$, the probability of error approaches 1 exponentially fast. 

An outer bound may also satisfy a strong edge removal property, meaning that for some constant $K$ and any $\delta$,
\be\label{eq:outer_bound_strong}
\calC(\calN,0^+,\edit{(\delta n)_n})\subseteq \calR_{\text{out}}(\calN)+\edit{[0,K\delta]}.
\ee
We have no equivalence between the strong edge removal property and the extremely strong converse for general noisy networks, but we do for deterministic networks. Thus, applying Lemma~\ref{lemma_det}, if a deterministic network satisfies \eqref{eq:outer_bound_strong}, then the outer bound holds in an \emph{extremely strong} sense; that is, for any rate vector outside $\calR_{\text{out}}(\calN)$, the probability of error approaches 1 at an exponential rate linear in the distance to the outer bound.

For many outer bounds (indeed, almost every computable outer bound that we know of), \eqref{eq:outer_bound_continuity} can be proved without much difficulty, and in some cases the stronger statement \eqref{eq:outer_bound_strong} can be proved as well. This implies that most outer bounds for discrete memoryless networks hold in an exponentially strong sense, and many outer bounds for deterministic networks hold in an extremely strong sense. We illustrate this for several outer bounds (or weak converse arguments) in the next few subsections.

\subsection{Cut-set Bound}

Recall that the \emph{cut-set outer bound} \edit{\cite{ElGamal1981}} is given by $\calC(\calN,0^+)\subseteq \calR_{\text{cut-set}}(\calN)$ \edit{where} 
\begin{equation}
 \calR_{\text{cut-set}}(\calN)
=\bigcup_{\edit{P_{X_1,\ldots,X_d}}} \left\{\bR: \sum_{i\in\calS:\calD_i\cap\calS^c\ne\emptyset} R_i \le I(X_{\calS};Y_{\calS^c}|X_{\calS^c})\text{ for all }\calS\subseteq[1:d]\right\}.
\end{equation}
In the following, we prove \eqref{eq:outer_bound_strong} for this bound. This allows us to reproduce the result of \cite{Fong2017}, that the cut-set bound holds in an \emph{exponentially strong} sense: that is, for any rate vector outside $\calR_{\text{cut-set}}(\calN)$, the probaility of error goes to 1 exponentially fast. This further implies that any network with a tight cut-set bound (i.e., where $\calC(\calN,0^+)=\calR_{\text{cut-set}}(\calN)$) satisfies the exponentially strong converse. Furthermore, we conclude that for deterministic networks, the cut-set bound holds in an \emph{extremely strong} sense.

Fix some sequence $\edit{(k_n)_n}$, and let $\bR\in \calC(\calN,0^+,\edit{(k_n)_n})$. Consider a code achieving this rate vector, and let $Z_t$ be the symbol sent along edge $(a,b)$ at time $t$, or $\emptyset$ if there is no symbol at time $t$. Note $H(Z^n)\le k_n$. Fix any cut set $\calS\subseteq[1:d]$, and let $\calS^c=[1:d]\setminus\calS$. Also let $\calT$ be the set of message flows that cross the cut; that is, the set of $i\in\calS$ where $\calD_i\cap\calS^c\ne\emptyset$. We may write
\begin{align}
\sum_{i\in\calT} R_i
&= H(M_\calT)
\\&\le I(M_\calT; Y_{\calS^c}^n,Z^n)+n\eps_n\label{eq:cutset1}
\\&= \sum_{t=1}^n I(M_\calT; Y_{\calS^c,t},Z_t|Y_{\calS^c}^{t-1},Z^{t-1})+n\eps_n
\\&= \sum_{t=1}^n I(M_\calT; Y_{\calS^c,t},Z_t|Y_{\calS^c}^{t-1},Z^{t-1},X_{\calS^c,t})+n\eps_n\label{eq:cutset1a}
\\&\le \sum_{t=1}^n I(M_{\calT}, Y_{\calS^c}^{t-1},X_{\calS,t}; Y_{\calS^c,t},Z_t|Z^{t-1},X_{\calS^c,t})+n\eps_n
\\&\le \sum_{t=1}^n\big[ I(M_{\calT}, Y_{\calS^c}^{t-1},X_{\calS,t}; Y_{\calS^c,t}|Z^{t-1},X_{\calS^c,t})+H(Z_t|Z^{t-1})\big]+n\eps_n
\\&\le \sum_{t=1}^n I(X_{\calS,t}; Y_{\calS^c,t}|X_{\calS^c,t})+H(Z^n)+n\eps_n\label{eq:cutset2}
\\&\le n I(X_{\calS};Y_{\calS^c}|X_{\calS^c},Q)+k_n+n\eps_n\label{eq:cutset3}
\\&\le n I(X_{\calS};Y_{\calS^c}|X_{\calS^c})+k_n+n\eps_n\label{eq:cutset4}
\end{align}
where \eqref{eq:cutset1} follows from Fano's inequality, where $\eps_n\to 0$ as $n\to\infty$; \eqref{eq:cutset1a} follows since $X_{\calS^c,t}$ is a function of $Y_{\calS^c}^{t-1}$ and $Z^{t-1}$; \eqref{eq:cutset2} follows from the memorylessness and causality of the network model; and \eqref{eq:cutset3} follows by defining $Q\sim\text{Unif}[1:n]$, $X_i=X_{i,Q}$, and $Y_i=Y_{i,Q}$, and by the fact that $H(Z^n)\le k_n$. Recalling that $\eps_n\to 0$, we have
\be
\calC_{\calV}(\calN,0^+,\edit{(k_n)_n}) \subseteq \calR_{\text{cut-set}}(\calN)+\edit{\left[0,\lim_{n\to\infty} \frac{k_n}{n}\right]^d}.
\ee
In particular, \eqref{eq:outer_bound_strong} holds with $K=1$. This in turn implies \eqref{eq:outer_bound_continuity}. Therefore, for discrete memoryless stationary networks, the cut-set bound holds in an exponentially strong sense, and for deterministic networks, the cut-set bound holds in an extremely strong sense.

These facts allow us to immediately derive strong converse results for various problems for which the cut-set bound is tight. For example:
\begin{enumerate}
\item since the cut-set bound is tight for relay channels that are degraded, reversely degraded \cite{Cover1979}, or semideterministic \cite{ElGamal1982}, the exponentially strong converse holds.
\item since the cut-set bound is tight for linear finite-field deterministic multicast networks \cite{Avestimehr2011}, the extremely strong converse holds.
\end{enumerate}

\subsection{Broadcast Channel}\label{sec:broadcast}

A broadcast channel is a network where $\calY_1=\emptyset$, $\calX_i=\emptyset$ for all $i>1$, and we allow multiple messages to originate at node 1, each to be decoded at a subset of nodes in $[2:d]$. Note that this model includes scenarios where there are private messages, public messages, and/or messages intended for some decoders but not all. We claim that the weak edge removal property and the exponentially strong converse hold for discrete memoryless broadcast channels. Indeed, the $\calV$ set in Theorem~\ref{thm:DMN} is simply $\{1\}$. Thus, for any sequence $\edit{(k_n)_n}$ (whether or not it is $o(n)$), $\calC_{\{1\}}(\calN,0^+,\edit{(k_n)_n})=\calC(\calN,0^+)$, simply because if the extra nodes $a$ and $b$ can only communicate with node $1$, then any processing done at nodes $a$ and $b$ can simply be reproduced internally at node 1.  Theorem~\ref{thm:DMN} immediately proves the claim.

For \emph{degraded} broadcast channels, the strong converse was proved in \cite{Ahlswede1976}, and the exponentially strong converse in \cite{Oohama2015}. However, since the capacity of the broadcast channel in general is unknown, strong converses for general broadcast channels have received little attention. As far as we know, this is the first strong (or exponentially strong) converse that has been proved for a problem for which the capacity region \edit{has no known single-letter characterization. In \cite{Liu2016}, a strong converse was established for a common randomness generation problem for which a single-letter characterization was established in \cite{Ahlswede1998}; this strong converse generalizes to non-discrete alphabets, including sources where the single-letter characterization has no known computable characterization, because of an auxiliary random variable. Both the result of \cite{Liu2016} and our result on the broadcast channel are examples of strong converses for problems with no known computable rate region. The simplicity of the above proof on the broadcast channel}, once we have Theorem~\ref{thm:DMN}, is particularly noteworthy.

\subsection{Discrete 2-User Interference Channel with Strong Interference}

\begin{figure}
\begin{center}
\includegraphics[scale=.9]{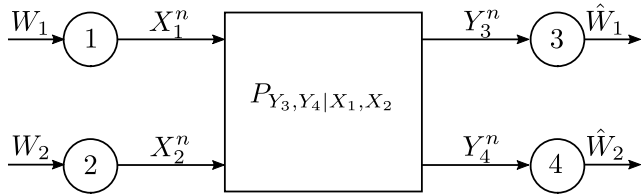}
\end{center}
\caption{The 2-user interference channel.}
\label{fig:IC}
\end{figure}

A 2-user interference channel, illustrated in Fig.~\ref{fig:IC}, is a network with 4 nodes, where $\calY_1=\calY_2=\calX_3=\calX_4=\emptyset$, $\calD_1=\{3\}$, and $\calD_2=\{4\}$. Note that, to be consistent with the notation in the rest of the paper, the received symbol by the node decoding the first message is $Y_3$, rather than $Y_1$\edit{, as it is typically denoted}.

Recall that an interference channel has \emph{strong interference} \edit{\cite{Sato1981}} if
\be
I(X_1;Y_3|X_2)\le I(X_1;Y_4|X_2),\quad
I(X_2;Y_4|X_1)\le I(X_2;Y_3|X_1)\label{eq:strong}
\ee
for all $\edit{P_{X_1}(x_1)P_{X_2}(x_2)}$. The capacity region of the interference channel in this regime \edit{was found in \cite{Costa1987} to be} the set of rate pairs $(R_1,R_2)$ such that
\begin{align}
R_1&\le I(X_1;Y_3|X_2,Q),\label{eq:strong_region1}\\
R_2&\le I(X_2;Y_4|X_1,Q),\\
R_1+R_2&\le \min\{I(X_1,X_2;Y_3|Q),\ I(X_1,X_2,Y_4|Q)\}\label{eq:strong_region3}
\end{align}
for some $\edit{P_Q(q)P_{X_1|Q}(x_1|q)P_{X_2|Q}(x_2|q)}$ \edit{with $|\calQ|\le 4$}.

The following proposition establishes the exponentially strong converse under strong interference. The strong converse for the interference channel with \emph{very strong interference} (in addition to fixed-error second-order results) was derived in\cite{Le2015}. The strong converse for the \emph{Gaussian} interference channel with strong interference was proved in \cite{Fong2016a}.

\begin{proposition}
For an interference channel with strong interference, weak edge removal and the exponentially strong converse hold.
\end{proposition}

\renewcommand{\c}{\nedit{,}}

\begin{IEEEproof}
Note that the only nodes $i$ in an interference channel where $\calX_i\ne\emptyset$ are the encoder nodes, i.e. nodes $1$ and $2$. Thus, by Theorem~\ref{thm:DMN}, to prove the proposition it is enough to show that for any $k_n\oin o(n)$, $\calC_{\{1,2\}}(\calN,0^+,\edit{(k_n)_n})\subseteq\calC(\calN,0^+)$, where $\calC(\calN,0^+)$ is the region defined in \eqref{eq:strong_region1}--\eqref{eq:strong_region3}.

We claim that an interference channel with strong interference also satisfies \eqref{eq:strong} for any joint distribution $\edit{P_{X_1,X_2}}$, even when $X_1,X_2$ are not independent. \edit{Consider any joint distribution $P_{X_1,X_2}$. For fixed $x_2$, define $\tilX_1,\tilX_2$ where $\tilX_1\sim P_{X_1|X_2=x_2}$ and $\tilX_2=x_2$ deterministically. Since $\tilX_2$ is deterministic, $\tilX_1$ and $\tilX_2$ are trivially independent, so by \eqref{eq:strong} we have
\be\label{eq:tilde_strong}
I(\tilX_1;\tilY_3|\tilX_2)\le I(\tilX_1;\tilY_4|\tilX_2)
\ee
where $\tilY_3,\tilY_4$ represent the outputs of the channel with $\tilX_1,\tilX_2$ as inputs. Note that $P_{\tilX_1,\tilY_3,\tilY_4}=P_{X_1,X_3,Y_4|X_2=x_2}$.
Thus $I(\tilX_1;\tilY_3|\tilX_2)=I(X_1;Y_3|X_2=x_2)$ and $I(\tilX_1;\tilY_4|\tilX_2)=I(X_1;Y_4|X_2=x_2)$, so by \eqref{eq:tilde_strong}
\be\label{eq:x2_conditioning}
I(X_1;Y_3|X_2=x_2)\le I(X_1;Y_4|X_2=x_2).
\ee
Since \eqref{eq:x2_conditioning} holds for any $x_2$,} we have
\begin{align}
I(X_1;Y_3|X_2)&= \sum_{x_2} \edit{P_{X_2}}(x_2) I(X_1;Y_3|X_2=x_2)
\\&\le \sum_{x_2} \edit{P_{X_2}}(x_2) I(X_1;Y_4|X_2=x_2)\label{eq:joint_strong}
\\&= I(X_1;Y_4|X_2)
\end{align}
\edit{Similar reasoning establishes the second inequality in \eqref{eq:strong} for any $P_{X_1,X_2}$. This proves the claim.}

Now, by the same proof as the lemma in \cite{Costa1987} for the independent case, for any $\edit{P_{X_1^n,X_2^n}}$,
\be\label{eq:nlength_strong_bound}
I(X_1^n;Y_3^n|X_2^n)\le I(X_1^n;Y_4^n|X_2^n),\quad I(X_2^n;Y_4^n|X_1^n)\le I(X_2^n;Y_3^n|X_1^n)
\ee
where 
\be
\edit{P_{Y_1^n,Y_2^n|X_1^n,X_2^n}}(y_1^n,y_2^n|x_1^n,x_2^n)=\prod_{t=1}^n \edit{P_{Y_1,Y_2|X_1,X_2}}(y_{1\c t},y_{2\c t}|x_{1\c t},x_{2\c t}).
\ee

Consider $(R_1,R_2)\in \calC_{\{1,2\}}(\calN,0^+,\edit{(k_n)_n})$ where $k_n\oin o(n)$. \edit{Thus, there exists a sequence of codes with rates $(R_1,R_2)$, with vanishing probability of error, on the modified network with an extra edge carrying $k_n$ bits as a function of the blocklength $n$. Given a code of blocklength $n$,} let $Z_t$ be the signal sent on the edge $(a,b)$ at time $t\edit{\in[1:n]}$. \edit{Note that, since $k_n\oin o(n)$, for most values of $t\in[1:n]$, no bit is transmitted across $(a,b)$ at time $t$ (cf. the transmission schedule in \eqref{eq:bit_pipe_schedule}); for these $t$ we simply take $Z_t$ to be null.} Certainly $H(Z^n)\le k_n$. Since for $j=1,2$, $X_j^n$ is a function of message $W_j$ and $Z^n$, we have
\begin{align}
I(X_1^n;X_2^n|Z^n)
&\le I(W_1;W_2|Z^n)
\\&\le  I(W_1;W_2,Z^n)
\\&= I(W_1;W_2)+I(W_1;Z^n|W_2)
\\&\le H(Z^n)\label{eq:W1W2_indep}
\\&\le k_n
\end{align}
where \eqref{eq:W1W2_indep} follows since the messages are assumed to be independent. Since node $a$ only has access to $W_1,W_2$, we have \edit{the Markov chain
\be\label{eq:IC_MC}
(W_1,W_2,Z^n)\to (X_1^n,X_2^n)\to (Y_3^n,Y_4^n).
\ee}%
We now write
\begin{align}
nR_1&=H(W_1|W_2)
\\&=I(W_1;Y_3^n,Z^n|W_2)+H(W_1|Y_3^n,W_2,Z^n)
\\&\le I(W_1;Y_3^n|W_2,Z^n)+k_n+n\eps_n\label{eq:strong_proof1}
\\&\edit{\le I(W_1,W_2,X_1^n;Y_3^n|X_2^n,Z^n)}+k_n+n\eps_n
\\&\le I(X_1^n;Y_3^n|X_2^n,Z^n)+k_n+n\eps_n\label{eq:strong_proof1a}
\end{align}
where in \eqref{eq:strong_proof1} we have used the fact that $H(Z^n)\le k_n$, and Fano's inequality, where $\eps_n\to 0$ as $n\to\infty$, \edit{and \eqref{eq:strong_proof1a} holds by the Markov chain in \eqref{eq:IC_MC}}. Similarly
\be
nR_2\le n I(X_2^n;Y_4^n|X_1^n,Z^n)+k_n+n\eps_n.\label{eq:R2_bd}
\ee
We also have
\begin{align}
nR_1&=H(W_1)
\\&\le I(W_1;Y_3^n,Z^n)+n\eps_n
\\&\le I(W_1;Y_3^n|Z^n)+k_n+n\eps_n
\\&\le I(W_1,X_1^n;Y_3^n|Z^n)+k_n+n\eps_n
\\&=I(X_1^n;Y_3^n|Z^n)+I(W_1;Y_3^n|X_1^n,Z^n)+k_n+n\eps_n
\\&\le I(X_1^n;Y_3^n|Z^n)+I(W_1;Y_3^n,X_2^n|X_1^n,Z^n)+k_n+n\eps_n
\\&= I(X_1^n;Y_3^n|Z^n)+I(W_1;X_2^n|X_1^n,Z^n)+k_n+n\eps_n\label{eq:strong_proof2}
\\&\le I(X_1^n;Y_3^n|Z^n)+I(W_1;W_2|Z^n)+k_n+n\eps_n
\\&\le I(X_1^n;Y_3^n|Z^n)+2k_n+n\eps_n\label{eq:R1_bd}
\end{align}
\edit{where in \eqref{eq:strong_proof2} we have again used the Markov chain in \eqref{eq:IC_MC}.}
Combining \edit{\eqref{eq:R2_bd} with \eqref{eq:R1_bd}} gives
\begin{align}
n(R_1+R_2)&\le I(X_1^n;Y_3^n|Z^n)+I(X_2^n;Y_4^n|Z^n,X_1^n)+3k_n+n2\eps_n
\\&\le I(X_1^n;Y_3^n|Z^n)+I(X_2^n;Y_3^n|Z^n,X_1^n)+3k_n+n2\eps_n\label{eq:strong_proof9}
\\&= I(X_1^n,X_2^n;Y_3^n|Z^n)+3k_n+n2\eps_n\label{eq:R12_bd}
\end{align}
where \eqref{eq:strong_proof9} follows from \eqref{eq:nlength_strong_bound}. \edit{We may also repeat this argument to find \eqref{eq:R12_bd} with $Y_3$ replaced by $Y_4$.} To summarize,
\begin{align}
nR_1&\le I(X_1^n;Y_3^n|X_2^n,Z^n)+k_n+n\eps_n,\label{eq:strong_region_dependence1}\\
nR_2&\le I(X_2^n;Y_4^n|X_1^n,Z^n)+k_n+n\eps_n,\label{eq:strong_region_dependence2}\\
n(R_1+R_2)&\le \min\{ I(X_1^n,X_2^n;Y_3^n|Z^n),\ I(X_1^n,X_2^n;Y_4^n|Z^n)\}+3k_n+n2\eps_n,\label{eq:strong_region_dependence3}\\
k_n&\ge I(X_1^n;X_2^n|Z^n).\label{eq:strong_region_dependence4}
\end{align}
One can see that this is precisely the region for the interference channel when both messages are required to be decoded at both decoders, except that we have close-to-independence instead of exact independence. The difficulty with condition \eqref{eq:strong_region_dependence4} is not just that $X_1^n,X_2^n$ are not perfectly independent, but that the dependence between individual letters $X_{1\c t},X_{2\c t}$ may vary depending on $t$. The method of Dueck in \cite{Dueck1981}  (also similar to Ahlswede's ``wringing'' technique \cite{Ahlswede1982}) allows us to show that for most $t\in[1:n]$, the letters $X_{1\c t},X_{2\c t}$ are nearly independent. This will allow single-letterization of the region in \eqref{eq:strong_region_dependence1}--\eqref{eq:strong_region_dependence4}. In particular, there exist some $m\le \sqrt{n k_n}$ and $t_1,\ldots,t_m\in[1:n]$, where for all $t\in[1:n]$
\be\label{eq:wringing_independence}
I(X_{1\c t};X_{2\c t}|Q')\le\sqrt{\frac{k_n}{n}}.
\ee
where
\be
Q'=(Z^n,X_{1\c t_1},\ldots,X_{1\c t_m},X_{2\c t_1},\ldots,X_{2\c t_m}).
\ee
We reproduce the essential proof of this fact from \cite{Dueck1981} as follows. First, \edit{let
\be\label{eq:calT1}
\calT_1=\left\{t\in[1:n]: I(X_{1\c t};X_{2\c t}|Z^n)>\sqrt{\frac{k_n}{n}}\right\}.
\ee
If $\calT_1$ is empty, then we may take $m=0$ and we are done. Otherwise, let $t_1$ be any element of $\calT_1$.}
We may write
\begin{align}
I(X_1^n;X_2^n|Z^n,X_{1\c t_1},X_{2\c t_1})
&=I(X_1^n;X_2^n|Z^n)-I(X_1^n;X_{2\c t_1}|Z^n)-I(X_{1\c t_1};X_2^n|Z^n,X_{2\c t_1})\label{eq:dueck1}
\\&\le I(X_1^n;X_2^n|Z^n)-I(X_{1\c t_1};X_{2\c t_1}|Z^n)
\\&\le k_n-\sqrt{\frac{k_n}{n}}.\label{eq:dueck2}
\end{align}
\edit{where \eqref{eq:dueck2} follows from \eqref{eq:strong_region_dependence4} and the fact that $t_1\in\calT_1$ as defined in \eqref{eq:calT1}. Next, let
\be
\calT_2=\left\{t\in[1:n]: I(X_{1\c t};X_{2\c t}|Z^n,X_{1\c t_1},X_{2\c t_1})>\sqrt{\frac{k_n}{n}}\right\}.
\ee
If $\calT_2$ is empty, then we may take $m=1$ and again we are done. Otherwise, take $t_2$ to be any element of $\calT_2$, and proceed as above.}
This process must terminate after a finite number (say $m$) of steps, at which point \eqref{eq:wringing_independence} must hold for all $t$. By a similar argument as in \eqref{eq:dueck1}--\eqref{eq:dueck2}, \edit{for each $i\in[1:m]$
\be
I(X_1^n;X_2^n|Z^n,X_{1\c t_1},\ldots,X_{1\c t_i},X_{2\c t_1},\ldots,X_{2\c t_i})\le k_n-i\sqrt{\frac{k_n}{n}}
\ee
and in particular}
\be\label{eq:m_bound}
I(X_1^n;X_2^n|Q')
\le k_n-m\sqrt{\frac{k_n}{n}}.
\ee
Since the mutual information is nonnegative, we have $m\le \sqrt{n k_n}$.

We now have
\begin{align}
I(X_1^n;Y_3^n|X_2^n,Z^n)
&\le I(X_1^n;Y_3^n|X_2^n,Q')+H(X_{1\c t_1},\ldots,X_{1\c t_m},X_{2\c t_1},\ldots,X_{2\c t_m})
\\&\le I(X_1^n;Y_3^n|X_2^n,Q')+m \log |\calX_1|\cdot|\calX_2|
\\&\le I(X_1^n;Y_3^n|X_2^n,Q')+\sqrt{nk_n} \log |\calX_1|\cdot|\calX_2|
\\&= \sum_{t=1}^n I(X_1^n;Y_{3\c t}|Y_3^{t-1},X_2^n,Q')+n\sqrt{\edit{nk_n}} \log |\calX_1|\cdot|\calX_2|
\\&\le \sum_{t=1}^n I(X_{1t};Y_{3\c t}|X_{2\c t},Q')+n\sqrt{\edit{nk_n}} \log |\calX_1|\cdot|\calX_2|
\\&= n I(X_1;Y_3|X_2,Q)+n\sqrt{\edit{nk_n}} \log |\calX_1|\cdot|\calX_2|
\end{align}
where
\be
Q''\sim \text{Unif}[1:n],\ 
Q=(Q',Q''),\ 
X_1=X_{1\c Q''},\ 
X_2=X_{2\c Q''},\ 
Y_3=Y_{3\c Q''},\ 
Y_4=Y_{4\c Q''}.
\ee
Applying \eqref{eq:strong_region_dependence1}, and performing similar analyses for \eqref{eq:strong_region_dependence2}--\eqref{eq:strong_region_dependence3}, combined with \eqref{eq:wringing_independence}, we have
\begin{align}
R_1&\le I(X_1;Y_3|X_2,Q)+\frac{k_n}{n}+\eps_n+\sqrt{\frac{k_n}{n}} \log |\calX_1|\cdot|\calX_2|,\label{eq:final_strong_region1}\\
R_2&\le I(X_2;Y_4|X_1,Q)+\frac{k_n}{n}+\eps_n+\sqrt{\frac{k_n}{n}} \log |\calX_1|\cdot|\calX_2|,\\
R_1+R_2&\le \min\{I(X_1,X_2;Y_3|Q),\ I(X_1,X_2,Y_4|Q)\}+\frac{3k_n}{n}+2\eps_n+\sqrt{\frac{k_n}{n}} \log |\calX_1|\cdot|\calX_2|,\\
\sqrt{\frac{k_n}{n}}&\ge I(X_1;X_2|Q).\label{eq:final_strong_region4}
\end{align}
\edit{Using standard tools to bound the cardinality of auxiliary random variables (e.g., \cite[Appendix C]{ElGamalKim:11Book}), for each $n$, there exists a joint distribution $\nedit{P^{(n)}_{QX_1X_2}}$ with $|\calQ|\le 5$ that preserves the value of each mutual information quantity in \eqref{eq:final_strong_region1}--\eqref{eq:final_strong_region4}. \nedit{Recall that we started with a different code for each blocklength $n$, so the above procedure results in a different joint distribution $P^{(n)}_{QX_1X_2}$ for each $n$.} This constitutes a sequence of joint distributions on a compact set, so there exists a convergent subsequence, with limit $P_{QX_1X_2}$. Since $k_n\oin o(n)$, $\eps_n\to 0$, and mutual information is continuous for fixed alphabets, this limiting distribution must satisfy \eqref{eq:strong_region1}--\eqref{eq:strong_region3}; moreover, in the limit \eqref{eq:final_strong_region4} implies that $I(X_1;X_2|Q)=0$, we may factor the joint distribution as $P_Q P_{X_1|Q}P_{X_2|Q}$. Finally, we may further reduce the cardinality of the auxiliary random variable in \eqref{eq:strong_region1}--\eqref{eq:strong_region3} to $|\calQ|\le 4$.}
\end{IEEEproof}

\section{Conclusions}\label{sec:conclusion}

This paper explored the relationship between edge removal properties and strong converses. Our main results are summarized in Fig.~\ref{fig:properties}. We found three main levels of properties for both edge removal and strong converse, and showed that for a very large class of networks, the strong converse property implies the corresponding edge removal property. Implications in the opposite direction hold for deterministic networks and sometimes for memoryless stationary networks.

Our strongest results are those for the ``middle'' level in Fig.~\ref{fig:properties}, connecting the weak edge removal property to the exponentially strong converse. In particular, we showed that these properties are equivalent for all discrete memoryless stationary networks. Thus, if an existing weak converse or outer bound can be strengthened to show that it still holds in the presence of an extra link carrying a sub-linear number of bits, then the converse or outer bound also holds in an exponentially strong sense, meaning that for any rate vector outside the region, the probability of error converges to 1 exponentially fast. It appears that many existing arguments can be strengthened in this sense with relatively little effort, thereby proving exponentially strong results. We believe that this middle level deserves more focus than it has received so far, because exponentially strong converses and weak edge removal properties seem to hold for so many problems (at least under average probability of error). Therefore, one should always ask whether a given converse proof can be strengthened in this sense.

Several open problems remain:
\begin{enumerate}
\item The most important question is whether edge removal and strong converse properties hold in general. In particular, we know of no memoryless stationary network for which the weak edge removal property or the exponentially strong converse does not hold under average probability of error. The techniques of Sec.~\ref{sec:applications} seem to allow one to prove a weak edge removal property (and thus an exponentially strong converse) for most (perhaps all) existing single-letter outer bounds, but there is no apparent way to do this without an existing single-letter result. Our observation that the properties hold for the discrete broadcast channel suggest that it may be possible to prove such results even for problems \edit{without known single-letter characterizations of the capacity region}, but we know of no other cases for which this has been done.
\item Many of our results (particularly those showing that edge removal implies a strong converse) apply only for discrete channel coding problems; generalizing these results to continuous systems, \edit{channel cost constraints,} source coding contexts, and random channel state would allow applicability to many other important network information theory problems. 
\item We conjecture that an equivalence holds for discrete memoryless networks on the ``lower layer'' in Fig.~\ref{fig:properties}, between very weak edge removal and the ordinary strong converse, but we have only been able to prove this result for deterministic networks and acyclic networks of
independent point-to-point links. 
\item Finally, it would be interesting to find a strong converse property equivalent to the extremely weak edge removal property.
\end{enumerate}

\section*{Acknowledgements}

The authors would like to thank Vincent Y.~F.~Tan, Michelle Effros, and Silas L.~Fong for helpful discussions and feedback.

\appendices

\section{Proof of Proposition~\ref{prop:error_sequences}}\label{appendix:error_sequences}

We will show that $\calC(\calN,\edit{(\eps_n)_n})\subseteq\calC(\calN,\edit{(\tilde\eps_n)_n})$; the opposite direction follows by reversing the roles of $\eps_n$ and $\tilde\eps_n$. Fix any rate vector
\be
\bR\in \bigcup_{n_0\in\bbN}\ \bigcap_{n\ge n_0} \calR\edit{(\calN,n,\eps_n)}.
\ee
We aim to show that $\bR\in\calC(\calN,\edit{(\tilde\eps_n)_n})$. There exists $n_0\in\bbN$ such that for all $n\ge n_0$, $\bR\in\calR\edit{(\calN,n,\eps_n)}$. By the assumption of the lemma, there exists a subsequence $n_i$ such that
\beq\label{eq:eps_limit}
\lim_{i\to\infty} -\frac{1}{n_i} \log(1-\eps_{n_i})=\alpha.
\eeq
For sufficiently large $i$, we have $n_i\ge n_0$, so  $\bR\in\calR\edit{(\calN,n_i,\eps_{n_i})}$. That is, there exists an $n_i$-length code with rate $\bR$ and probability of error at most $\eps_{n_i}$. Fix integer $N$, and form a new code on network $\calN$ of length $n_iN$ and rate $\frac{N-2}{N}\bR$ as follows. Roughly, \edit{reduce the overall probability of error by repeating} the original code $N$ times, and \edit{introducing a small amount of error correction in the form of} an outer \edit{maximum distance separable (MDS)} code \edit{\cite[Chap.~4]{Roth2006}}. In particular, for each node $v\in[1:d]$ where $R_v>0$, form a $(N,N-2)$ MDS code on symbols from the finite field of order $2^{\lfloor n_i R_v\rfloor}$. This code exists for sufficiently large $i$ \edit{(e.g., a Reed-Solomon code \cite[Chap.~5]{Roth2006})}. Let the MDS codeword be denoted by $(W_v(1),\ldots,W_v(N))$. Repeat the original code $N$ times, where on the $\ell$th repetition $W_v(\ell)$ is treated as the message originating at node $v$. Because each outer code is MDS, one error can be corrected, so if it most one of the $N$ repetitions results in an error, the full code will decode correctly. Because the network is memoryless and stationary, each repetition is independent and results in error with probability $\eps_{n_{\edit{i}}}$, so the probability of error for the full code is given by
\begin{align}
\nedit{\mathrm{P}_\mathrm{e}}
&=1-(1-\eps_{n_i})^N\edit{-}N\eps_{n_i}(1-\eps_{n_i})^{N-1}
\\&=1-(1-\eps_{n_i})^{N-1}\left[ 1-\eps_{n_i}+N\eps_{n_i}\right].
\end{align}
Note that \eqref{eq:eps_limit} and the assumption that $\alpha>0$ imply that $\eps_{n_i}\to 1$, meaning $1-\eps_{n_i}+N\eps_{n_i}\to N$. Thus
\begin{align}
\lim_{i\to\infty}  \frac{1}{n_i}\log(1-\nedit{\mathrm{P}_\mathrm{e}})
&=\lim_{i\to\infty} \frac{1}{n_i}\big[ (N-1)\log (1-\eps_{n_i})+N\big]
\\&= -(N-1)\alpha.\label{eq:i_limit}
\end{align}
In particular, for sufficiently large $i$, we have
\be
1-\nedit{\mathrm{P}_\mathrm{e}}\ge \exp\{-n_i (N-1/2)\alpha\}
\ee
Hence, for any $N$ and sufficiently large $i$,
\be
\frac{N-2}{N}\bR\in \calR\edit{(\calN,n_iN,1-\exp\{-n_i(N-1/2)\alpha\})}.
\ee

Consider any blocklength $m$ where $n_iN\le m\le n_i(N+1)$. We may convert a code with blocklength $n_iN$ to one with blocklength $m$ simply by ignoring the additional $m-n_iN$ symbols. This reduces the rate by a factor of $\frac{n_iN}{m}\ge \frac{N}{N+1}$, but does not change the probability of error. Thus we have
\be
\frac{N-2}{N+1} \bR\in \calR\edit{(\calN,m,1-\exp\{-n_i(N-1/2)\alpha\})}.
\ee
By the liminf assumption on $\tilde\eps_n$ \edit{in \eqref{eq:liminf_equality}}, for sufficiently large $m$ we have
\beq\label{eq:m_assumption}
-\frac{1}{m} \log (1-\tilde\eps_m)\ge \frac{N-1/2}{N}\alpha.
\eeq
Thus, if $m \ge n_i N$, we have
\begin{align}
\tilde\eps_m
&\ge 1-\exp\left\{-m\frac{N-1/2}{N} \alpha\right\}\label{eq:m_derive1}
\\&\ge 1-\exp\{-n_i (N-1/2)\alpha\}
\end{align}
where \eqref{eq:m_derive1} holds by \eqref{eq:m_assumption} for sufficiently large $i$. Hence, for any $N$, for all $m$ sufficiently large we have
\be
\frac{N-2}{N+1}\bR\in\calR\edit{(\calN,m,\tilde\eps_m)}.
\ee
Thus
\be\label{eq:prop1_almost_done}
\frac{N-2}{N+1}\bR\in\calC(\calN,\edit{(\tilde\eps_n)_n}).
\ee
\edit{Since \eqref{eq:prop1_almost_done}} holds for all $N$, \edit{and $\calC(\calN,(\tilde\eps_n)_n)$ is closed}, we have $\bR\in\calC(\calN,\edit{(\tilde\eps_n)_n})$. \edit{Note that both $i$ and $N$ must go to infinity, but $i$ converges to infinity first for fixed $N$ in \eqref{eq:i_limit}.}

\section{Proof of Proposition~\ref{prop:strong_converse_defs}}\label{appendix:strong_converse_defs}

\emph{Extremely strong converse $\Leftrightarrow$ (1b)}: By taking $\gamma=K\alpha$, the extremely strong converse holds if and only if, for any $\alpha\ge 0$,
\be
\calC(\calN,\edit{(1-2^{-n\alpha})_n})\subseteq \calC(\calN,0^+)+\edit{[0,K\alpha]}.
\ee
By Proposition~\ref{prop:error_sequences}, $\calC(\calN,\edit{(\eps_n)_n})=\calC(\calN,\edit{(1-2^{-n\alpha})_n})$ if $1-\eps_n\doteq 2^{-n\alpha}$. This proves that the extremely strong converse is equivalent to the condition in (1b).

\emph{(1a) $\Rightarrow$ (1b)}. Consider any $\eps_n$ where $1-\eps_n\doteq 2^{-n\alpha}$, and any $\bR\in\calC(\calN,\edit{(\eps_n)_n})$. If $\bR\in\calC(\calN,0^+)$, then obviously $\bR\in\calC(\calN,0^+)+\edit{[0,K\alpha]^d}$. If $\bR\notin\calC(\calN,0^+)$, then by condition (1a) we have $\alpha\ge \beta/K$, and $\bR\in\calC(\calN,0^+)+\edit{[0,\beta]^d}$. Thus $\bR\in\calC(\calN,0^+)+\edit{[0,K\alpha]^d}$. This proves (1b).

\emph{(1b) $\Rightarrow$ (1a)}. Consider any $\bR\notin\calC(\calN,0^+)$, and any sequence of $(\bR,n)$ codes with probability of error $\eps_n$. By Proposition~\ref{prop:error_sequences}, this implies $\bR\in\calC(\calN,\edit{(1-2^{-n\alpha})_n})$, where 
\be
\alpha=\liminf_{n\to\infty} -\frac{1}{n}\log(1-\eps_n).
\ee
Hence, by condition (1b), $\bR\in\calC(\calN,0^+)+\edit{[0,K\alpha]^d}$. If $\beta$ is the smallest number such that $\bR\in\calC(\calN,0^+)+\edit{[0,\beta]^d}$, then we have $\beta\le K\alpha$. This proves \eqref{eq:extremely_strong_converse_alternative}, and hence (1c).

\emph{Exponentially strong converse $\Rightarrow$ (2b).} Let $\eps_n$ be a sequence where $-\log(1-\eps_n)\oin o(n)$. By the exponentially strong converse, for any $\gamma>0$ there exists $\eps'_n$ where $-\log(1-\eps'_n)\oin\Theta(n)$ where \eqref{eq:strong_converse_condition} holds. For sufficiently large $n$, $-\log(1-\eps_n)\le \edit{-}\log(1-\eps'_n)$, meaning $\eps_n\le \eps'_n$. Thus
\be
\calC(\calN,\edit{(\eps_n)_n})\subseteq\calC(\calN,\edit{(\eps'_n)_n})\subseteq\calC(\calN,0^+)+\edit{[0,\gamma]^d}.
\ee
As this holds for all $\gamma>0$, we have $\calC(\calN,\edit{(\eps_n)_n})\edit{\subseteq}\calC(\calN,0^+)$. This proves condition (2b).

\emph{(2b) $\Rightarrow$ Exponentially strong converse.} Specifically, we prove that if the exponentially strong converse does not hold, then condition (2b) does not hold. Suppose there exist $\gamma>0$ such that for all $\eps_n$ where $-\log(1-\eps_n)\oin\Theta(n)$, $\calC(\calN,\edit{(\eps_n)_n})\not\subseteq\calC(\calN,0^+)+\edit{[0,\gamma]^d}$. Specifically, for any integer $r$, $\calC(\calN,\edit{(1-\exp\{-n/r\})_n})\not\subseteq\calC(\calN,0^+)+\edit{[0,\gamma]^d}$. Since the sets $\calC(\calN,\edit{(1-\exp\{-n/r\})_n})$ are sorted (decreasing as $r$ grows), there exists $\bR$ in the interior of $\calC(\calN,\edit{(1-\exp\{-n/r\})_n})$ for all integers $r$ such that $\bR\notin\calC(\calN,0^+)$. For all $r$, there exists $n_0(r)$ such that for all $n\ge n_0(r)$,
\be\label{eq:exp_SC_nr}
\bR\in \calR\edit{(\calN,n,1-\exp\{-n/r\})}.
\ee
Define a sequence
\be
\eps_n=\min_{r:n\ge n_0(r)} 1-\exp\{-n/r\}.
\ee
Note that $-\log(1-\eps_n)\le n/r$ for $n\ge n_0(r)$, so $-\log(1-\eps_n)\oin o(n)$. Moreover, for any $n$, there is some $r$ such that $n\ge n_0(r)$ and $\eps_n=1-\exp\{-n/r\}$, so by \eqref{eq:exp_SC_nr}, $\bR\in\calR\edit{(\calN,n,\eps_n)}$ for all $n$. Thus $\bR\in\calC(\calN,\edit{(\eps_n)_n})$. But since $\bR\notin\calC(\calN,0^+)$, (2b) does not hold.

\emph{(2a) $\Rightarrow$ (2b)}. By (2a), for any $\bR\notin\calC(\calN,0^+)$, the probability of \edit{correct decoding} must vanish exponentially fast, so $\bR\notin\calC(\calN,\edit{(\eps_n)_n})$ for any sequence  $\eps_n$ such that $-\log(1-\eps_n)\oin o(n)$. Therefore $\calC(\calN,\edit{(\eps_n)_n})\edit{\subseteq}\calC(\calN,0^+)$, which proves (2b). 

\emph{(2b) $\Rightarrow$ (2a).} For any $\bR\notin\calC(\calN,0^+)$ and any sequence $\eps_n$ for which $\bR\in\calC(\calN,\edit{(\eps_n)_n})$, it cannot be that $-\log(1-\eps_n)\oin o(n)$, or else by (2b) we would have $\bR\in\calC(\calN,\edit{0^+})$. Therefore $\eps_n$ must approach 1 exponentially fast, which proves (2a).

\emph{Strong converse $\Rightarrow$ (3b).} Note that the condition in the definition of the strong converse that $-\log(1-\eps_n)\to\infty$ can be more simply written as $\eps_n\to 1$. Consider any $\eps\in(0,1)$. By the strong converse, for any $\gamma>0$, there exists a sequence $\eps_n\to 1$ where $\calC(\calN,\edit{(\eps_n)_n})\subseteq\calC(\calN,0^+)+\edit{[0,\gamma]^d}$. Noting that $\eps\le \eps_n$ for sufficiently large $n$, we have $\calC(\calN,\edit{(\eps)_n})\subseteq\calC(\calN,\edit{(\eps_n)_n})\subseteq\calC(\calN,0^+)+\edit{[0,\gamma]^d}$. As this holds for all $\gamma>0$, we have $\calC(\calN,\edit{(\eps)_n})=\calC(\calN,0^+)$, which proves (3b).

\emph{(3b) $\Rightarrow$ (3c).} By (3b), for any integer $r$, $\calC(\calN,\edit{(1-1/r)_n})=\calC(\calN,0^+)$. In particular, there exists $n_0(r)$ such that for all $n\ge n_0(r)$, 
\be
\calR\edit{\left(\calN,n,1-\frac{1}{r}\right)}\subseteq \calC(\calN,0^+)+\edit{\left[0,\frac{1}{r}\right]^d}.
\ee
Define a sequence 
\be
\eps_n=\sup_{r: n\ge n_0(r)} 1-\frac{1}{r}.
\ee
Certainly $\eps_n\ge 1-1/r$ for $n\ge n_0(r)$, meaning $\eps_n\to 1$. Moreover, if $n,r$ are such that $\eps_n=1-\frac{1}{r}$, then 
\be
\calR\edit{(\calN,n,\eps_n)}=\calR\left(\calN,\edit{n},1-\frac{1}{r}\right)\subseteq \calC(\calN,0^+)+\edit{\left[0,\frac{1}{r}\right]^d}
=\calC(\calN,0^+)+\edit{[0,1-\eps_n]^d}.
\ee
Since $1-\eps_n\to 0$, we have
\be
\calC(\calN,\edit{(\eps_n)_n})=\calC(\calN,0^+).
\ee
This proves (3c).

\emph{(3c) $\Rightarrow$ Strong converse.} By (3c), there exists a sequence $\eps_n\to\edit{1}$ where $\calC(\calN,\edit{(\eps_n)_n})=\calC(\calN,0^+)\subseteq\calC(\calN,0^+)+\edit{[0,\gamma]^d}$ for all $\gamma>0$. This proves the strong converse. 

\emph{(3c) $\Rightarrow$ (3a).} By (3c), there exists $\eps_n\to 1$ where $\bR\notin\calC(\calN,\edit{(\eps_n)_n})$ for any $\bR\notin\calC(\calN,0^+)$. This implies that any sequence of $(\bR,n)$ codes must have probability of error exceeding $\eps_n$ for sufficiently large $n$, so the probability of error must approach 1, which proves (3a).

\emph{(3a) $\Rightarrow$ (3b).} For any $\eps\in(0,1)$, by (3a) any $\bR\notin\calC(\calN,0^+)$ has probability of error approaching 1, so $\bR\notin\calC(\calN,\edit{(\eps)_n})$. Therefore, $\calC(\calN,\edit{(\eps)_n})=\calC(\calN,0^+)$, which proves (3b).

\section{Proof of Proposition~\ref{prop:above_capacity}}\label{appendix:above_capacity}


Consider a channel where \eqref{eq:above_capacity_condition} holds. For any $Q_{X,Y}$, we may write
\begin{align}
D(Q_{Y|X}\|P_{Y|X}|Q_X)
&= \sum_{x,y} Q_{X,Y}(x,y)\log \frac{Q_{Y|X}(y|x)}{P_{Y|X}(y|x)}\label{eq:rel_entr_derive1}
\\&= \sum_{x,y} Q_{X,Y}(x,y)\left[\log \frac{Q_{Y|X}(y|x)}{Q_Y(y)}-\log \frac{P_{Y|X}(y|x)}{P_Y(y)}+\log \frac{Q_Y(y)}{P_Y(y)}\right]
\\&= I_{Q_{X,Y}}(X;Y)-\sum_{x,y} Q_{X,Y}(x,y) \log \frac{P_{Y|X}(y|x)}{P_Y(y)}+D(Q_Y\|P_Y)\label{eq:rel_entr_derive3}
\\&\ge I_{Q_{X,Y}}(X;Y)-C\label{eq:alpha_bd1}
\end{align}
where \eqref{eq:alpha_bd1} follows from \eqref{eq:above_capacity_condition}, and the fact that relative entropy is non-negative. Thus, we may lower bound $\alpha(R)$ by
\begin{align}
\alpha(R)&\ge \min_{Q_{X,Y}} \Big[I_{Q_{X,Y}}(X;Y)-C+|R-I_{Q_{X,Y}}(X;Y)|^+\Big]
\\&\ge R-C\label{eq:alpha_bd2}
\end{align}
where \eqref{eq:alpha_bd2} holds because $x+|y-x|^+\ge y$ for any real numbers $x,y$. This lower bound is achievable by setting $Q_{X,Y}=P_X \times P_{Y|X}$, where $P_X$ is any capacity-achieving input distribution, so indeed $\alpha(R)=R-C$.

Now consider a channel where \eqref{eq:above_capacity_condition} does not hold. That is, there exists some $x_0,y_0$ where 
\be\label{eq:x0y0}
\log \frac{P_{Y|X}(y_0|x_0)}{P_Y(y_0)}>C.
\ee
Let $P_X$ be any capacity-achieving input distribution. Thus, 
\be
\sum_{x,y} P_X(x) P_{Y|X}(y|x)\log \frac{P_{Y|X}(y|x)}{P_Y(y)}=C.
\ee
In particular, there exists some $x_1,y_1$ where 
\be\label{eq:x1y1}
\log\frac{P_{Y|X}(y_1|x_1)}{P_Y(y_1)}\le C
\ee 
and $P_X(x_1) P_{Y|X}(y_1|x_1)>0$. \nedit{For parameter $\lambda\ge 0$,} define a joint distribution $Q_{X,Y}^{(\lambda)}$ where
\be
Q_{X,Y}^{(\lambda)}(x,y)=P_X(x)P_{Y|X}(y|x)+\lambda 1(x=x_0,y=y_0)-\lambda 1(x=x_1,y=y_1).
\ee
 As long as $0\le\lambda\le P_X(x_1) P_{Y|X}(y_1|x_1)$, this is a valid distribution. \nedit{If we marginalize out $X$, we see that
\be
Q_Y^{(\lambda)}(y)=P_Y(y)+\lambda 1(y=y_0)-\lambda 1(y=y_1).
\ee 
By \cite[Lemma~17.3.3]{CoverThomas:Book}, the first term in the Taylor expansion for $D(Q_Y^{(\lambda)}\|P_Y)$ around $\lambda=0$ is
\be\label{eq:first_term_taylor}
\frac{1}{2}\sum_y \frac{(Q_Y^{(\lambda)}(y)-P_Y(y))^2}{P_Y(y)}=
\frac{\lambda^2}{2}\left(\frac{1}{P_Y(y_0)}+\frac{1}{P_Y(y_1)}\right).
\ee
By \cite[Cor.~1 in Sec.~4.5]{Gallager1968}, $P_Y(y)>0$ for all $y$ that are reachable from some input symbol. Note that \eqref{eq:x0y0} implies that $P_{Y|X}(y_0|x_0)>0$, and also by assumption $P_{Y|X}(y_1|x_1)>0$. That is, both $y_0$ and $y_1$ are reachable output symbols, so $P_Y(y_0),P_Y(y_1)>0$. Thus in \eqref{eq:first_term_taylor} the coefficient on $\lambda^2$ is finite, and so
\be\label{eq:rel_entr_der}
\frac{d}{d \lambda} D(Q_Y^{(\lambda)}\|P_Y)\Big|_{\lambda=0}=0
\ee}%
Noting that
\be
\frac{\partial}{\partial Q_{XY}(x,y)} I_{Q_{XY}}(X;Y)=\log \frac{Q_{Y|X}(y|x)}{Q_Y(y)}-1
\ee
we have
\be\label{eq:zeta_def}
\zeta:=\frac{d}{d \lambda} I_{Q^{(\lambda)}_{X,Y}}(X;Y)\Big|_{\lambda=0}=\log \frac{P_{Y|X}(y_0|x_0)}{P_Y(y_0)}-\log \frac{P_{Y|X}(y_1|x_1)}{P_Y(y_1)}>0
\ee
where we have used the assumptions in \eqref{eq:x0y0} and \eqref{eq:x1y1}. 
\nedit{
Applying the derivation in \eqref{eq:rel_entr_derive1}--\eqref{eq:rel_entr_derive3}, we have
\begin{align}
&\frac{d}{d\lambda} D(Q^{(\lambda)}_{Y|X}\|P_{Y|X}|Q_X^{(\lambda)})\Big|_{\lambda=0}
\\&=\frac{d}{d\lambda}
\left[ I_{Q^{(\lambda)}_{X,Y}}(X;Y)-\sum_{x,y} Q^{(\lambda)}_{X,Y}(x,y)\log \frac{P_{Y|X}(y|x)}{P_Y(y)}+D(Q_Y^{(\lambda)}\|P_Y)\right]_{\lambda=0}\label{eq:lambda_three_terms}
\\&=0\label{eq:D_diff_0}
\end{align}
where we have used \eqref{eq:rel_entr_der}, \eqref{eq:zeta_def}, and the fact that $\zeta$ is also the derivative of the second term in \eqref{eq:lambda_three_terms}.
}
Given $\lambda$ small enough so that $Q^{(\lambda)}_{X,Y}$ is a valid distribution, we may upper bound 
\be
\alpha(C+\zeta\lambda)\le D(Q_{Y|X}^{(\lambda)}\|P_{Y|X}|Q_{\nedit{X}}^{(\lambda)})+|C+\zeta\lambda-I_{Q^{(\lambda)}_{X,Y}}(X;Y)|^+.
\ee
Thus,
\begin{align}
\frac{d\alpha(R)}{dR}\Big|_{R=C}&=\lim_{\lambda\to 0} \frac{\alpha(C+\zeta\lambda)}{\zeta\lambda}
\\&\le \lim_{\lambda \to 0} \frac{1}{\zeta\lambda}\Big[D(Q_{Y|X}^{(\lambda)}\|P_{Y|X}|Q_Y^{(\lambda)})+|C+\zeta\lambda-I_{Q^{(\lambda)}_{X,Y}}(X;Y)|^+\Big]
\\&=\frac{1}{\zeta} \frac{d}{d\lambda} D(Q_{Y|X}^{(\lambda)}\|P_{Y|X}|Q_Y^{(\lambda)})\Big|_{\lambda=0} +\left|1-\frac{1}{\zeta}\frac{d}{d\lambda} I_{Q^{(\lambda)}_{X,Y}}(X;Y)\Big|_{\lambda=0}\right|^+\label{eq:zeta_derive3}
\\&=0\label{eq:zeta_derive4}
\end{align}
where \nedit{in \eqref{eq:zeta_derive3} we have used the fact that $Q_{X,Y}^{(0)}=P_X\times P_{Y|X}$, so $I_{Q^{(0)}_{X,Y}}(X;Y)=C$; and \eqref{eq:zeta_derive4}} follows from the definition of $\zeta$ in \eqref{eq:zeta_def}, as well as \eqref{eq:D_diff_0}. Note also that this derivation is valid only because $\zeta>0$, as shown in \eqref{eq:zeta_def}. Since $\alpha(R)$ is non-decreasing in $R$, we must have $\frac{d\alpha(R)}{dR}\big|_{R=C}=0$.

\section{Proof of Proposition~\ref{prop:edge_removal_defs}}\label{appendix:edge_removal_defs}

Statement 1 follows immediately from the definition of the strong edge removal property.

We now prove statement 2. Suppose the weak edge removal property holds. Thus, for any $\gamma>0$, there exists a sequence $k_n\oin\edit{\Theta}(n)$ satisfying \eqref{eq:edge_removal_condition}. Let
\be
\delta'=\liminf_{n\to\infty} \frac{k_n}{n}.
\ee
Note that $\delta'$, and so for any $0<\delta<\delta'$, we have $\delta n\le k_n$ for sufficiently large $n$. Thus
\be
\calC(\calN,0^+,\edit{(\delta n)_n})\subseteq\calC(\calN,0^+,\edit{(k_n)_n})\subseteq\calC(\calN,0^+)+\edit{[0,\gamma]^d}.
\ee
Hence, the \edit{LHS} of \eqref{eq:weak_alternative1} is contained in $\calC(\calN,0^+)+\edit{[0,\gamma]^d}$. Since this holds for all $\gamma>0$, this proves \eqref{eq:weak_alternative1}.

Now we show that \eqref{eq:weak_alternative1} implies the weak edge removal property. For any $\gamma>0$, by \eqref{eq:weak_alternative1} there exists $\delta>0$ such that $\calC(\calN,0^+,\edit{(\delta n)_n})=\calC(\calN,0^+)+\edit{[0,\gamma]^d}$. Thus, setting $k_n=\delta n$ satisfies \eqref{eq:edge_removal_condition}. This proves the weak edge removal property.

To prove that the weak edge removal property is also equivalent to \eqref{eq:weak_alternative2}, we will show that
\beq\label{eq:kn_vs_dn}
\bigcup_{k_n\oin o(n)} \calC(\calN,0^+,\edit{(k_n)_n}) = \bigcap_{\delta>0} \calC(\calN,0^+,\edit{(\delta n)_n}).
\eeq
To show $\subseteq$ in \eqref{eq:kn_vs_dn}, we need to show that for all $k_n\oin o(n)$, $\calC(\calN,0^+,\edit{(k_n)_n})$ is contained in the RHS of \eqref{eq:kn_vs_dn}, or that $\calC(\calN,0^+,\edit{(k_n)_n})\subseteq\calC(\calN,0^+,\edit{(\delta n)_n})$ for all $\delta>0$. Indeed this holds because for any $k_n\oin o(n)$ and any $\delta>0$, $k_n\le \delta n$ for sufficiently large $n$. To show $\supseteq$ in \eqref{eq:kn_vs_dn}, let $\bR$ be in the RHS of \eqref{eq:kn_vs_dn}. Thus, for all $\eps,\delta,\gamma>0$, for sufficiently large $n$ we have $\bR\in\calR(\calN,\edit{n},\eps,n\delta)\edit{+[0,\gamma]^d}$. In particular, \edit{for any fixed integer $r$, we may let $\eps=\delta=\gamma=1/r$, so there exists $n_0(r)$ such that for all $n\ge n_0(r)$ we have
\be\label{eq:any_r}
\bR\in\calR\left(\calN,n,\frac{1}{r},\frac{n}{r}\right)+\left[0,\frac{1}{r}\right]^d.
\ee}%
\edit{Let
\be
r_n=\max\{r: n_0(r)\le n\}.
\ee
By \eqref{eq:any_r}, for any $n$ we have
\be\label{eq:rn_form}
\bR\in \calR\left(\calN,n,\frac{1}{r_n},\frac{n}{r_n}\right)+\left[0,\frac{1}{r_n}\right]^d.
\ee
Letting $k_n=\frac{n}{r_n}$, we may rewrite \eqref{eq:rn_form} as
\be\label{eq:kn_form}
\bR\in\calR\left(\calN,n,\frac{k_n}{n},k_n\right)+\left[0,\frac{k_n}{n}\right]^d.
\ee
Note that for any integer $r$, if $n\ge n_0(r)$, then $r_n\ge r$, so $k_n\le n/r$. Thus $k_n/n\to 0$; i.e., $k_n\oin o(n)$. From \eqref{eq:kn_form}, we have}
$\bR\in \calC(\calN,0^+,\edit{(k_n)_n})$. This proves $\supseteq$ in \eqref{eq:kn_vs_dn}.

We now prove statement 3. Note that the very weak edge removal property is equivalent to the statement that for all $\gamma>0$,
\be
\bigcap_{k_n:k_n\to\infty} \calC(\calN,0^+,\edit{(k_n)_n})\subseteq \calC(\calN,0^+)+\edit{[0,\gamma]^d}.
\ee
This is easily seen to be equivalent to \eqref{eq:very_weak_alternative1}.

To show that the very weak edge removal property is also equivalent to \eqref{eq:very_weak_alternative2}, we show  that
\be
\bigcap_{k_n:k_n\to\infty} \calC(\calN,0^+,\edit{(k_n)_n})=\bigcap_{\eps>0}\  \overline{\bigcup_{k\in\bbN} \calC(\calN,\edit{(\eps)_n},\edit{(k)_n})}.
\ee
Noting that
\be
\bigcap_{k_n:k_n\to\infty} \calC(\calN,0^+,\edit{(k_n)_n})
=\bigcap_{k_n:k_n\to\infty}\ \bigcap_{\eps>0} \calC(\calN,\edit{(\eps)_n,(k_n)_n})
=\bigcap_{\eps>0}\ \bigcap_{k_n:k_n\to\infty} \calC(\calN,\edit{(\eps)_n,(k_n)_n})
\ee
it is enough to show that for all $\eps>0$,
\beq\label{eq:kn_vs_k}
\bigcap_{k_n:k_n\to\infty} \calC(\calN,\edit{(\eps)_n,(k_n)_n})= \overline{\bigcup_{k\in\bbN} \calC(\calN,\edit{(\eps)_n,(k)_n})}.
\eeq
For any $k\in\bbN$ and any sequence $k_n\to\infty$, $k\le k_n$ for sufficiently large $n$. Thus
\be
\bigcap_{k_n:k_n\to\infty} \calC(\calN,\edit{(\eps)_n,(k_n)_n})\supseteq \bigcup_{k\in\bbN} \calC(\calN,\edit{(\eps)_n,(k)_n}).
\ee
Taking a closure yields $\supseteq$ in \eqref{eq:kn_vs_k}, since the LHS of \eqref{eq:kn_vs_k} is already closed. To prove the opposite direction, let $\gamma_k$ be a positive sequence where $\lim_{k\to\infty} \gamma_k\to 0$. \edit{For fixed $\eps\in(0,1)$ and $k\in\bbN$, by the definition of $\calC(\calN,(\eps)_n,(k)_n)$ in \eqref{eq:edge_capacity_region_def}, there exists} $n_0(k)$ such that for all $n\ge n_0(k)$, we have
\beq\label{eq:gammak}
\calR(\calN,\edit{n},\eps,k)\subseteq \calC(\calN,\edit{(\eps)_n,(k)_n})+\edit{[0,\gamma_k]^d}.
\eeq
Now define a sequence
\be
k_n=\max\{k: n\ge n_0(k)\}.
\ee
\edit{Note that for any $k\in\bbN$, $k_n\ge k$ for all $n\ge n_0(k)$, so} $k_n\to\infty$ as $n\to\infty$, because for any $k$, $k_n\ge k$ for all $n\ge n_0(k)$. Thus the LHS of \eqref{eq:kn_vs_k} is contained in $\calC(\calN,\edit{(\eps)_n,(k_n)_n})$. Moreover
\begin{align}
\calC(\calN,\edit{(\eps)_n,(k_n)_n})&=\overline{\bigcup_{n_0\in\bbN}\ \bigcap_{\edit{n'}\ge n_0}\calR(\calN,\edit{n'},\eps,k_{\edit{n'}})}\label{eq:kn0}
\\&\subseteq \overline{\bigcup_{n_0\in\bbN}\ \bigcap_{\edit{n'}\ge n_0} \big(\calC(\calN,\edit{(\eps)_n,(k_{\edit{n'}})_n})+\gamma_{k_{\edit{n'}}}\big)}\label{eq:kn1}
\\&= \overline{\bigcup_{n_0\in\bbN}\ \bigcap_{\edit{n'}\ge n_0} \calC(\calN,\edit{(\eps)_n,(k_{\edit{n'}})_n})}\label{eq:kn2}
\\& \subseteq \overline{\bigcup_{k\in\bbN} \calC(\calN,\edit{(\eps)_n,(k)_n})}\label{eq:kn4}
\end{align}
where \eqref{eq:kn0} holds by definition, \eqref{eq:kn1} follows from \eqref{eq:gammak}, \eqref{eq:kn2} holds because $\gamma_k\to 0$, \edit{and \eqref{eq:kn4} holds because for any $n'$, $k_{n'}$ is some integer}. This proves $\subseteq$ in \eqref{eq:kn_vs_k}.

We now prove statement 4. The definition of the extremely weak edge removal property may be equivalently written
\be\label{eq:extremely_weak_rewrite}
\bigcup_{\text{bounded }k_n} \calC(\calN,0^+,\edit{(k_n)_n})\subseteq\bigcap_{\gamma>0}\calC(\calN,0^+)+\edit{[0,\gamma]^d}.
\ee
Note that for any bounded $k_n$, $\calC(\calN,0^+,\edit{(k_n)_n})\subseteq\calC(\calN,0^+,\edit{(k)_n})$ for some constant integer $k$. Thus the LHS \eqref{eq:extremely_weak_rewrite} can be written
\be
\bigcup_{k\in\bbN} \calC(\calN,0^+,\edit{(k)_n}).
\ee
Moreover, the RHS of \eqref{eq:extremely_weak_rewrite} is simply $\calC(\calN,0^+)$. Therefore the extremely weak edge removal property is equivalent to \eqref{eq:extremely_weak_alternative}.

\section{Proof of Theorem~\ref{thm:p2p}}\label{appendix:p2p}

A significant technical tool in proving network equivalence \edit{(cf. see the discussion in Sec.~\ref{sec:p2p}, and the original result in \cite{Koetter2011})}  is the idea of channel simulation, in which a point-to-point channel is accurately simulated by any other with higher capacity. This idea was at the heart of the proof in \cite{Koetter2011}. A version of this idea was stated in \cite{Xiang2014} as the \emph{universal channel simulation lemma}, stated as follows. This lemma states that two nodes with shared randomness (represented by $U$) can use a noiseless link to accurately simulate a noisy channel, as long as the capacity of the noiseless link is greater than the capacity of the noisy channel. While \cite{Xiang2014} did not provide a proof, we presented a proof in the appendix of \cite{Kosut2017a}.

\begin{lemma}\label{lemma:sim_error_exp}
Let $(\calX,\edit{Q_{Y|X}},\calY)$ be a discrete memoryless channel with capacity $C$. Given a rate $R>C$, a channel simulation code $(f,g)$ consists of
\begin{itemize}
\item $f:\calX^n\times[0,1]\to \{0,1\}^{nR}$,
\item $g:\{0,1\}^{nR}\times[0,1]\to \calY^n$.
\end{itemize}
Let \edit{$P_{Y^n|X^n}$} be the conditional pmf of $Y^n$ given $X^n$ where $U\sim\text{Unif}[0,1]$ and
\be
Y^n=g(f(X^n,U),U).
\ee
There exists a sequence of length-$n$ simulation codes where
\beq
\lim_{n\to\infty}\, \max_{x^n}\, d_{\text{TV}}(\edit{P_{Y^n|X^n=x^n},Q_{Y^n|X^n=x^n}})=0.\label{eq:TV_bound}
\eeq
\end{lemma}

We now proceed to prove Theorem~\ref{thm:p2p}. By Theorem~\ref{thm:1direction}, we only need to show that the very weak edge removal property  implies the ordinary strong converse. The basic approach is to use network equivalence to convert a code for noisy network $\calN$ into a code on the noiseless version, then apply Lemma~\ref{lemma_det} on this noiseless network, and then again use network equivalence to convert back to the noisy network.

Let $\calE\subset[1:d]\times[1:d]$ be the set of pairs of nodes connected by point-to-point links. Recall that by assumption, the directed graph $([1:d],\calE)$ is acyclic. Thus\edit{, by \cite[Prop.~19.1]{Yeung2008}} we may assign each node $i$ a distinct integer $\pi_i\in[1:d]$ where $\pi_i<\pi_j$ if $(i,j)\in\calE$. For any $(i,j)\in\calE$, let $C_{i\to j}$ be the capacity of the link from $i$ to $j$. Assume without loss of generality that $C_{i\to j}>0$ for all $(i,j)\in\calE$. Let $C_{\min}=\min_{(i,j)\in\calE} C_{i\to j}$, so in particular $C_{\min}>0$. Denote $X_{i\to j}$ and $Y_{i\to j}$ as the input and output respectively of the link $(i,j)$. Thus the transmitted symbol from node $i$ can be written
\be
X_i=(X_{i\to j}:(i,j)\in\calE)
\ee
and the received symbol at node $j$ can be written
\beq\label{eq:Y_notation}
Y_j=(Y_{i\to j}:(i,j)\in\calE).
\eeq

Let $\bR$ be achievable with respect to fixed $\eps\in(0,1)$. Thus, for sufficiently large $n$, there exists a  length-$n$ code for network $\calN$ with rate $\bR$ and probability of error $\eps$. \edit{By \eqref{eq:encoding_function}--\eqref{eq:decoding_function}, this code is defined by encoding functions $\phi_{it}$ for each node $i\in[1:d]$ and time $t\in[1:n]$, and decoding functions $\psi_i$ for each node $i\in[1:d]$. It will be useful to work with coding functions on $n$-length blocks rather than single time instances, so we} define the block-wise encoding function at node $i$
\be\label{eq:blockwise_encoding}
\phi_i^n:[1:2^{nR_i}]\times \calY_i^n\to \calX_i^n
\ee
as
\be\label{eq:blockwise_encoding_def}
\phi_i^n(\edit{w_i},y_i^n) = (\phi_{i1}(\edit{w_i}),\phi_{i2}(\edit{w_i},y_{i1}),\ldots,\phi_{in}(\edit{w_i},y_{i}^{n-1})).
\ee
Using the notation in \eqref{eq:Y_notation}, we may \edit{notate the arguments to this function as}
\be
\phi_i^n(\edit{w_i},y_{\edit{k\to i}}^n:(\edit{k,i})\in\calE).
\ee

Due to the network being acyclic, we may form a pipelined block-Markov version of this code as follows.  Given integer $N$, we form a code with length $n(N+d)$ and rate $\frac{N}{N+d}\bR$. The outer blocklength $N$ serves a similar function as it did for network stacking, but here it represents the number of message blocks transmitted subsequently, rather than the number of stacks. Note that message $i$ consists of $NnR_i$ bits, which we denote $W_i(1),\ldots,W_i(N)$, each consisting of $nR_i$ bits. We then pipeline $N$ copies of the original code, encoding $n$-length blocks at a time. In particular, we introduce notation 
\begin{align}
X_j^{n(N+d)}&=(X_j^n(1),\ldots,X_j^n(N+d)),\\
Y_{i\to j}^{n(N+d)}&=(Y_{i\to j}^n(1),\ldots,Y_{i\to j}^n(N+d)).
\end{align}
Now, we define the coding operations at node $j$ by, for all $\ell\in[1:N]$,
\beq\label{eq:pipelining}
X_j^n(\ell+\pi_j)=\phi_j^n(W_j(\ell),Y_{i\to j}^n(\ell+\pi_i):(i,j)\in\calE).
\eeq
Recall that if $(i,j)\in\calE$, then $\pi_i<\pi_j$, meaning that the arguments of $\phi_j^n$ in \eqref{eq:pipelining} are causally available. Note that \eqref{eq:pipelining} does not specify all channel inputs, namely $X_j^n(\ell')$ for $\ell'\in[1:\pi_j]\cup[N+\pi_j+1:N+d]$; these channel inputs can be arbitrary, as the corresponding channel outputs will be ignored. To decode at node $i$, for all $\ell\in[1:N]$ let
\be
(\hatW_{ji}(\ell):i\in\calD_j)=\psi_i(W_i(\ell),Y_{k\to i}^n(\ell+\pi_k):(k,i)\in\calE).
\ee
Observe that the variables associated with a given index $\ell\in[1:N]$ associate only with themselves, and behave exactly like the original $n$-length code. Thus, an error occurs on this pipelined code if and only if any of the $N$ copies make an error, so the probability of error is
\be
1-(1-\eps)^N.
\ee
Thus we have
\be
\frac{N}{N+d}\,\bR\in\calR(\calN,\edit{n(N+d)},1-(1-\eps)^N).
\ee
Note that in this pipelined code, encoding operations are performed on $n$-length blocks at a time. Thus, the pipelined code on $\calN$ can be converted to one on a deterministic network using channel simulation codes. In particular, fix $\Delta\in(0,C_{\min})$ and let $\bar\calN_{\Delta}$ be the network of noiseless links where link $(i,j)$ is replaced by a noiseless link with capacity $C_{i\to j}+\Delta$. By Lemma~\ref{lemma:sim_error_exp}, for each link $(i,j)$ there exists a channel simulation code for link $(i,j)$ of rate $C_{i\to j}+\Delta$ and total variational distance at most $d^{(i\to j)}_n$, where $d^{(i\to j)}_n\to 0$ as $n\to\infty$. \edit{For each link $(i,j)\in\calE$,} we use $N$ copies of \edit{the associated} channel simulation code to simulate the behavior of link $(i,j)$ in network $\calN$ using the corresponding link on $\bar\calN_{\Delta}$. \edit{We analyze the impact on the overall probability of error from replacing these noisy channels by channel simulation codes as follows. Let $P_{\bX,\bY,\bW,\hat{\bW}}$ by the joint distribution of all channel inputs $\bX$, channel outputs $\bY$, messages $\bW$, and message estimates $\hat{\bW}$ for the pipelined code on noisy network $\calN$. Similarly, let $Q_{\bX,\bY,\bW,\hat{\bW}}$ be the joint distribution of the same random variables on the code on noiseless network $\bar\calN_{\Delta}$ constructed out of channel simulation codes. Note that in the latter, $\bX$ and $\bY$ are not real channel inputs and outputs, but rather simulated inputs and outputs that feed into the channel simulation codes, used to simulate noisy links with noiseless links. Since each channel simulation code used on an $n$-length block for link $(i,j)$ results in total variational distance at most $d^{(i\to j)}_n$, we may bound 
\be\label{eq:dtv_noisy_noiseless}
d_{\text{TV}}(P_{\bX,\bY,\bW,\hat{\bW}},Q_{\bX,\bY,\bW,\hat{\bW}})\le \sum_{(i,j)\in\calE} N d^{(i\to j)}_n.
\ee
The probability of error for the code on the noiseless network $\bar\calN_{\Delta}$ differs from that on the original noisy network by at most the quantity in \eqref{eq:dtv_noisy_noiseless}.
Because total variational distance is an upper bound on the difference in the probability of any event between the two distributions,} the probability of error of the resulting code on $\bar\calN_{\Delta}$ is at most
\be\label{eq:simulation_network_prob_error}
1-(1-\eps)^N+\sum_{(i,j)\in\calE} N d^{(i\to j)}_n\le 1-\frac{1}{2}(1-\eps)^N
\ee
where the inequality holds for sufficiently large $n$, since each sequence $d^{(i\to j)}_n$ vanishes with $n$. Recall that the channel simulation codes described in Lemma~\ref{lemma:sim_error_exp} employ common randomness $U$ between the transmitter and receiver of each link. Thus, a direct application of Lemma~\ref{lemma:sim_error_exp} implies only the existence of a code achieving the probability in \eqref{eq:simulation_network_prob_error} if nodes are allowed common randomness. However, we may treat this common randomness as a randomized codebook, and employ a usual random coding argument to show that there exists at least one deterministic code achieving \eqref{eq:simulation_network_prob_error}. Hence, for sufficiently large $n$,
\be\label{eq:simulation_network_rate}
\frac{N}{N+d}\,\bR\in\calR\left(\bar\calN_\Delta,\edit{n(N+d)},1-\frac{1}{2}(1-\eps)^N\right).
\ee
We now apply Lemma~\ref{lemma_det} on $\bar\calN_\Delta$, to find that for any $\tilde\eps>0$ and for sufficiently large $n$, we have
\be
\frac{N}{N+d}\,\bR\in \calR(\bar\calN_\Delta,\edit{n(N+d)},\tilde\eps,\eta(\tilde\eps\edit{,d})-\edit{3d}N\log(1-\eps)+\edit{3d})
\ee
\edit{where $\eta(\tilde\eps,d)$ is defined in \eqref{eq:eta_def}.}

Let $\bar\calN_{-\Delta}$ be the noiseless network where each link $(i,j)$ is replaced by a noiseless one with capacity $C_{i\to j}-\Delta$. By the assumption that $\Delta<C_{\min}$, we always have $C_{i\to j}-\Delta>0$. We may convert the code on $\bar\calN_\Delta$ to one on $\bar\calN_{-\Delta}$ by stretching each block of $n$ to one of length
\be
n'=\frac{C_{\min}+\Delta}{C_{\min}-\Delta}n.
\ee
Thus
\be
\frac{N}{N+d}\cdot \frac{C_{\min}-\Delta}{C_{\min}+\Delta}\,\bR\in \calR(\bar\calN_{-\Delta},\edit{n'(N+d)},\tilde\eps,\eta(\tilde\eps\edit{,d})-\edit{3d}N\log(1-\eps)+\edit{3d}).
\ee
Now we use ordinary noisy channel codes to convert this code back to one on $\calN$, again one block (now of length $n'$) at a time. For any $N$ and sufficiently large $n$, the probability of an error occurring on any of these channel codes can be made at most $\tilde\eps$. Thus we have
\be
\frac{N}{N+d}\cdot \frac{C_{\min}-\Delta}{C_{\min}+\Delta}\,\bR\in\calR(\calN,\edit{n'(N+d)},2\tilde\eps,\eta(\tilde\eps\edit{,d})-\edit{3d}N\log(1-\eps)+\edit{3d}).
\ee
As the above holds for any $\tilde\eps>0$, we may write
\begin{align}
\frac{N}{N+d}\cdot \frac{C_{\min}-\Delta}{C_{\min}+\Delta}\,\bR
&\in \bigcap_{\tilde\eps>0} \calC(\calN,\edit{(2\tilde\eps)_n,(\eta(\tilde\eps\edit{,d})-\edit{3d}N\log(1-\eps)+\edit{3d})_n})
\\&\subseteq \bigcap_{\tilde\eps>0}\  \overline{\bigcup_{k\in\bbN} \calC(\calN,\edit{(\tilde\eps)_n,(k)_n})}.
\end{align}
Since we may take $N$ to be arbitrarily large, and $\Delta$ arbitrarily small, and we chose $\bR$ to be any achievable vector with respect to $\eps$, by closure we have
\be\label{eq:applying_very_weak}
\calC(\calN,\edit{(\eps)_n})\subseteq  \bigcap_{\tilde\eps>0}\  \overline{\bigcup_{k\in\bbN} \calC(\calN,\edit{(\tilde\eps)_n,(k)_n})}.
\ee
By the equivalent form of the very weak edge removal property in \eqref{eq:very_weak_alternative2} of Proposition~\ref{prop:edge_removal_defs}, if very weak edge removal holds, then the \edit{RHS} of \eqref{eq:applying_very_weak} equals $\calC(\calN,0^+)$, so the strong converse holds.



\begin{thebibliography}{10}
\providecommand{\url}[1]{#1}
\csname url@samestyle\endcsname
\providecommand{\newblock}{\relax}
\providecommand{\bibinfo}[2]{#2}
\providecommand{\BIBentrySTDinterwordspacing}{\spaceskip=0pt\relax}
\providecommand{\BIBentryALTinterwordstretchfactor}{4}
\providecommand{\BIBentryALTinterwordspacing}{\spaceskip=\fontdimen2\font plus
\BIBentryALTinterwordstretchfactor\fontdimen3\font minus
  \fontdimen4\font\relax}
\providecommand{\BIBforeignlanguage}[2]{{%
\expandafter\ifx\csname l@#1\endcsname\relax
\typeout{** WARNING: IEEEtran.bst: No hyphenation pattern has been}%
\typeout{** loaded for the language `#1'. Using the pattern for}%
\typeout{** the default language instead.}%
\else
\language=\csname l@#1\endcsname
\fi
#2}}
\providecommand{\BIBdecl}{\relax}
\BIBdecl

\bibitem{ho10}
T.~Ho, M.~Effros, and S.~Jalali, ``On equivalence between network topologies,''
  in \emph{Proc.~Forty-Eighth Annual Allerton Conference}, Monticello, IL, Oct.
  2010.

\bibitem{jalali11}
S.~Jalali, M.~Effros, and T.~Ho, ``On the impact of a single edge on the
  network coding capacity,'' in \emph{Proc.~Information Theory and Applications
  Workshop (ITA)}, San Diego, CA, Feb. 2011, pp. 1--5.

\bibitem{lee13:_outer}
E.~J. Lee, M.~Langberg, and M.~Effros, ``Outer bounds and a functional study of
  the edge removal problem,'' in \emph{Proc.~IEEE Information Theory Workshop},
  Sevilla, Spain, Sep. 2013, pp. 1--5.

\bibitem{kamath11:_gener}
S.~U. Kamath, {D. N. C. Tse}, and V.~Anantharam, ``Generalized network sharing
  outer bound and the two-unicast problem,'' in \emph{Proc.~International
  Symposium on Network Coding (NetCod)}, Beijing, China, Jul. 2011.

\bibitem{yeung97}
R.~W. Yeung, ``A framework for linear information inequalities,'' \emph{IEEE
  Trans. Inf. Theory}, vol.~43, no.~6, pp. 1924--1934, Nov. 1997.

\bibitem{langberg11:_networ_codin}
M.~Langberg and M.~Effros, ``Network coding: {I}s zero error always possible?''
  in \emph{Proc.~Forty-Nine Annual Allerton Conference}, Monticello, IL, Sep.
  2011, pp. 1--8.

\bibitem{Chan2014}
T.~H. Chan and A.~Grant, ``Network coding capacity regions via entropy
  functions,'' \emph{IEEE Trans. Inf. Theory}, vol.~60, no.~9, pp. 5347--5374,
  Sept 2014.

\bibitem{Wong2013}
M.~F. Wong, M.~Langberg, and M.~Effros, ``On a capacity equivalence between
  network and index coding and the edge removal problem,'' in \emph{2013 IEEE
  International Symposium on Information Theory}, July 2013, pp. 972--976.

\bibitem{Noorzad2014}
P.~Noorzad, M.~Effros, M.~Langberg, and T.~Ho, ``On the power of cooperation:
  {Can} a little help a lot?'' in \emph{2014 IEEE International Symposium on
  Information Theory}, June 2014, pp. 3132--3136.

\bibitem{Noorzad2015}
P.~Noorzad, M.~Effros, and M.~Langberg, ``On the cost and benefit of
  cooperation,'' in \emph{2015 IEEE International Symposium on Information
  Theory (ISIT)}, June 2015, pp. 36--40.

\bibitem{Noorzad2016}
------, ``Can negligible cooperation increase network reliability?'' in
  \emph{2016 IEEE International Symposium on Information Theory (ISIT)}, July
  2016, pp. 1784--1788.

\bibitem{Noorzad2016a}
------, ``The unbounded benefit of encoder cooperation for the $k$-user
  {MAC},'' in \emph{2016 IEEE International Symposium on Information Theory
  (ISIT)}, July 2016, pp. 340--344.

\bibitem{Noorzad2018}
------, ``Can negligible rate increase network reliability?'' \emph{IEEE Trans.
  Inf. Theory}, vol.~64, no.~6, pp. 4282--4293, June 2018.

\bibitem{Noorzad2017a}
------, ``The benefit of encoder cooperation in the presence of state
  information,'' in \emph{2017 IEEE International Symposium on Information
  Theory (ISIT)}, 2017.

\bibitem{Langberg2016}
M.~Langberg and M.~Effros, ``On the capacity advantage of a single bit,'' in
  \emph{2016 IEEE Globecom Workshops (GC Wkshps)}, Dec 2016, pp. 1--6.

\bibitem{Gu2009}
W.~Gu, ``On achievable rate regions for source coding over networks,'' Ph.D.
  dissertation, California Institute of Technology, 2009.

\bibitem{Polyanskiy2010}
Y.~Polyanskiy, H.~V. Poor, and S.~Verd\'{u}, ``Channel coding rate in the
  finite blocklength regime,'' \emph{IEEE Trans. Inf. Theory}, vol.~56, pp.
  2307--2359, 2010.

\bibitem{Wolfowitz1957}
J.~Wolfowitz, ``The coding of messages subject to chance errors,''
  \emph{Illinois Journal of Mathematics}, vol.~1, no.~4, pp. 591--606, 1957.

\bibitem{Winter1999}
A.~Winter, ``Coding theorem and strong converse for quantum channels,''
  \emph{IEEE Trans. Inf. Theory}, vol.~45, no.~7, pp. 2481--2485, Nov 1999.

\bibitem{Ogawa1999}
T.~Ogawa and H.~Nagaoka, ``Strong converse to the quantum channel coding
  theorem,'' \emph{IEEE Trans. Inf. Theory}, vol.~45, no.~7, pp. 2486--2489,
  Nov 1999.

\bibitem{Fong2017}
S.~L. Fong and V.~Y.~F. Tan, ``Strong converse theorems for discrete memoryless
  networks with tight cut-set bound,'' in \emph{2017 IEEE International
  Symposium on Information Theory (ISIT)}, June 2017, pp. 933--937.

\bibitem{Arimoto1973}
S.~Arimoto, ``On the converse to the coding theorem for discrete memoryless
  channels (corresp.),'' \emph{IEEE Trans. Inf. Theory}, vol.~19, no.~3, pp.
  357--359, May 1973.

\bibitem{Dueck1979}
G.~Dueck and J.~Korner, ``Reliability function of a discrete memoryless channel
  at rates above capacity (corresp.),'' \emph{IEEE Trans. Inf. Theory},
  vol.~25, no.~1, pp. 82--85, Jan 1979.

\bibitem{Oohama2015a}
Y.~Oohama, ``Strong converse exponent for degraded broadcast channels at rates
  outside the capacity region,'' in \emph{2015 IEEE International Symposium on
  Information Theory (ISIT)}, June 2015, pp. 939--943.

\bibitem{Oohama2015b}
------, ``Exponent function for one helper source coding problem at rates
  outside the rate region,'' in \emph{2015 IEEE International Symposium on
  Information Theory (ISIT)}, June 2015, pp. 1575--1579.

\bibitem{Oohama2016}
------, ``Exponent function for asymmetric broadcast channels at rates outside
  the capacity region,'' in \emph{2016 International Symposium on Information
  Theory and Its Applications (ISITA)}, Oct 2016, pp. 537--541.

\bibitem{Oohama2016a}
------, ``Exponent function for {Wyner-Ziv} source coding problem at rates
  below the rate distortion function,'' in \emph{2016 International Symposium
  on Information Theory and Its Applications (ISITA)}, Oct 2016, pp. 171--175.

\bibitem{Marton1986}
K.~Marton, ``A simple proof of the blowing-up lemma (corresp.),'' \emph{IEEE
  Trans. Inf. Theory}, vol.~32, no.~3, pp. 445--446, May 1986.

\bibitem{ElGamalKim:11Book}
A.~El~Gamal and Y.~Kim, \emph{Network Information Theory}.\hskip 1em plus 0.5em
  minus 0.4em\relax Cambridge University Press, 2011.

\bibitem{Gu2008}
W.~Gu, M.~Effros, and M.~Bakshi, ``A continuity theory for lossless source
  coding over networks,'' in \emph{2008 46th Annual Allerton Conference on
  Communication, Control, and Computing}, Sept 2008, pp. 1527--1534.

\bibitem{Langberg2012}
M.~Langberg and M.~Effros, ``Source coding for dependent sources,'' in
  \emph{Information Theory Workshop (ITW), 2012 IEEE}, Sept 2012, pp. 70--74.

\bibitem{Ahlswede1976}
R.~Ahlswede, P.~G\'{a}cs, and J.~K\"{o}rner, ``Bounds on conditional
  probabilities with applications in multi-user communication,'' \emph{Z.
  Wahrscheinlichkeitstheorie verw. Gebiete}, vol.~34, pp. 157--177, 1976.

\bibitem{Raginsky2013}
M.~Raginsky and I.~Sason, ``Concentration of measure inequalities in
  information theory, communications, and coding,'' \emph{Foundations and
  Trends in Communications and Information Theory}, vol.~10, no. 1-2, pp.
  1--246, 2013.

\bibitem{Strassen1965}
V.~Strassen, ``The existence of probability measures with given marginals,''
  \emph{Ann. Math. Statist.}, vol.~36, pp. 423--439, 1965.

\bibitem{Koetter2011}
R.~Koetter, M.~Effros, and M.~M\'edard, ``A theory of network
  equivalence---{Part I}: Point-to-point channels,'' \emph{IEEE Trans. Inf.
  Theory}, vol.~57, no.~2, pp. 972--995, 2011.

\bibitem{Cover1979}
T.~Cover and A.~E. Gamal, ``Capacity theorems for the relay channel,''
  \emph{IEEE Trans. Inf. Theory}, vol.~25, no.~5, pp. 572--584, September 1979.

\bibitem{ElGamal1981}
A.~El~Gamal, ``On information flow in relay networks,'' in \emph{Proc. IEEE
  National Telecomm. Conf.}, vol.~2, New Orleans, LA, Nov. 1981, pp.
  D4.1.1--D4.1.4.

\bibitem{ElGamal1982}
A.~El~Gamal and M.~Aref, ``The capacity of the semideterministic relay channel
  (corresp.),'' \emph{IEEE Trans. Inf. Theory}, vol.~28, no.~3, pp. 536--536,
  May 1982.

\bibitem{Avestimehr2011}
A.~S. Avestimehr, S.~N. Diggavi, and D.~N.~C. Tse, ``Wireless network
  information flow: A deterministic approach,'' \emph{IEEE Trans. Inf. Theory},
  vol.~57, no.~4, pp. 1872--1905, April 2011.

\bibitem{Oohama2015}
Y.~Oohama, ``Strong converse theorems for degraded broadcast channels with
  feedback,'' in \emph{2015 IEEE International Symposium on Information Theory
  (ISIT)}, June 2015, pp. 2510--2514.

\bibitem{Liu2016}
J.~Liu, T.~A. Courtade, P.~Cuff, and S.~Verd\'u, ``Smoothing {Brascamp-Lieb}
  inequalities and strong converses for common randomness generation,'' in
  \emph{2016 IEEE International Symposium on Information Theory (ISIT)}, July
  2016, pp. 1043--1047.

\bibitem{Ahlswede1998}
R.~Ahlswede and I.~Csisz\'{a}r, ``Common randomness in information theory and
  cryptography. {II. CR} capacity,'' \emph{IEEE Trans. Inf. Theory}, vol.~44,
  no.~1, pp. 225--240, Jan 1998.

\bibitem{Sato1981}
H.~Sato, ``The capacity of the {Gaussian} interference channel under strong
  interference (corresp.),'' \emph{IEEE Trans. Inf. Theory}, vol.~27, no.~6,
  pp. 786--788, Nov 1981.

\bibitem{Costa1987}
M.~Costa and A.~El~Gamal, ``The capacity region of the discrete memoryless
  interference channel with strong interference (corresp.),'' \emph{IEEE Trans.
  Inf. Theory}, vol.~33, no.~5, pp. 710--711, Sep 1987.

\bibitem{Le2015}
S.~Q. Le, V.~Y.~F. Tan, and M.~Motani, ``A case where interference does not
  affect the channel dispersion,'' \emph{IEEE Trans. Inf. Theory}, vol.~61,
  no.~5, pp. 2439--2453, May 2015.

\bibitem{Fong2016a}
S.~L. Fong and V.~Y.~F. Tan, ``A proof of the strong converse theorem for
  {Gaussian} multiple access channels,'' \emph{IEEE Trans. Inf. Theory},
  vol.~62, no.~8, pp. 4376--4394, Aug 2016.

\bibitem{Dueck1981}
G.~Dueck, ``The strong converse to the coding theorem for the multiple-access
  channel,'' \emph{J. Combinat., Inf. Syst. Sci}, vol.~6, no.~3, pp. 187--196,
  1981.

\bibitem{Ahlswede1982}
R.~Ahlswede, ``An elementary proof of the strong converse theorem for the
  multiple access channel,'' \emph{J. Combinat., Inf. Syst. Sci.}, vol.~7,
  no.~3, pp. 216--230, 1982.

\bibitem{Roth2006}
R.~Roth, \emph{Introduction to Coding Theory}.\hskip 1em plus 0.5em minus
  0.4em\relax New York, NY, USA: Cambridge University Press, 2006.

\bibitem{Gallager1968}
R.~G. Gallager, \emph{Information Theory and Reliable Communication}.\hskip 1em
  plus 0.5em minus 0.4em\relax New York, NY, USA: John Wiley \& Sons, Inc.,
  1968.

\bibitem{CoverThomas:Book}
T.~M. Cover and J.~Thomas, \emph{{Elements of Information Theory}}.\hskip 1em
  plus 0.5em minus 0.4em\relax John Wiley, 1991.

\bibitem{Borade2008}
S.~Borade and L.~Zheng, ``Euclidean information theory,'' in \emph{2008 IEEE
  International Zurich Seminar on Communications}, March 2008, pp. 14--17.

\bibitem{Xiang2014}
Y.~Xiang and Y.-H. Kim, ``A few meta-theorems in network information theory,''
  in \emph{Information Theory Workshop (ITW), 2014 IEEE}, Nov 2014, pp. 77--81.

\bibitem{Kosut2017a}
O.~Kosut and J.~Kliewer, ``Equivalence for networks with adversarial state,''
  \emph{IEEE Trans. Inf. Theory}, vol.~63, no.~7, pp. 4137--4154, July 2017.

\bibitem{Yeung2008}
R.~W. Yeung, \emph{Information Theory and Network Coding}, 1st~ed.\hskip 1em
  plus 0.5em minus 0.4em\relax Springer Publishing Company, Incorporated, 2008.

\end{thebibliography}
\end{document}